\documentclass[aps,prd,longbibliography]{revtex4-2}
\usepackage{graphicx}
\usepackage{dcolumn}
\usepackage{amsmath,amssymb,epsfig}
\usepackage{paralist}
\usepackage{comment} 
\usepackage{siunitx}
\usepackage{graphicx}
\usepackage{multirow}
\usepackage{color,soul}
\usepackage{suffix}
\usepackage{mathtools}
\usepackage{tabularx}
\usepackage{tabulary}
\usepackage{booktabs,makecell}
\usepackage[caption=false]{subfig} 

\usepackage{pgfplots}  
\pgfplotsset{compat=1.18}

\pgfmathsetmacro{\etaZero}{0.25}
\pgfmathsetmacro{\etaInf}{0.98}
\pgfmathsetmacro{\Fc}{0.10}
\pgfmathsetmacro{\pexp}{1.4}

\pgfmathsetmacro{\betaDense}{1.5}
\pgfmathsetmacro{\betaAper}{2.5}
\pgfmathsetmacro{\betaSparse}{3.5}

\usepackage[breaklinks=True,colorlinks=True, linkcolor=magenta,citecolor=blue,debug=True]{hyperref}

\usepackage{wasysym}

\renewcommand{\vec}[1]{\boldsymbol{\mathrm{#1}}}

\def\beqn{\begin{eqnarray*}}
\def\eeqn{\end{eqnarray*}}

\newcommand{\be}{\begin{equation}}
\newcommand{\ee}{\end{equation}}
\newcommand{\ba}{\begin{eqnarray}}
\newcommand{\ea}{\end{eqnarray}}

\begin{document}

\title{Phased-Array Laser Power Beaming from Cislunar Space to the Lunar Surface}

\author{Slava G. Turyshev}   

\affiliation{ 
Jet Propulsion Laboratory, California Institute of Technology,\\
4800 Oak Grove Drive, Pasadena, CA 91109-0899, USA
}

\begin{abstract}

We present a time-dependent, end-to-end framework for laser power beaming from cislunar orbits to the lunar surface. The model links on-orbit generation (solar arrays and wall-plug to optical), terrain-masked visibility and range, beam propagation with realistic divergence and jitter, and surface conversion with thermal and dust limits, returning delivered daily energy. Baseline loads for early polar activities (habitat survival, mobility, comm/nav, pilot ISRU) set target Wh\,day$^{-1}$ and are used consistently in scaling laws and design maps. A near-rectilinear halo orbit (NRHO) to a Shackleton-rim site provides a worked example: for a 2\,m-class phased array at 1064\,nm the reference geometry yields $\sim$0.6--0.8\,kWh\,day$^{-1}$ to a 1\,m$^2$ receiver (about 28\,W averaged over the day). We place this result in context by comparing on the same daily-energy metric to surface photovoltaics (PV) with storage and to compact fission, and by showing how delivered energy scales nearly linearly with transmitted power and as $D_{\rm eff}^{2}$ via encircled-energy capture, with a multiplicative gain from visibility (constellations). The same framework indicates practical regimes already within reach: e.g., a 10\,m effective-aperture optical phased array at $P_{\rm tx}=100$\,kW delivers $\sim$30--50\,kWh\,day$^{-1}$ at polar sites with typical single-orbiter visibility, as quantified by the delivered-energy and sizing maps. Thus, laser beaming is mass-competitive where darkness or permanent shadow forces deep storage for PV, or where distributed and duty-cycled users can amortize a shared transmitter; compact fission retains advantage for continuous multi-kW baseload at fixed sites.

\end{abstract}

\date{\today}
\maketitle

\tableofcontents

\section{Introduction}
\label{sec:Intro}

Ensuring continuous, reliable power on the lunar surface remains a primary engineering challenge for both near-term exploration and long-duration missions  \cite{NASA_ArtemisPlan_2020}. A single lunar night\footnote{Here we use “lunar day” to mean the synodic period $T_{\mathrm{syn}}\approx 29.53$ d; the “lunar night” is approximately $T_{\mathrm{syn}}/2\approx 14.77$ d.} at mid-latitudes spans roughly 14.77 Earth days (i.e.,  $\sim$\,354.48 hours), and permanently shadowed regions (PSRs) near the poles receive little or no direct sunlight \cite{Bussey:1999}. Conventional surface-based solutions, such as photovoltaic (PV) arrays paired with large-scale batteries, must carry extensive energy storage to bridge these multi-week gaps, causing significant mass and cost overhead \cite{McKinney2025_SBSP}. Nuclear systems provide continuous power, but different technologies entail distinct trade-offs: fission reactors deliver high power densities at the expense of substantial development complexity, mass, and stringent radiological‐safety controls, whereas radioisotope thermoelectric generators (RTGs) offer lower continuous power with relatively simple, lightweight designs and well-understood radiological‐safety protocols \cite{McClure2019Kilopower}.

A potentially transformative alternative would be to locate one or more power-generation spacecraft in lunar orbit, where continuous or near-continuous solar illumination is available. These spacecraft would convert incident solar radiation ($\sim$\SI{1361}{W/m^2} near the Earth-Moon system) into electrical power, then beam it to the lunar surface via high-efficiency lasers \cite{Shaik2015LaserBeaming, Jaffe2013SpaceSolar,Turyshev-LLR-CW:2025}. By choosing orbits with minimal eclipses, such as low lunar polar or near-rectilinear halo orbits (NRHOs), the spacecraft could harness sunlight almost uninterruptedly \cite{Montenbruck:2000}. On the surface, receivers equipped with specialized PV cells or other photonic-to-electrical conversion methods could reliably produce usable electrical power \cite{Gou2022OE1064}, mitigating the need for massive on-site energy storage. Such systems could be scaled to support diverse applications, ranging from robotic prospecting and ice extraction \cite{Colaprete2010Water} in polar PSRs to habitat operations during the extended lunar night. For context, representative, order-of-magnitude loads for early polar activities span habitat survival (1--3\,kW; 24--72\,kWh/day), local comm/nav nodes (50--200\,W; 1.2--4.8\,kWh/day), rover traverse (50--300\,W while driving with duty 0.2--0.5; 0.24--3.6\,kWh/day), and pilot ISRU (0.5--5\,kW with duty 0.1--0.5; 1.2--60\,kWh/day). Table~\ref{tab:baseline_loads} provides a summary of baseline loads for early polar activities and daily-energy targets.

The basic concept of space-based collection and beaming of solar power traces back to Glaser's seminal proposal of solar power satellites \cite{Glaser1968Science}. Extensive microwave wireless-power transmission heritage \cite{Brown1984MTT} established the beaming paradigm and safety frameworks later adopted by optical wireless power transfer (OWPT). Recent NASA and agency studies have revisited space-based solar power and cislunar infrastructure planning \cite{NASA_SBSP_2024,ipnpr_pnt_2024,Turyshev-2025-GNSS-Moon}, while device-level progress in laser-PV converters supports $>40\%$ wall-to-electric efficiency at 1.06--1.55\,$\mu$m under multi-kW\,m$^{-2}$ irradiance. 

Optical power beaming for lunar operations has been studied in surface--to--surface configurations (crater-rim $\rightarrow$ PSRs) and orbit--to--surface links. Foundational PSR link budgets and rover-support concepts are presented in \cite{Landis2020LunarLaser,Grandidier2021LunarPSR}; orbiting beamcraft and relay/tower architectures are analyzed in \cite{OlesonLandis2024OrbitBeaming,Oleson2025BEACON} and earlier mission studies~\cite{Brandhorst2009Acta,Takeda2002SPIE}. In parallel, microwave power beaming provides decades of heritage~\cite{Brown1984MicrowaveReview,Jaffe2013SpaceSolar}, while subsystem maturity and safety framing for OWPT and $\mu$rad-class beam control are reviewed in \cite{Mohsan2023OWPT,NRL2023SWELL,JAXA_LSSPS}. Surveys and demonstrations indicate end-to-end performance with $\mu$rad-class acquisition, tracking, and pointing (ATP)~\cite{Kaymak2018Comst,Rodenbeck2021JMW}. Device-level photovoltaic conversion at 1064\,nm with efficiencies $\gtrsim 40$--$44\%$ under multiwatt CW irradiance supports the receiver assumptions adopted here~\cite{Gou2022OE1064,Algora2022Joule,Landis1993JPP}. Environmental constraints relevant to sustained OWPT---radiation effects, dust optical properties, and deposition/lofting rates---bound optical throughput and maintenance cadence~\cite{Srour:2003,Messenger:2015,EscobarCerezo2018ApJS,Gaier2005,Katzan1991,ChangE3Dust2019,Immer2011,Szalay2015GRL,NASA_OTPS_SBSP_2024}.

\begin{table*}[t]
\caption{Illustrative baseline loads for early polar activities and daily-energy targets. Values are order-of-magnitude bands to anchor delivered-energy maps; exact loads are mission-specific.}
\label{tab:baseline_loads}
\centering
\begin{tabular}{lcc}
\hline
Activity & Continuous/effective power & $E_{24\mathrm{h}}$ target \\
\hline\hline
Habitat survival (thermal + life support) & 1--3 kW & 24--72 kWh \\
Rover traverse (mobility + avionics) & 50--300 W (duty~0.2--0.5) & 0.24--3.6 kWh \\
Local comm/nav node (relay/beacon) & 50--200 W & 1.2--4.8 kWh \\
Pilot ISRU (drill, volatile handling) & 0.5--5 kW (duty~0.1--0.5) & 1.2--60 kWh \\
\hline
\end{tabular}
\end{table*}

Building on these efforts, we present an end-to-end, time-dependent model of cislunar laser power beaming. The framework couples five effects that set delivered power at the surface: (i) diffraction and array fill factor, (ii) loss of optical concentration from residual phase and pointing errors, (iii) attenuation by lofted and deposited dust, (iv) receiver conversion efficiency, and (v) thermal-rejection constraints across representative LLO, NRHO, and L1 geometries. For any specified transmitter/receiver configuration and orbit, the model produces time-resolved surface irradiance and electrical-power profiles, daily energy limited by contact duty cycle, and delivered-energy maps. To avoid overestimation, captured power is conservatively capped by the geometric overlap between the beam footprint and the receiver aperture. These outputs enable quantitative trades over transmitter aperture and wavelength, pointing stability, receiver area, and orbital parameters, providing concrete design guidance without optimistic assumptions.

This paper makes five contributions: (1) a time-domain, end-to-end link from solar array to net electrical energy at a specified site with terrain-masked visibility; (2) a bookkeeping-clean optical link that uses Gaussian encircled-energy capture $\eta_{\mathrm{cap}}$ with jitter-broadened $w_{\mathrm{eff}}$ and separates residual pointing losses, avoiding double counting; (3) a bounded main-lobe efficiency model $\eta_{\mathrm{main}}(F)$ that connects phased-array fill factor to an effective $M^2_{\mathrm{eff}}$; (4) a PSD-based jitter budget with explicit FSM/AOCS loop shaping; and (5) parametric delivered-energy maps and a terrain-masked NRHO$\to$South Pole example that quantify daily Wh and constellation scaling. A quantitative comparison with surface photovoltaics plus storage and with compact fission is given later on the same daily-energy metric, so that design choices for the beaming link can be read directly against alternative provisioning at a specific site.

The paper is organized as follows: Section~\ref{sec:spacecraft_power} details on-orbit solar power generation, including the solar-irradiance model with eclipse geometry, photovoltaic conversion and balance-of-plant losses, and the spacecraft power budget that sets available laser power. Section~\ref{sec:orbit_geometry} explores orbital dynamics and visibility, defining the relevant orbital elements, constructing spacecraft and surface-receiver state vectors, and deriving slant range, line-of-sight and terrain masking, coverage/repeat ground tracks, and contact-window statistics for power delivery. Section~\ref{sec:laser_link} develops the optical transmission link budget, from diffraction-limited divergence and far-field spot size to pointing/Strehl penalties and time-varying surface power flux. Section~\ref{sec:phased_arrays} focuses on phased-array transmitter architectures, quantifying effective-aperture scaling via the array factor and outlining coherent sub-aperture beamforming requirements and resulting end-to-end efficiency gains. Section~\ref{sec:receiver_model} examines surface reception and conversion, including beam--receiver overlap and geometric capture, conversion efficiency, thermal constraints, and dust considerations. Section~\ref{sec:summary_equations} consolidates the governing equations and scalings for end-to-end power delivery. Finally, Section~\ref{sec:conclusion} discusses scalability, trade-offs, and mission-level implications. The Appendices provide detailed numerical case studies for three representative scenarios: a low lunar orbiter  (LLO) (Appendix~\ref{sec:illustrative_example}), a spacecraft at the Earth--Moon L1 point (Appendix~\ref{sec:illustrative_example_L1}), and a near-rectilinear halo orbit (NRHO) illuminating a south-polar site (Appendix~\ref{sec:illustrative_example_nrho}). 

\section{Power Generation Aboard the Spacecraft}
\label{sec:spacecraft_power}

\subsection{Incident Solar Power}
\label{subsec:incident_solar_power}

Solar power generation in lunar orbit benefits from stable and intense solar irradiance, $I_{\text{sun}}$, which is measured near Earth at approximately:
\begin{equation}
    I_{\text{sun}} \approx 1361 \, \text{W/m}^2,
\end{equation}
with variations of $\pm 3\%$ resulting from Earth's orbital eccentricity and fluctuations in Earth-Moon distance \cite{Moe:2000, Kopp2017}. This stability simplifies spacecraft power system design by eliminating interruptions caused by atmospheric or weather-related factors encountered on planetary surfaces.

The total incident solar power, $P_{\text{incident}}$, on the spacecraft's solar arrays is determined by the array area, $A_{\text{SA}}$, and the angle of incidence, $\theta_{\text{inc}}$, which is defined as the angle between the incoming sunlight and the normal to the array:
\begin{equation}
    P_{\text{incident}} = I_{\text{sun}} A_{\text{SA}} \cos \theta_{\text{inc}}.
\end{equation}
The $\cos \theta_{\text{inc}}$ factor accounts for the effective projection area of the solar array under varying spacecraft orientations.

The photovoltaic efficiency, $\eta_{\text{pv}}$, defines the fraction of incident solar power converted into electrical power. The net electrical power output is expressed as:
\begin{equation}
    P_{\text{array}} = I_{\text{sun}} A_{\text{SA}} \eta_{\text{pv}} \cos \theta_{\text{inc}}.
\end{equation}

Modern triple-junction photovoltaic cells achieve efficiencies between 28\% and 33\%, with experimental designs reaching up to 35\% under concentrated illumination \cite{NREL:2019}. These cells, typically based on III-V semiconductors (e.g., GaAs, InGaP), are highly suitable for space applications due to their radiation resistance and thermal stability. However, operational challenges such as degradation from radiation damage, thermal cycling, and manufacturing imperfections can reduce their long-term performance \cite{Srour:2003, Messenger:2015}.

Variations in $\theta_{\text{inc}}$ due to orbital dynamics and spacecraft attitude require active control mechanisms to optimize solar power generation. Periodic eclipses caused by the Earth or Moon necessitate energy storage systems to bridge power gaps. Over the 27.3-day lunar orbital period, spacecraft systems must account for instantaneous variations and compute time-averaged $P_{\text{array}}$ to ensure consistent power availability.

\subsection{Spacecraft Power Budget and Laser Power}
\label{subsec:spacecraft_power_budget}

The spacecraft's total power budget includes consumption by ancillary systems such as thermal management, attitude control, and communications. The baseline power required for these systems is denoted as $P_{\text{SC,aux}}$. The power available for laser transmission is given by:
\begin{equation}
    P_{\text{available}} = P_{\text{array}} - P_{\text{SC,aux}}.
\end{equation}

Conversion of electrical power to optical power involves efficiency losses encapsulated in the electrical-to-laser conversion efficiency, $\eta_{\ell}$. The resulting laser output power is:
\begin{equation}
    P_{\text{tx}} = P_{\text{available}} \, \eta_{\ell}
    \equiv \Big(P_{\text{array}} - P_{\text{SC,aux}}\Big)\, \eta_{\ell}.
    \label{eq:p_tx}
\end{equation}

State-of-the-art diode and fiber lasers achieve wall-plug efficiencies of 30\% to 40\% at the system level, with laboratory demonstrations exceeding 50\% under ideal conditions \cite{Hitz:2012}. However, system-level inefficiencies arise due to thermal management, driver electronics, and other operational constraints. GaAs-based diode lasers offer a promising solution for space applications, providing compact size, high thermal tolerance, and robust performance under radiation exposure \cite{Soref:2010}.

Energy storage systems, such as lithium-ion batteries or supercapacitors, play a critical role in maintaining consistent laser output during periods of reduced solar power availability, such as during eclipses. These systems allow spacecraft to sustain operations even when $P_{\text{array}}$ is less than $P_{\text{SC,aux}}$.

\subsection{Baseline lunar power needs and daily-energy targets}

We adopt representative, order-of-magnitude loads for early polar activities to anchor delivered-energy targets (also summarized in Table~\ref{tab:baseline_loads}). Habitat survival (thermal control plus essential life support) is $\mathcal{O}$(1--3) kW continuous; mobile robotics are $\mathcal{O}$(50--300)~W while driving with burst loads for actuation and thermal control; local comm/nav beacons and relay nodes are $\mathcal{O}$(50--200) W; and pilot ISRU (e.g., drilling, volatile extraction, water handling) spans $\mathcal{O}$(0.5--5)\,kW depending on duty cycle. For a 24\,h reference window, a continuous load $P_{\mathrm{cont}}$ corresponds to a daily energy target $E_{24\mathrm{h}} \approx 24\,\mathrm{h}\times P_{\mathrm{cont}}$, while intermittent users can be satisfied by $E_{24\mathrm{h}}$ delivered in short passes and buffered locally. (For brevity we use $E_{24h}\equiv E_{\rm day}$ throughout.) These targets are used consistently in the sizing maps and the polar case study to translate transmitter/receiver choices into operationally meaningful outcomes (e.g., Wh\,day$^{-1}$ that map to hours of rover traverse, or habitat survival margins during lunar night).

\subsection{When surface PV+storage or compact fission is competitive}

The beaming architecture is competitive when the mass and complexity to deliver $E_{24\mathrm{h}}$ at the site beat (i) surface PV sized for the mean insolation and storage sized for the site’s night/PSR duty or (ii) compact fission sized for $P_{\mathrm{cont}}$. In practice the crossover depends on four scalars: range to the transmitter, effective transmit aperture, receiver area, and the realized visibility fraction over the reference window. Increasing effective aperture and visibility tends to push the crossover in favor of beaming at polar sites (especially PSRs), whereas longer range, small receivers, and poor visibility favor local PV with storage or fission. The parametric maps and the polar case study quantify these trends by reporting delivered Wh\,day$^{-1}$; dividing by 24\,h yields the continuous-equivalent power that can be compared directly to a habitat or ISRU load. In the following sections, we use $P_{\text{tx}}$ from \eqref{eq:p_tx} in the link budget and reception models to determine how much of this power ultimately arrives at a user station on the lunar surface.

To make the comparison explicit we parameterize the alternatives on the same daily-energy scale $E_{\rm day}$. For PV with storage the mass burden can be written
\begin{equation}
M_{\mathrm{PV+bat}}
\simeq \frac{E_{\rm day}}{\eta_{\mathrm{PV}}\,H_{\mathrm{site}}}\,m_{\mathrm{PV}}
\;+\;\Bigl(\frac{\Delta E}{\mathrm{kWh}}\Bigr)m_{\mathrm{bat}}
\;+\;M_{\mathrm{BOP}},
\label{eq:mpv}
\end{equation}
where $\eta_{\mathrm{PV}}$ is panel efficiency, $H_{\mathrm{site}}$ is the site’s daily insolation (in the same energy units as $E_{\mathrm{day}}$, e.g., kWh\,m$^{-2}$\,day$^{-1}$), $m_{\mathrm{PV}}$ and $m_{\mathrm{bat}}$ are the areal panel mass (kg\,m$^{-2}$) and the storage specific mass (kg\,kWh$^{-1}$), $\Delta E$ is the energy required to bridge darkness, and $M_{\mathrm{BOP}}$ collects balance-of-plant. For compact fission a fair first proxy is the deployed specific mass $\alpha_{\mathrm{fis}}$ (kg\,kW$^{-1}$), giving
\begin{equation}
M_{\mathrm{fis}}\simeq \alpha_{\mathrm{fis}}\;P_{\mathrm{eq}},
\qquad
P_{\mathrm{eq}}=\frac{E_{\rm day}}{24\,\mathrm{h}}.
\label{eq:mfis}
\end{equation}
These relations are used purely as decision aids: where $\Delta E$ is large (night/PSR) the storage term dominates Eq.~(\ref{eq:mpv}); where continuous multi-kW baseload is required at a fixed site, Eq.~(\ref{eq:mfis}) tends to favor a reactor; between these regimes the beaming link is sized by $(D_{\rm eff},P_{\rm tx},A_{\rm rx})$ and by the visibility fraction.

For site-specific decisions, one reads the required $E_{\rm day}$ directly from the delivered-energy maps (Sec.~\ref{sec:receiver_model} and Fig.~\ref{fig:parametric-2x2}) and substitutes it into Eqs.~(\ref{eq:mpv})--(\ref{eq:mfis}) to compare mass on the same $P_{\rm eq}=E_{\rm day}/(24\,\mathrm{h})$ basis. This makes the choice among LPB, PV+storage, and compact fission an explicit function of $(D_{\rm eff},P_{\rm tx},A_{\rm rx},N)$ and the site’s visibility fraction.

\section{Orbit Geometry and Link Visibility}
\label{sec:orbit_geometry}

This section analyzes the orbital geometry and link visibility for lunar power beaming missions using a lunicentric coordinate reference system (LCRS) \cite{Turyshev-etal:2024,Turyshev-scales:2025}. Time-dependent variations in orbital parameters, driven by gravitational perturbations, lunar libration, and relative motion between the spacecraft and receiver, must be incorporated to accurately model power delivery efficiency and link visibility.

\subsection{Orbital Parameters and Time Variability}
\label{subsec:orbital_parameters}

A spacecraft's orbit is described by classical Keplerian elements, which evolve over time in the LCRS (i.e., lunicentric frame) due to perturbations caused by the Moon’s irregular gravitational field, third-body interactions, and libration. Table~\ref{tab:orbital_elements} summarizes the orbital elements and their sources of variability.

\begin{table}[h!]
\centering
\caption{Orbital Elements and Time Variability in a Lunicentric Frame.}
\label{tab:orbital_elements}
\begin{tabular}{ll}
\hline
{ Orbital Element} & { Time-Dependent Factors} \\
\hline\hline
Semi-major axis, $a(t)$ & Variations due to mascons and tidal forces altering orbital energy. \\
Eccentricity, $e(t)$ & Changes from perturbations in the Moon’s non-uniform gravity. \\
Inclination, $i(t)$ & Precession driven by the Moon's oblateness and third-body interactions. \\
RAAN, $\Omega(t)$ & Regresses over time due to lunar gravitational harmonics (e.g., $J_2$). \\
Argument of periapsis, $\omega(t)$ & Rotates due to perturbations, including mascons and tidal effects. \\
True anomaly, $\nu(t)$ & Instantaneous position of the spacecraft along its orbit. \\
\hline
\end{tabular}
\end{table}

The dynamic evolution of these parameters directly influences the power beaming system by affecting visibility windows, slant range, and pointing accuracy. For example, variations in $\Omega(t)$ and $\omega(t)$ alter the orientation of the spacecraft’s orbit relative to the receiver, while changes in $a(t)$ and $e(t)$ modify the slant range.

Table~\ref{tab:representative_orbits} summarizes common orbital configurations used for power beaming missions, highlighting their altitude ranges, key features, and approximate orbital periods. NRHOs are particularly well-suited for providing continuous power delivery to polar regions, such as permanently shadowed craters. In contrast, LLOs offer frequent revisits but shorter visibility windows, requiring precise beam steering during each pass.

\begin{table}[h!]
\centering
\caption{Representative Lunar Orbits and Configurations for Power Beaming Missions.}
\label{tab:representative_orbits}
\begin{tabular}{p{4.0cm}|p{3.0cm}|p{6.3cm}|p{2.5cm}}
\hline
{Configuration} & {Altitude/Distance} & {Key Features for Power Beaming} & {Orbital Period} \\
\hline\hline
Low Lunar Orbit (LLO) & 100--200\,km
& Very high revisit frequency; short but frequent visibility windows
& $\sim$2 hours \\
\hline
Near-Rectilinear Halo Orbit (NRHO) & 3,000--70,000\,km (from Moon's center) & 
Long continuous visibility, especially for polar sites; minimal eclipses
& $\sim$6.5 days \\
\hline
Highly Elliptical Orbit (HEO) & 500--10,000\,km
& Extended dwell time over target at apoapsis; longer visibility per pass
& 7--15 hours \\
\hline
Polar Orbit & 100--300\,km 
& Near-global coverage over multiple passes; good polar access
& 2--3 hours \\
\hline
Transmitter at Earth--Moon L1 & $\sim$58,000\,km  & 
Continuous solar illumination; stable line-of-sight geometry to poles; diffraction-limited at long range without large apertures
& Quasi-static orbit (halo/Lissajous, Earth--Moon L1) \\
\hline
\end{tabular}
\end{table}

\subsection{Orbital Dynamics and Perturbations}
\label{subsec:orbital_dynamics}

The evolution of orbital elements in a lunicentric frame is governed by perturbation theory. Key effects include:
\begin{enumerate}
\item \textit{Precession of RAAN ($\Omega$):}
   \begin{equation}
       \frac{\mathrm{d}\Omega}{\mathrm{d}t} = -\frac{3}{2} J_2\,n(t)\,\Big(\frac{R_{\text{moon}}}{a(t)}\Big)^2
       \frac{\cos i(t)}{\bigl(1 - e(t)^2\bigr)^2},
   \end{equation}
   where $n(t)=\sqrt{\mu_{\text{moon}}/a(t)^3}$ is the mean motion, $\mu_{\text{moon}}=GM_{\text{moon}}$ is the lunar gravitational constant and $J_2$ is the Moon's second zonal harmonic.

\item \textit{Rotation of the Argument of Periapsis ($\omega$):}
   \begin{equation}
       \frac{\mathrm{d}\omega}{\mathrm{d}t} = \frac{3}{4} J_2\,n(t)\,\Big(\frac{R_{\text{moon}}}{a(t)}\Big)^2
       \frac{\left(4 - 5 \sin^2 i(t)\right)}{\bigl(1 - e(t)^2\bigr)^2}.
   \end{equation}

\item \textit{Orbital Period:}
$ T = 2\pi \sqrt{{a(t)^3}/{\mu_{\text{moon}}}},$
   where $T$ is affected by changes in $a(t)$ due to gravitational perturbations.
\end{enumerate} 

\subsection{Slant Range and Dynamic Visibility}
\label{subsec:slant_range}

The positions of the spacecraft and surface receiver are represented as time-dependent vectors in the LCRS frame:
{}
\begin{enumerate}
\item \textit{Spacecraft Position Vector:}
$  \vec{r}_{\text{sc}}(t) = \big(x_{\text{sc}}(t), y_{\text{sc}}(t), z_{\text{sc}}(t)\big)$ where the components are derived from the spacecraft's orbital parameters, including $a(t)$, $e(t)$, $i(t)$, $\Omega(t)$, and $\omega(t)$. The true anomaly $\nu(t)$ determines the instantaneous position along the orbit.

\item \textit{Surface Receiver Position Vector:}
$ \vec{r}_{\text{rx}}(t) = R_{\text{moon}} \big(
   \cos\phi \cos\bigl(\lambda + \omega_{\text{moon}} t\bigr),
   \cos\phi \sin\bigl(\lambda + \omega_{\text{moon}} t\bigr),
   \sin\phi\big)$, where $\phi$ is the receiver's latitude, $\lambda$ is the longitude, and $\omega_{\text{moon}}$ is the Moon's rotation rate.
\end{enumerate}

The slant range $d(t)$, which determines the power flux density at the receiver, is calculated dynamically as:
\begin{equation}
    d(t) = \|\vec{r}_{\text{sc}}(t) - \vec{r}_{\text{rx}}(t)\|,
        \label{eq:slant-dist}
\end{equation}
where $\vec{r}_{\text{sc}}(t)$ and $\vec{r}_{\text{rx}}(t)$ are the spacecraft and receiver position vectors, respectively. For a spherical Moon,  the slant range defined by (\ref{eq:slant-dist}) may be approximated as: $d(t) = \sqrt{{r}^2_{\text{sc}}(t)+ {r}^2_{\text{rx}}(t) - 2 {r}_{\text{sc}}(t){r}_{\text{rx}}(t)\cos\alpha(t)}$, where $r_{\rm sc}(t)=|\vec r_{\rm sc}(t)|$, $r_{\rm rx}(t)=|\vec r_{\rm rx}(t)|$ and $\alpha(t)$ is the central angle between the spacecraft and receiver, given  by:
\begin{equation}
    \cos\alpha(t) = \frac{\vec{r}_{\text{sc}}(t) \cdot \vec{r}_{\text{rx}}(t)}{\|\vec{r}_{\text{sc}}(t)\| \|\vec{r}_{\text{rx}}(t)\|}.
     \label{eq:slant-angle}   
\end{equation}

The time variability of orbital elements in a lunicentric frame necessitates detailed trajectory modeling, dynamic beam steering, and high-resolution topographic data integration. The choice of orbit, such as LLO, NRHO, or HEO, must be optimized to balance visibility durations, slant range, and mission-specific requirements for power delivery.

\subsection{Line-of-Sight Visibility, Terrain Masking and Local Horizon Effects}
\label{subsec:terrain_masking}
\label{subsec:line_of_sight}

Line-of-sight (LoS) visibility is satisfied when:
\begin{equation}
    \alpha(t) < \alpha_{\text{crit}}(t),
\end{equation}
where $\alpha(t)$ is the central angle between the spacecraft and the receiver (\ref{eq:slant-angle}). The critical angle $\alpha_{\text{crit}}(t)$ is defined by:
\begin{equation}
    \alpha_{\text{crit}}(t) = \arccos\left(\frac{R_{\text{moon}}}{\|\vec{r}_{\text{sc}}(t)\|}\right),
\end{equation}
where $\|\vec{r}_{\text{sc}}(t)\|$ is the time-dependent distance of the spacecraft from the Moon’s center.

The time dependence of $\alpha(t)$ and $\alpha_{\text{crit}}(t)$ necessitates dynamic updates based on the spacecraft’s orbital evolution and the receiver’s location. These variations are particularly important for missions targeting polar regions, where libration effects can introduce significant visibility fluctuations.

Lunar topography imposes significant constraints on visibility, especially in regions with rugged terrain such as the poles. In the ideal spherical model, the LoS condition is determined by the central angle \(\alpha(t)\), defined by (\ref{eq:slant-angle}), and the corresponding geometric horizon. However, local terrain features---such as crater rims, mountain peaks, and other irregularities---can elevate the effective horizon at the receiver’s location.

For a receiver located on the lunar surface (with \(r_{\text{rx}}(t) = R_{\text{moon}}\)) and a spacecraft at a radial distance \(r_{\text{sc}}(t)\) from the Moon’s center, with a central angle \(\alpha(t)\) between the two, one common expression for the effective elevation angle \(\epsilon(t)\) (i.e., the angle above the local horizontal as seen at the receiver) is
\[
\epsilon(t) = \arcsin\!\bigg(\frac{r_{\text{sc}}(t)\cos\alpha(t) - R_{\text{moon}}}{\sqrt{r_{\text{sc}}(t)^2 + R_{\text{moon}}^2 - 2\,r_{\text{sc}}(t)\,R_{\text{moon}}\,\cos\alpha(t)}}\bigg).
\]

The local terrain imposes an additional constraint on LoS: the effective elevation angle must exceed the local horizon elevation angle, \(\beta_{\text{mask}}(\phi,\lambda,t)\), which accounts for features such as crater rims and mountains. That is,
\[
\epsilon(t) > \beta_{\text{mask}}(\phi,\lambda,t),
\]
where \(\phi\) and \(\lambda\) denote the latitude and longitude of the receiver, and \(t\) accounts for temporal variations (e.g., due to libration or local illumination conditions).

High-resolution digital elevation maps (DEMs) are required to accurately determine \(\beta_{\text{mask}}(\phi,\lambda,t)\), especially in permanently shadowed regions and at high latitudes. In many polar regions, the local horizon may be elevated to values such as \(\beta_{\text{mask}} > 30^\circ\), thereby reducing the available visibility windows for power transmission \cite{Messenger:2015}.

\subsection{Visibility Windows and Power Delivery}
\label{subsec:visibility_windows}

Visibility windows are defined as the periods during which the spacecraft maintains line-of-sight (LoS) with a surface receiver, enabling uninterrupted power transmission. These windows are determined by the orbital geometry, lunar topography, and terrain masking. Accurate modeling of visibility windows allows mission planners to optimize power delivery schedules, minimize outages, and size surface energy storage systems appropriately.

A useful scalar measure of coverage is the \emph{visibility fraction} referenced to a user-chosen interval $T_{\mathrm{ref}}$:
\begin{equation}
f_{\mathrm{vis},1}(T_{\mathrm{ref}}) = \frac{1}{T_{\mathrm{ref}}}{\sum_{k} \Delta t_k},
\label{eq:fvis1}
\end{equation}
where $\Delta t_k$ are the durations of individual LoS visibility windows within the reference interval $T_{\mathrm{ref}}$. We use $T_{\mathrm{ref}}=24~\mathrm{h}$ when reporting “daily” energy (Sec.~\ref{sec:receiver_model}), $T_{\mathrm{ref}}=T_{\mathrm{syn}}\!\approx\!29.53~\mathrm{d}$ (one synodic lunar day, sunrise-to-sunrise) to discuss \emph{lunar-day} availability, and $T_{\mathrm{ref}}=T_{\mathrm{night}}\!\approx\!14.77~\mathrm{d}$ for \emph{lunar-night} (dark-only) availability. For clarity we denote these by $f_{\mathrm{vis},1}^{24\mathrm{h}}$, $f_{\mathrm{vis},1}^{\mathrm{day}}$, and $f_{\mathrm{vis},1}^{\mathrm{night}}$, respectively. Unless stated otherwise, “daily energy” results in this paper use $T_{\mathrm{ref}}=24~\mathrm{h}$.

For polar sites, NRHO configurations often increase link availability relative to LLO, but the realized $f_{\mathrm{vis},1}$ depends strongly on the specific NRHO family and the local horizon mask. In our Shackleton-rim case study (Sec.~\ref{sec:parametric_maps}, Table~\ref{tab:NRHO-visibility}), terrain-masked access over $T_{\mathrm{ref}}=24~\mathrm{h}$ totals $\sim 116$ minutes, i.e., $f_{\mathrm{vis},1}^{24\mathrm{h}}\approx 0.08$. Accordingly, throughout this paper we report $f_{\mathrm{vis},1}$ together with the chosen $T_{\mathrm{ref}}$ and the assumed horizon mask, rather than quoting a universal high coverage value.

Given a single--orbiter “daily” net electrical energy $E_{24\mathrm{h},1}$ (computed by integrating $P_{\mathrm{rx,elec}}(t)$ over $T_{\mathrm{ref}}=24~\mathrm{h}$), the visibility fraction \eqref{eq:fvis1} provides the link between orbital geometry and delivered energy. This parameter will be used in Sec.~\ref{subsec:constellation} to quantify the scaling of coverage and energy delivery for constellations of multiple orbiters.

Incorporating dynamic orbital modeling and terrain constraints ensures accurate predictions of visibility and enhances the overall efficiency of lunar power delivery systems.

\section{Laser Transmission Link Budget}
\label{sec:laser_link}

The laser transmission link budget for power beaming from orbit to a lunar surface receiver requires precise modeling of beam divergence, far-field spot size, temporal variations, and efficiency losses. This section provides a technical and quantitative analysis, including practical considerations, numerical examples, and key design parameters for optimizing energy delivery.

\subsection{Beam Divergence and Far-Field Spot Size}
\label{subsec:beam_div}

The divergence of a laser beam transmitted through an optical system, such as a telescope or collimator, is governed by diffraction and system-specific imperfections (see \cite{Turyshev-LLR-CW:2025}). In this analysis, we assume that the laser beam has a Gaussian intensity profile, which is common for many practical laser systems. For a diffraction-limited Gaussian beam, the theoretical half-angle divergence is given by
\begin{equation}
    \theta_{\rm diff} \approx \frac{\lambda}{\pi w_0}
    \approx \frac{2\lambda}{\pi D},
    \label{eq:theta_ideal}
\end{equation}
where \(\lambda\) is the laser wavelength, \(w_0\simeq \frac{1}{2}D\) is the radius of the beam waist, and \(D\) is the effective aperture diameter of the transmitting optics. This expression assumes an ideal wavefront with no distortions.

In practical systems, wavefront imperfections, pointing instabilities, and non-Gaussian
beam profiles all degrade the ideal (diffraction-limited) performance. This degradation
is quantified by the beam-quality factor, $M^2$, which typically exceeds 1 in real-world
applications. For instance, laboratory fiber or diode lasers may achieve $M^2\approx1.1-1.5$,
while fully integrated flight hardware subjected to thermal gradients and launch/operational
vibrations might see $M^2$ climb toward 2 or higher unless actively stabilized.

For a beam of wavelength $\lambda$ emitted through an aperture of diameter $D$, 
the half-angle divergence becomes
\begin{equation}
    \theta_{\text{tx}}
    \;=\;
    M^2 \frac{2\lambda}{\pi D},
    \label{eq:theta_practical}
\end{equation}
where $M^2=1$ corresponds to a perfect Gaussian beam (diffraction-limited). Thus, $M^2$
represents the ratio of the measured beam divergence to the ideal case. Beyond divergence,
$M^2$ also scales other beam parameters. For example, if the ideal ($M^2=1$) waist radius
is $w_0$, then the actual waist radius is
\[
    w_{0}^\prime
    \;=\;
    M^2 w_0,
\]
and the Rayleigh range (distance over which the beam radius grows by a factor
of $\sqrt{2}$) \emph{decreases} from the diffraction-limited $z_R$ to
\[
    z_R^\prime
    \;=\;
    \frac{z_R}{M^2}
    \;=\;
    \frac{\pi w_0^2}{M^2 \lambda}.
\]
Hence, any $M^2>1$ indicates a departure from the ideal Gaussian scenario, capturing
real-world imperfections such as optical aberrations, mechanical misalignment,
thermal distortions, or higher-order transverse modes.

\begin{table}[h]
\centering
\caption{Illustrative Ranges of $M^2$ for Space-Based Lasers}
\begin{tabular}{lc}
\hline
{Laser Configuration} & {Typical $M^2$ Range} \\
\hline\hline
Well-aligned laboratory diode/fiber lasers & 1.1--1.3 \\
Spaceflight hardware with basic stabilization & 1.3--2.0 \\
Deployable apertures or minimal wavefront control ~& 2.0+ \\
\hline
\end{tabular}
\label{tab:m2_ranges}
\end{table}

As summarized in Table~\ref{tab:m2_ranges}, actual $M^2$ values can vary significantly depending on design details, thermal and mechanical control, and the degree of on-board phase correction. Lower $M^2$ translates directly to a narrower far-field divergence, smaller beam footprints, and higher power densities at the receiver---critical considerations for long-range lunar laser power beaming.

In the far field, the beam radius expands linearly with the time-varying slant range \(d(t)\). Specifically, the beam radius at a distance \(d(t)\) is given by
\begin{equation}
    w(t) = \theta_{\text{tx}} d(t),
        \label{eq:slant-range}
\end{equation}
where \(\theta_{\text{tx}}\) is the practical half-angle divergence of the beam. Consequently, the spot area \(A_{\text{spot}}(t)\) is
\begin{equation}
    A_{\text{spot}}(t) = \pi  [w(t)]^2 = \pi \bigl[\theta_{\text{tx}} d(t)\bigr]^2.
\end{equation}
Minimizing \(\theta_{\text{tx}}\) is crucial for reducing the spot size and maintaining high power densities at the receiver over time.

\subsection{Power Flux on the Lunar Surface with Temporal Variations}
\label{subsec:power_flux}

The power flux at the lunar surface, defined as the power per unit area delivered by the laser, depends on the time-varying spot size. For a transmitted power \(P_{\text{tx}}\) and a spot area \(A_{\text{spot}}(t)\), the ideal power flux is given by
\begin{equation}
    I_{\text{ideal}}(t) = \frac{P_{\text{tx}}}{A_{\text{spot}}(t)} = \frac{P_{\text{tx}}}{\pi \bigl[w(t)\bigr]^2} = \frac{P_{\text{tx}}}{\pi \bigl[\theta_{\text{tx}}d(t)\bigr]^2},
    \label{eq:top_hat_flux}
\end{equation}
where the beam radius at the slant range \(d(t)\) is defined by (\ref{eq:slant-range}). (Note that this expression is a bookkeeping flux for the \emph{top-hat} limit. When using the Gaussian capture model, use \eqref{eq:Ieff_def} with $\eta_{\rm cap}$ and do not also multiply by a flux$\times$area bracket, to avoid double counting.)

Unless noted otherwise, we use a Gaussian beam with jitter broadening. The $1/e^2$ radius at range is $w(d,t)=\theta_{\rm tx}\,d(t)$ and the effective radius is $w_{\rm eff}^2(d,t)=w^2(d,t)+[\sigma_\theta(t)\,d(t)]^2$ \eqref{eq:weff_def}. \emph{Geometric capture} is handled by the encircled--energy fraction $\eta_{\rm cap}(d,t)$ \eqref{eq:eta_cap_def}, while \emph{residual pointing/loop losses not representable by encircled energy} are collected in $\eta_{\rm point}(t)\in(0,1]$ \eqref{eq:etapoint_redef}.  The master link used by default is \eqref{eq:Prx_opt_main} with $\eta_{\rm cap}$ from \eqref{eq:eta_cap_def}; where a top-hat bookkeeping approximation is used, it is stated explicitly.

\subsubsection{Efficiency Losses with Temporal Dependence}

In practical systems, time-varying losses arise due to pointing, optical-path changes, and operational conditions. We decouple \emph{geometric capture} at the receiver from \emph{residual pointing/loop} losses.

In the case of a small-aperture (top-hat) approximation, assuming top-hat bookkeeping for the beam-plane irradiance, the unobscured irradiance at the receiver plane is
\begin{equation}
I_{\text{beam}}(t) \;=\; \eta_{\text{opt}}(t)
\frac{P_{\text{tx}}(t)}{\pi [\theta_{\text{tx}}d(t)]^{2}},
\label{eq:Ibeam_def}
\end{equation}
where $\eta_{\text{opt}}(t)$ represents time-varying optical transmission efficiency.

Unless noted otherwise, we use a Gaussian beam with jitter broadening. The $1/e^2$ radius at range is $w(d,t)=\theta_{\rm tx}\,d(t)$ in the absence of jitter, and an RMS line-of-sight jitter $\sigma_{\theta}(t)$ broadens the effective spot in quadrature:
\begin{equation}
w_{\rm eff}^2(d,t) \;\equiv\; w^2(d,t) + \big[2\,\sigma_{\theta}(t)\,d(t)\big]^2 .
\label{eq:weff_def}
\end{equation}

The \emph{geometric capture} (encircled-energy fraction) by a circular receiver of radius $r_{\rm rx}$ is
\begin{equation}
\eta_{\text{cap}}(d,t) \equiv
1 - \exp\!\Big[-\frac{2\, r_{\rm rx}^2}{w_{\rm eff}^2(d,t)}\Big].
\label{eq:eta_cap_def}
\end{equation}
All other pointing-related losses that are not representable by encircled energy (e.g., control-loop latency, rare large excursions, ephemeris/boresight bias) are lumped into
\begin{equation}
\eta_{\text{point}}(t) \in (0,1] ,
\label{eq:etapoint_redef}
\end{equation}
which is treated as a multiplicative factor independent of $\eta_{\text{cap}}$.

The captured optical power and the \emph{effective mean optical irradiance} on a receiver of area $A_{\rm rx}=\pi r_{\rm rx}^2$ are
\begin{align}
  P_{\rm rx,opt}(t) &= \eta_{\rm point}(t)\,\eta_{\rm main}\,\eta_{\rm cap}(d,t)\,\eta_{\rm opt}(t)\,P_{\rm tx}(t),
  \label{eq:Prx_opt_main}\\[2mm]
I_{\text{eff}}(t) &\equiv \frac{P_{\rm rx,opt}(t)}{A_{\rm rx}}
= \eta_{\rm point}(t)\,\eta_{\rm main}\,\eta_{\rm cap}(d,t)\,\eta_{\rm opt}(t)\,\frac{P_{\rm tx}(t)}{A_{\rm rx}}.
\label{eq:Ieff_def}
\end{align}
Here $\eta_{\rm main}\in(0,1]$ is the \emph{main-lobe efficiency} (fraction of total transmit power in the synthesized main lobe at boresight); for single apertures or dense arrays with negligible sidelobes set $\eta_{\rm main}=1$. 

Note that under zero-mean Gaussian line-of-sight jitter with RMS $\sigma_\theta$, the time-averaged beam is the convolution of two Gaussians, giving the jitter-broadened 1/e$^2$ radius
\begin{equation}
w_\mathrm{eff}^2(d) = (\theta_\mathrm{tx} d)^2 + \big(2\sigma_\theta d\big)^2,
\end{equation}
so that a notional ``effective divergence''
\begin{equation}
\theta_\mathrm{eff}^2 \equiv \frac{w_\mathrm{eff}^2}{d^2} = \theta_\mathrm{tx}^2 + 4\sigma_\theta^2
\end{equation}
recovers the familiar root-sum-square form often used as a quick estimate. In this work we retain (\ref{eq:weff_def})--(\ref{eq:Ieff_def}) (Gaussian capture with $w_\mathrm{eff}$) for accurate link budgeting and to avoid double-counting when computing intercepted power.

Our end-to-end link, uses the Gaussian encircled-energy
capture for a finite receiver, $\eta_{\rm cap}(d,t)$, from (\ref{eq:eta_cap_def}) and treats residual tracking losses not representable by encircled energy via a separate factor $\eta_{\rm point}$ from \eqref{eq:etapoint_redef}. This avoids double counting and properly handles regimes where the receiver size is  comparable to the beam radius.

As an example, we consider  $D=0.5$\,m, $\lambda=1064$\,nm, $M^2\!\approx\!1.3$, and $d=200$\,km, for which the divergence is $\theta_{\text{tx}}\!\approx\!1.76~\mu$rad, so $w\!\approx\!0.35$\,m at the receiver plane. For a $1\,\mathrm{m}^2$ receiver ($r_{\rm rx}\!\approx\!0.564$\,m), the capture requirement
$\eta_{\text{cap}}\!\ge\!0.90$ implies $w_{\rm eff}\!\leq\!0.526$\,m, i.e.,
\[
\sigma_{\theta} \;\lesssim\; \frac{\sqrt{w_{\rm eff}^2-w^2}}{2d}
\;\approx\; 1.0~\mu\text{rad}\quad\text{at } d=200~\text{km}.
\]
Tighter capture targets map to tighter jitter: $\eta_{\text{cap}}\!\ge\!0.98 \Rightarrow \sigma_{\theta}\!\lesssim\!0.50~\mu$rad, and
$\eta_{\text{cap}}\!\ge\!0.99 \Rightarrow \sigma_{\theta}\!\lesssim\!0.31~\mu$rad.

The LoS jitter is formed by quadrature of independent contributors:
\begin{equation}
\sigma_{\theta}^2 \;=\; \sigma_{\rm RW}^2 + \sigma_{\rm FSM}^2 + \sigma_{\rm struct}^2 + \sigma_{\rm know}^2,
\label{eq:sigma_budget}
\end{equation}
where $\sigma_{\rm RW}$ is residual reaction-wheel/CMG disturbance after isolation, $\sigma_{\rm FSM}$ is fast-steering-mirror residual, $\sigma_{\rm struct}$ is structural/thermoelastic jitter, and $\sigma_{\rm know}$ is the line-of-sight knowledge error. This breakdown governs $\eta_{\text{cap}}$ via \eqref{eq:eta_cap_def}, while $\eta_{\text{point}}$ captures non-encircled-energy losses (e.g., infrequent dropouts).

\subsubsection{Pointing and Jitter Budget}
\label{subsec:pointing_budget_psd}

We make the end-to-end LoS jitter claim explicit by decomposing $\sigma_\theta$ into power spectral densities (PSDs) from dominant disturbance, sensor, and knowledge terms, each shaped by the appropriate closed-loop transfer functions. Let $S_x(f)$ denote the one-sided PSD of contribution $x$ (rad$^2$/Hz).
Using the cascaded architecture (inner fast-steering-mirror (FSM) loop and outer AOCS/gimbal loop), the LoS variance integrates as
\begin{equation}
  \sigma_\theta^2 = \int_{0}^{\infty}
  \Big(
     |T_{\tt FSM}(f)|^2 S_{\rm RW}(f)
   + |T_{\tt FSM}(f)|^2 S_{\rm struct}(f)
   + |T_{\tt AOCS}(f)|^2 S_{\rm ext}(f)
   + |S_{\tt FSM}(f)|^2 S_{\rm sens}(f)
   + S_{\rm know}(f)
  \Big) df,
  \label{eq:psd_budget_master}
\end{equation}
where $T(f)=L(f)/(1{+}L(f))$ and $S(f)=1/(1{+}L(f))$ are the complementary and sensitivity functions of the relevant loop with open loop $L(f)$. Here: $S_{\tt RW}$ is the reaction-wheel/CMG micro vibration input as seen at the FSM, $S_{\rm struct}$ is the thermoelastic/structural jitter floor at the optical head, $S_{\rm ext}$ is low-frequency rigid-body torque/torque noise rejected by AOCS, $S_{\rm sens}$ is fine-sensor noise referred to LOS, and $S_{\rm know}$ is the knowledge PSD (star tracker, metrology, ephemeris) that adds directly when used for feed-forward and bias removal.

We use loop models consistent with $\,\sim$kHz fine sensing and $\sim$10\,ms outer-loop latency, with representative numerical parameters (zero/pole locations, latencies, and gains) listed in Table~\ref{tab:loop_params} for the NRHO design case:
\begin{align}
  L_{\tt FSM}(f) &= \frac{K_{\tt FSM}}{(1+jf/f_z)}\frac{1}{(1+jf/f_p)} e^{-j2\pi f \tau_{\tt FSM}},
  \qquad \qquad \,\,
  f_{c,\rm FSM}\approx 80\text{--}120~\text{Hz},\ \ \tau_{\tt FSM}\sim1~\text{ms},\\
  L_{\tt AOCS}(f) &= \frac{K_{\tt AOCS}}{(1+jf/f_{z,\rm A})}\frac{1}{(1+jf/f_{p,\rm A})^2}e^{-j2\pi f \tau_{\tt AOCS}},
  \qquad f_{c,\tt AOCS}\approx 10\text{--}15~\text{Hz},\ \ \tau_{\tt AOCS}\sim10~\text{ms}.
\end{align}

The controller gains $K_{\tt FSM}$ and $K_{\tt AOCS}$ are tuned such that the magnitude of $L(f)$ crosses unity at the target closed-loop bandwidth $f_c$ for each loop, i.e.
\[
|L_{\tt loop}(f_c)| = 1.
\]
For a nominal plant model $G_{\tt plant}(f)$ (mechanical + actuator dynamics), this implies a first estimate
\begin{equation}
  K_{\tt loop} \ \approx\ \frac{1}{|G_{\tt plant}(f_c)|},
\end{equation}
with $f_c \approx f_{c,\rm FSM}$ or $f_{c,\rm AOCS}$ as appropriate. If, for example, the FSM actuator has a torque-to-angle transfer of $G_{\tt FSM}(f) \simeq (2\pi f_{\rm nat})^{-2}$~[rad/N·m] with $f_{\rm nat}$ in the 200--300\,Hz range, and the gimbal/AOCS stage has $G_{\tt AOCS}(f)$ dominated by rigid-body dynamics below $\sim 5$\,Hz, these $K$ values directly scale with the required angular control authority. After setting the crossover to the desired $f_c$, gains are iteratively adjusted to achieve at least $6$\,dB gain margin and $45^\circ$ phase margin when latencies $\tau_{\tt FSM}$ and $\tau_{\tt AOCS}$ are included.

For the NRHO example in Table~\ref{tab:los_budget}, representative values are $K_{\tt FSM}$ chosen to yield $f_{c,\rm FSM} \approx 100$~Hz and $K_{\tt AOCS}$
set for $f_{c,\rm AOCS} \approx 12$~Hz, given the assumed actuator constants and plant transfer functions.

\begin{table}[t]
\centering
\caption{Control and sensing parameters used in the pointing budget.}
\label{tab:loop_params}
\renewcommand\arraystretch{1.0}
\setlength{\tabcolsep}{4pt}
\begin{tabular}{l c c l}
\hline
Element & Sample Rate & Latency & Loop Bandwidth \\
\hline\hline
Fine sensor (quad-cell / FPA) & $\geq$1\,kHz & $\sim$1\,ms & FSM: 80--120\,Hz \\
FSM actuator (tip/tilt) & Analog/PWM & $<$1\,ms & Inner loop \\
Star tracker / metrology & 10--20\,Hz & $\sim$10\,ms & AOCS: 10--15\,Hz \\
Coarse gimbal / reaction wheels & 50--200\,Hz cmd & $\sim$10\,ms & Outer loop \\
\hline
\end{tabular}
\end{table}

The split-band design avoids excessive phase lag from $\tau_{\tt AOCS}$ by pushing
microvibration rejection into the low-latency FSM loop. For budgeting, we adopt
simple parametric PSDs that can be refined with hardware data:
\begin{align}
S_{\tt RW}(f)     &= A_{\tt RW} \,(1+(f/f_0)^2)^{-1},
&& \text{wheel lines 40--120 Hz; $1/f^2$ tail},\\[2pt]
S_{\rm struct}(f) &= A_{\rm struct},
&& \text{white within 20--300 Hz},\\[2pt]
S_{\rm ext}(f)    &= A_{\rm ext}\,(1+f/f_b)^{-2},
&& \text{LF rigid-body},\\[2pt]
S_{\rm sens}(f)   &= A_{\rm sens},
&& \text{fine-sensor noise, flat to Nyquist},\\[2pt]
S_{\rm know}(f)   &= A_{\rm know}\,(1+(f/f_k)^2)^{-1},
&& \text{LF-dominated knowledge}.
\end{align}

Gains are chosen for $\geq 6$\,dB gain margin and $\geq 45^\circ$ phase margin. The inner loop provides $>20$\,dB attenuation at wheel-line bands (40--120\,Hz);
the outer loop provides $>20$\,dB attenuation below 1--2\,Hz while handing over microvibrations to the inner loop. Assumed $\mu$rad-class steering stability and ATP closure are compatible with mobile FSO ATP capabilities and power-beaming demonstrations surveyed in~\cite{Kaymak2018Comst,Rodenbeck2021JMW}.

\begin{table}[t]
\centering
\caption{End-to-end LoS jitter budget (NRHO example). Each entry is the $1\sigma$ contribution from integrating Eq.~\eqref{eq:psd_budget_master} over the stated band with the indicated loop shaping.}
\label{tab:los_budget}
\renewcommand\arraystretch{1.12}
\setlength{\tabcolsep}{2pt}
\begin{tabular}{l p{4.9cm} p{5.1cm} r}
\hline
Source & Model / Bandwidth & Loop Shaping & $\sigma_\theta$ [$\mu$rad] \\
\hline\hline
Reaction-wheel microvibes & Peaks 40--120\,Hz; $1/f^2$ tail & FSM comp.\ $|T_{\tt FSM}(f)|$, $f_c\!\approx\!100$\,Hz & 0.050 \\
FSM residual / sensor quantization & Flat to 500\,Hz (1\,kHz readout) & FSM sens.\ $|S_{\tt FSM}(f)|$ & 0.040 \\
Structural / thermoelastic & White 20--300\,Hz & FSM comp.\ $|T_{\tt FSM}(f)|$ & 0.030 \\
Rigid-body and external torques & LF $<2$\,Hz; $(1+f/f_b)^{-2}$ roll-off & AOCS comp.\ $|T_{\tt AOCS}(f)|$, $f_c\!\approx\!12$\,Hz & 0.025 \\
Knowledge (star tracker, metrology) & 10--20\,Hz update; LF-dominated & Adds directly & 0.040 \\
Ephemeris / boresight bias & Quasi-static; drift corrected hourly & Adds directly (post-cal) & 0.030 \\
\hline
\multicolumn{3}{r}{RSS (no margin)} & 0.090 \\
\multicolumn{3}{r}{Programmatic margin (25\%)} & 0.023 \\
\hline
\multicolumn{3}{r}{Total with margin} & 0.113 \\
\hline
\end{tabular}
\end{table}

\subsubsection{Pointing-Budget Sensitivity}

\begin{figure}[t]
  \centering
  \includegraphics[width=0.62\linewidth]{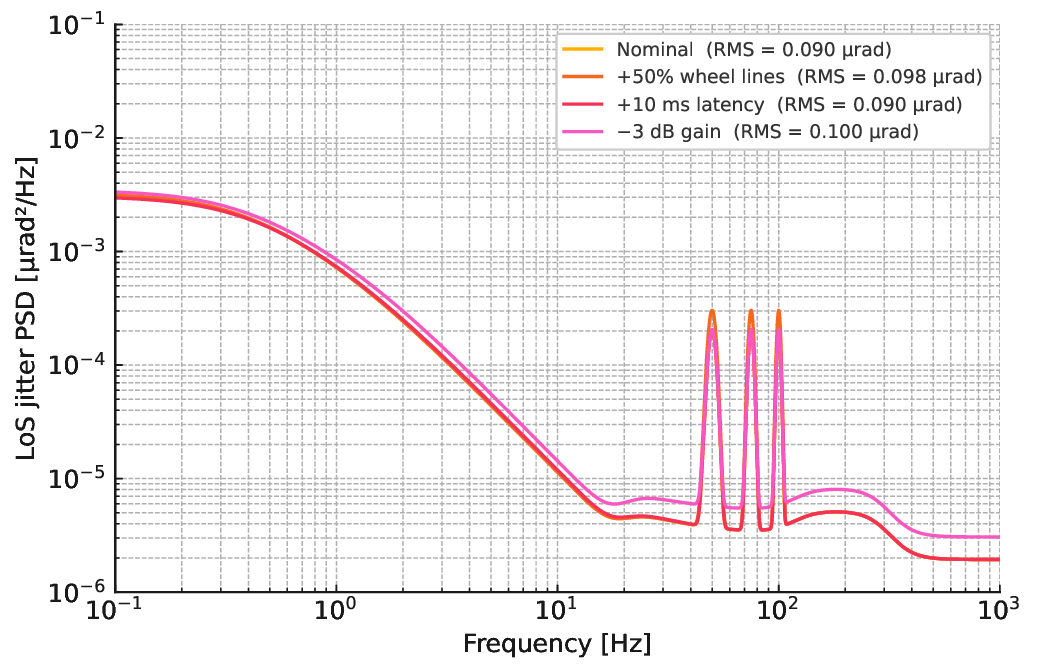}
  \caption{Closed-loop one-sided LoS jitter PSD for a 2\,m-class phased-array optical transmitter in NRHO (0.1--1000\,Hz).  The nominal case includes low-frequency rigid-body residuals, reaction-wheel-line features at 50, 75, 100\,Hz, a structural/thermoelastic floor (20--300\,Hz), and fine-sensor/FSM noise. Perturbations applied individually: +50\% wheel-line amplitudes, +10\,ms outer-loop latency, and $-3$\,dB loop gain. Numerical integration of each PSD yields the total RMS LoS jitter in $\mu$rad (see the legend).}
  \label{fig:psd_sensitivity}
\end{figure}

Figure~\ref{fig:psd_sensitivity} shows the one-sided line-of-sight (LoS) jitter power spectral density (PSD) for a 2\,m-class phased-array transmitter in NRHO, together with three perturbation cases applied to the same closed-loop model. The nominal PSD comprises (i) low-frequency rigid-body residuals shaped by the AOCS loop, 
(ii) reaction-wheel microvibration lines at 50, 75, 100\,Hz, 
(iii) a structural/thermoelastic floor spanning 20--300\,Hz, 
and (iv) a fine-sensor/FSM residual that rolls on above the FSM bandwidth. The frequency range shown is 0.1--1000\,Hz, with PSD units of $\mu$rad$^2$/Hz.

\begin{table}[h]
\centering
\caption{Integrated RMS LoS jitter from PSD (0.1--1000\,Hz).}
\label{tab:psd_rms}
\renewcommand{\arraystretch}{1.0}
\setlength{\tabcolsep}{10pt}
\begin{tabular}{lc}
\hline
Case & RMS LoS jitter ($\mu$rad) \\
\hline\hline
Nominal & 0.090 \\
+50\% wheel-line amplitudes & 0.098 \\
+10\,ms latency & 0.090 \\
$-3$\,dB loop gain & 0.100 \\
\hline
\end{tabular}
\end{table}

Perturbations are applied individually: (1) wheel-line amplitudes increased by 50\% (raising the narrowband peaks), (2) outer-loop latency increased by 10\,ms (slightly degrading low-frequency rejection), and (3) loop gain reduced by 3\,dB (increasing broadband residuals in both the AOCS and FSM bands). Numerical integration of each PSD over 0.1--1000\,Hz yields the total RMS LoS jitter reported in the legend and summarized in Table~\ref{tab:psd_rms}: 
\mbox{Nominal $=0.090~\mu$rad}, 
\mbox{$+50\%$ wheel lines $=0.098~\mu$rad}, 
\mbox{$+10$\,ms latency $=0.090~\mu$rad}, 
\mbox{$-3$\,dB gain $=0.100~\mu$rad}. 
All cases remain at or below the $0.1~\mu$rad design target, with $0.15~\mu$rad as the conservative upper bound for system budgeting.

The PSD-based budget meets the $\lesssim 0.1~\mu{\rm rad}$ RMS requirement in NRHO in the nominal case: the RSS without margin is $0.090~\mu {\rm rad}$; with a conservative $25\%$ programmatic margin, the total is $0.113~\mu{\rm rad}$. The split-band FSM/AOCS control structure maintains gain and phase margins in the presence of micro-vibration lines and structural modes, and the sensitivity analysis here shows that moderate degradations in wheel-line levels, control-loop latency, or gain still preserve compliance. This also supports the development of advanced navigation capabilities for cislunar space, enabling precision navigation and attitude control \cite{Bar-Sever-etal:2024}.

\subsubsection{Acquisition and Tracking Considerations}

Initial link acquisition in NRHO scenarios can be achieved using a beacon‐receiver pair, where the surface receiver emits an optical or RF beacon of power $\sim$1--10~W into a narrow beam aligned to the predicted spacecraft trajectory. The spacecraft’s acquisition sensor, with a field‐of‐view (FOV) of $\pm0.5^\circ$ and a frame rate of $\geq$10~Hz, detects this beacon and performs coarse pointing. Fine tracking uses a high‐rate ($\geq$1~kHz) quad‐cell or focal‐plane sensor with a $\pm0.5$--1~mrad FOV.

Closed‐loop end‐to‐end performance combines knowledge error, control loop latency, and mechanical stability. For NRHO dynamics, simulations show that with $\leq$1~Hz ephemeris updates and $\sim$10~ms control latency, total closed‐loop pointing error can be held to $\lesssim$1~$\mu$rad RMS, consistent with the $\eta_{\rm point}$ values used in (\ref{eq:Ieff_def}).

\subsubsection{Optical Losses from Dust and Self-Induced Lofting}

In addition to static optical losses from coatings, windows, and structure, the optical path efficiency $\eta_{\text{opt}}(t)$ can be reduced by lunar dust contamination and scattering. These effects arise from (i) background dust deposition in permanently shadowed regions and near‐rim sites, and (ii) dust lofted or regolith disturbed by high‐flux beam heating.

We include these effects via a multiplicative dust term:
\[
\eta_{\text{opt}}(t) = \eta_{\text{opt,base}}(t) \times \eta_{\text{dust}}(t),
\]
where $\eta_{\text{opt,base}}$ accounts for all non‐dust losses. For a receiver with dust surface coverage fraction $f_{\mathrm{cov}}$ and Mie‐regime single‐scattering albedo $\omega_s$ at $\lambda \approx 1.0~\mu\mathrm{m}$, a first‐order exponential attenuation model is
\[
\eta_{\text{dust}} \approx e^{-\tau_{\lambda}}, \qquad
\tau_{\lambda} \approx Q_{\mathrm{ext}}(\lambda)\frac{m_{\mathrm{dust}}}{\rho_{\mathrm{dust}}A_{\mathrm{rx}}\delta},
\]
where $Q_{\mathrm{ext}}(\lambda)$ is the extinction efficiency at wavelength $\lambda$ (with $Q_{\mathrm{ext}} \approx 1.5$--$2$ for $d \approx 1$--$10~\mu$m grains), $\rho_{\mathrm{dust}}\approx 3000~\mathrm{kg/m^3}$ is the particle bulk density, and $\delta$ is the dust‐layer thickness. This formulation makes the dependence on material properties and coverage explicit: $Q_{\mathrm{ext}}$ encodes the scattering/absorption cross‐section relative to the projected area, $f_{\mathrm{cov}}$ controls the fractional obscuration of the aperture, and $\delta$ sets the effective column density for multi‐layer coatings. The model assumes random, uniformly distributed particles in the geometric‐optics or Mie regime. Laboratory Mie scattering for basaltic analogs at $\lambda=1.064~\mu$m indicates $\approx 5$--10\% throughput loss for $f_{\mathrm{cov}}\approx 0.01$ and $\delta \approx 10~\mu$m \citep{Gaier2005}.

To evaluate the dust impact at $\lambda=1064$\,nm, we take $Q_{\mathrm{ext}}=1.8$ for silicate grains, $f_{\mathrm{cov}}=0.05$ (5\% areal coverage), $\delta=20~\mu$m, $\rho_{\mathrm{dust}}=3000~\mathrm{kg/m^3}$, and $A_{\mathrm{rx}}=\pi(0.05~\mathrm{m})^2$. The dust mass is 
$
m_{\mathrm{dust}} = f_{\mathrm{cov}} A_{\mathrm{rx}} \rho_{\mathrm{dust}} \delta \simeq 2.36\times 10^{-5}~\mathrm{kg}.
$
Substituting gives $\tau_{\lambda} \approx Q_{\mathrm{ext}} f_{\mathrm{cov}} \approx 0.09$, so
$
\eta_{\mathrm{dust}} \approx e^{-0.09} \simeq 0.914,
$
an $\sim 8.6\%$ instantaneous throughput loss. For context, a $D_{\mathrm{ap}}=10~\mathrm{cm}$ aperture with these parameters implies $\approx 1.3\times 10^{6}$ particles on the optic. Sensitivity is linear in $f_{\mathrm{cov}}$ and $Q_{\mathrm{ext}}$: for $f_{\mathrm{cov}}=0.01$, $\eta_{\mathrm{dust}}\approx 0.982$ (1.8\% loss); for $f_{\mathrm{cov}}=0.10$, $\eta_{\mathrm{dust}}\approx 0.835$ (16.5\% loss), so finer grains with higher $Q_{\mathrm{ext}}$ at 1064\,nm could impose significantly larger penalties.

Measurements from the Apollo Dust Detector and Chang’E‐3, combined with micrometeoroid ejecta models, give background deposition rates of $\sim 0.1$--$1~\mathrm{g/m^2/yr}$ in PSRs \citep{Katzan1991,Gaier2005,ChangE3Dust2019}. For $A_{\mathrm{rx}}=1~\mathrm{m^2}$ and $\delta \approx 10~\mu$m, this yields $\tau_{\lambda}\sim 0.01$--0.05, corresponding to a few percent annual loss without cleaning. Near‐rim sites may experience $\lesssim 0.1~\mathrm{g/m^2/yr}$, but cumulative degradation over multi‐year missions remains significant.

Under high optical fluxes ($I_{\mathrm{rx,opt}}\gtrsim 5~\mathrm{kW/m^2}$), local regolith temperatures can exceed $\sim 400$ K in seconds, triggering volatile release and lofting of $\mu$m‐scale grains \citep{Immer2011}. A conservative case assumes $f_{\mathrm{cov}}\approx 0.01$ after minutes of exposure, giving an instantaneous penalty of $\eta_{\text{dust}}\approx 0.95$. This is most severe for unprotected optics at near‐normal incidence.

For link‐budget implementation, we parameterize dust losses as
\[
\eta_{\text{dust}}(t) = \exp\!\Big[-\alpha_{\mathrm{bg}}t - \beta_{\mathrm{loft}}(I_{\mathrm{rx,opt}},\theta_{\mathrm{inc}})\Big],
\]
where $\alpha_{\mathrm{bg}}$ (s$^{-1}$) is the secular fouling rate and $\beta_{\mathrm{loft}}$ is an instantaneous penalty term dependent on incident flux and beam incidence angle $\theta_{\mathrm{inc}}$. For example, $\alpha_{\mathrm{bg}}\approx 10^{-9}~\mathrm{s^{-1}}$ corresponds to $\sim 3\%$/yr loss, and $\beta_{\mathrm{loft}}\approx 0.05$ represents a 5\% instantaneous loss during high‐flux operation.

This formulation explicitly integrates environmental deposition and beam‐induced lofting into $\eta_{\text{opt}}(t)$, linking the optical loss budget directly to site conditions, operational flux, and wavelength‐specific scattering. The adopted $\eta_{\mathrm{dust}}$ parameterization is consistent with LADEE/LDEX constraints on lofted grains near the terminator and with measured scattering matrices for lunar simulants at visible wavelengths \cite{Szalay2015GRL,EscobarCerezo2018ApJS}.
We augment $\eta_{\rm dust}(t)$ with a threshold $I_{\rm loft}$, above which beam‐induced heating liberates regolith fines. For $I_{\rm rx,opt}>I_{\rm loft}$, an instantaneous penalty $\beta_{\rm loft}$ is applied; below threshold, only secular background deposition $\alpha_{\rm bg}$ acts. A cleaning/recovery duty cycle with fraction $f_{\rm clean}$ per day can be modeled as $\eta_{\rm dust,eff} = (1-f_{\rm clean})\,\eta_{\rm dust}$, allowing operations to be tuned against optical loss.

\section{Phased Array Architectures for Lunar-Orbiting Laser Power Beaming}
\label{sec:phased_arrays}

Laser power beaming from a spacecraft in lunar orbit to a surface receiver requires a highly collimated beam to maximize the delivered power flux. Beam divergence must be minimized to reduce energy losses due to beam spreading, ensuring that a sufficient fraction of the transmitted power reaches the receiver. At the  distance of \( d \simeq 1000 \) km, a small transmitting aperture results in rapid beam expansion, significantly reducing the received power. Consequently, increasing the effective aperture size is critical for ensuring practical power densities at the receiver.

A monolithic telescope with a large aperture could, in principle, achieve diffraction-limited performance. However, practical limitations related to mass, deployment complexity, and structural stability render such designs infeasible for spacecraft applications. Instead, optical phased arrays provide a scalable alternative by coherently combining multiple smaller apertures, forming an effective synthetic aperture while maintaining modularity and redundancy.

The selection of a phased array architecture is governed by key mission constraints, including spacecraft mass, pointing accuracy, power requirements, and beam quality. This section examines four primary phased array configurations---dense (filled) arrays, sparse arrays, sparse fill-factor-corrected arrays, and optimized aperiodic arrays. Each architecture is evaluated in terms of beam quality, divergence, sidelobe suppression, and system complexity, providing insight into their suitability for different mission scenarios.

\subsection{Beam Divergence and Effective Aperture}
\label{subsec:beam_divergence}

A laser power beaming system must maintain a highly collimated beam to ensure that the majority of transmitted energy reaches the lunar surface receiver. Beam divergence directly impacts the power flux density at the receiver and must be minimized to reduce energy losses due to spreading. The primary constraint on beam collimation is diffraction, which imposes a fundamental limit on achievable beam divergence as a function of the aperture size.

For a single-aperture transmitter with diameter \(D\), the diffraction-limited half-angle beam divergence is given by (\ref{eq:theta_ideal}) and (\ref{eq:theta_practical}) as below:
{}
\begin{equation}
    \theta_{\text{diff}} = \frac{2\lambda}{\pi D} \qquad {\rm and}\qquad
      \theta_{\text{pract}} = M^2\,\frac{2\lambda}{\pi D},
    \label{eq:theta_diff}
\end{equation}
where \(\lambda\) is the laser wavelength. Here $ \theta_{\text{diff}}$  represents the ideal case, assuming a perfect wavefront (\(M^2 = 1\)) with no phase errors, mechanical distortions, or optical aberrations. The expression $ \theta_{\text{pract}}$ is
for realistic case where wavefront imperfections, pointing instabilities, and non-Gaussian profiles degrade beam quality are present.  

A phased array synthesizes a large effective aperture by coherently combining multiple sub-apertures, forming an effective aperture diameter \(D_{\text{eff}}\). However, practical system imperfections introduce additional divergence, quantified by the beam quality factor \(M_{\text{eff}}^2\). The actual beam divergence, accounting for system imperfections, is:
{}
\begin{equation}
    \theta_{\text{actual}} = M_{\text{eff}}^2 \frac{2\lambda}{\pi D_{\text{eff}}}.
    \label{eq:theta_actual}
\end{equation}

For a spacecraft positioned at a slant range \(d\), the beam spot radius at the lunar surface is:
{}
\begin{equation}
    w = \theta_{\text{actual}} d = M_{\text{eff}}^2 \frac{2\lambda d}{\pi D_{\text{eff}}}.
    \label{eq:beam_radius}
\end{equation}

The corresponding spot area on the lunar surface is:
{}
\begin{equation}
    A_{\text{spot}} = \pi w^2 = \pi \Big(M_{\text{eff}}^2 \frac{2\lambda d}{\pi D_{\text{eff}}}\Big)^2.
    \label{eq:spot_area}
\end{equation}

The power flux density at the receiver, assuming a transmitted optical power \(P_{\text{tx}}\), is:
{}
\begin{equation}
    I_{\text{ideal}} = \frac{P_{\text{tx}}}{A_{\text{spot}}} = \frac{P_{\text{tx}}}{\pi \Big(M_{\text{eff}}^2 \frac{2\lambda d}{\pi D_{\text{eff}}}\Big)^2}.
    \label{eq:power_flux}
\end{equation}

To maximize \(I_{\text{ideal}}\), the system must minimize the beam spot area \(A_{\text{spot}}\). This requires increasing \(D_{\text{eff}}\) to reduce diffraction effects while simultaneously minimizing \(M_{\text{eff}}^2\), ensuring a high-quality beam with minimal power losses.

\subsubsection{Beam Quality and Perturbation Effects}

The beam quality factor \(M_{\text{eff}}^2\) quantifies deviations from an ideal diffraction-limited beam and accounts for various perturbation effects. These effects degrade wavefront coherence and contribute to increased divergence. The primary sources of degradation include:
(1) \textit{Wavefront errors}: Thermal distortions, structural misalignments, and optical aberrations introduce phase inconsistencies across sub-apertures, increasing effective divergence.
(2) \textit{Pointing jitter}: Spacecraft vibrations from reaction wheels, micro-thrusters, and solar radiation pressure induce angular deviations, reducing beam coherence.
(3) \textit{Fill factor effects}: Sparse arrays have incomplete aperture filling, leading to diffraction artifacts that degrade beam quality and increase sidelobe power.

We adopt a bounded main-lobe efficiency model to capture how the fraction of radiated power contained in the coherent main beam varies with the aperture fill fraction in a physically consistent way across \(0<F\leq 1\). The fill fraction \(F\) is defined as the ratio of the actively emitting or collecting physical area to the total geometric aperture area, so that \(F=1\) corresponds to a fully filled aperture and \(F\ll 1\) to a highly sparse array with large inactive gaps. Model parameters are calibrated for representative layouts in Table~\ref{tab:pa-cal}. From diffraction theory, reducing \(F\) increases the fraction of power radiated into grating lobes and sidelobes, lowering the main-lobe fraction \(\eta_{\mathrm{main}}\). Conversely, a densely filled aperture maximizes \(\eta_{\mathrm{main}}\) but still leaves a finite fraction of power in sidelobes due to finite-aperture diffraction and any amplitude taper. The empirical form
\begin{equation}
    \eta_{\mathrm{main}}(F) = \eta_{\infty} - \big(\eta_{\infty} - \eta_{0}\big)\exp\!\Big[-\Big(\frac{F}{F_{c}}\Big)^{p}\Big],
    \label{eq:eta_main_bounded}
\end{equation}
is constructed to satisfy these limits and remain bounded within \(0\leq\eta_{\mathrm{main}}\leq 1\). Here, \(\eta_{\infty}\) (\(0<\eta_{\infty}\leq 1\)) is the dense-array limit set by the intrinsic sidelobe floor of the geometry; \(\eta_{0}\) is the asymptotic main-lobe fraction for highly sparse apertures dominated by grating lobes; \(F_{c}\) is a geometry-dependent ``knee'' fill fraction marking the transition where grating lobes clear the main beamwidth; and \(p\) controls the sharpness of this transition, reflecting how rapidly sidelobe power is redistributed as element density increases. This bounded exponential rise from \(\eta_{0}\) to \(\eta_{\infty}\) is consistent with aperture theory and can be calibrated directly to measured or simulated far-field patterns for a given array layout. 

Extending this to beam quality, the effective beam-quality factor is written as
\begin{equation}
    M_{\mathrm{eff}}^{2} = M_{\mathrm{subap}}^{2} + \frac{\sigma_{\phi}^{2}}{(2\pi)^{2}} + \frac{\sigma_{\mathrm{pos}}^{2}}{D_{\mathrm{eff}}^{2}} + \beta\Big(\frac{1}{\eta_{\mathrm{main}}(F)} - 1\Big),
    \label{eq:m2_bounded}
\end{equation}
where \(M_{\text{subap}}^2\) is the intrinsic beam-quality factor of each sub-aperture (often \(\approx 1\) if near diffraction-limited), \(\sigma_{\phi}^2\) is the variance of wavefront error, \(\sigma_{\mathrm{pos}}\) is the RMS sub-aperture misalignment, and \(\beta\) is a layout-dependent scaling factor. The last term converts main-lobe loss into an equivalent broadening of the far-field spot. In this way, sparse filling not only reduces delivered power via \(\eta_{\mathrm{main}}\) but also enlarges the spot size at the receiver, further lowering capture efficiency. Parameters \(\{\eta_{0},\eta_{\infty},F_{c},p,\beta\}\) are obtained from 2-D wave-optics array-factor simulations for representative layouts (see Table~\ref{tab:pa-cal}).

\begin{figure}[t]
\centering

\begin{minipage}{0.48\linewidth}
\centering
\begin{tikzpicture}
\begin{axis}[
  width=\linewidth, height=0.5\linewidth, 
  xmin=0, xmax=1, ymin=0, ymax=1,
  xlabel={{\footnotesize Fill factor}, {\footnotesize $F$}}, ylabel={$\eta_{\rm main}$},
  tick align=outside,
  tick style={black, thin},
  major tick length=2.5pt,
  minor tick length=1.5pt,
  xtick distance=0.2, ytick distance=0.2,
  minor x tick num=1, minor y tick num=1,
  axis line style={black, thin},
  every axis plot/.append style={black,  thick},
  grid=major, grid style={line width=0.1pt, draw=gray!75},
  title={\footnotesize Main-lobe efficiency},
]
\addplot[domain=0.001:1, samples=400, no marks]
  ({x},{ \etaInf - (\etaInf - \etaZero) * exp( -pow(x/\Fc,\pexp) ) });

\addplot[only marks, mark=*, mark size=1.6pt]
  coordinates {(0.80, { \etaInf - (\etaInf - \etaZero) * exp(-pow(0.80/\Fc,\pexp)) })};
\addplot[only marks, mark=square*, mark size=1.6pt]
  coordinates {(0.20, { \etaInf - (\etaInf - \etaZero) * exp(-pow(0.20/\Fc,\pexp)) })};
\addplot[only marks, mark=triangle*, mark size=1.9pt]
  coordinates {(0.10, { \etaInf - (\etaInf - \etaZero) * exp(-pow(0.10/\Fc,\pexp)) })};
\end{axis}
\end{tikzpicture}
\end{minipage}
\hfill
\begin{minipage}{0.48\linewidth}
\centering
\begin{tikzpicture}
\begin{axis}[
  width=\linewidth, height=0.5\linewidth, 
  xmin=0, xmax=1,
  xlabel={{\footnotesize Fill factor}, {\footnotesize $F$}}, ylabel={$\beta\,\big(\eta_{\rm main}^{-1}-1\big)$},
  tick align=outside,
  tick style={black, semithick},
  major tick length=2.5pt,
  minor tick length=1.5pt,
  xtick distance=0.2,
  minor x tick num=1, minor y tick num=1,
  axis line style={black, thin},
  every axis plot/.append style={black,  thick},
  grid=major, grid style={line width=0.3pt, draw=gray!75},
  title={\footnotesize{Fill-factor penalty for $M^2_{\rm eff}$ in (\ref{eq:m2_bounded})}},
  legend style={draw=none, fill=none, font=\footnotesize, at={(0.98,0.98)}, anchor=north east},
  legend cell align=left,
]
\addplot[domain=0.05:1, samples=400, no marks]
  ({x},{ \betaDense * ( 1/(\etaInf - (\etaInf - \etaZero)*exp(-pow(x/\Fc,\pexp))) - 1 ) });
\addlegendentry{$\beta=1.5$}

\addplot[domain=0.05:1, samples=400, no marks, dashed]
  ({x},{ \betaAper * ( 1/(\etaInf - (\etaInf - \etaZero)*exp(-pow(x/\Fc,\pexp))) - 1 ) });
\addlegendentry{$\beta=2.5$}

\addplot[domain=0.05:1, samples=400, no marks, dotted]
  ({x},{ \betaSparse * ( 1/(\etaInf - (\etaInf - \etaZero)*exp(-pow(x/\Fc,\pexp))) - 1 ) });
\addlegendentry{$\beta=3.5$}
\end{axis}
\end{tikzpicture}
\end{minipage}

\caption{Left: $\eta_{\rm main}(F)$ from \eqref{eq:eta_main_bounded} using $(\eta_0,\eta_\infty,F_c,p)=(0.25,0.98,0.10,1.4)$; markers show reported fills ($F\!\approx\!0.80,0.20,0.10$). Right: the corresponding fill-factor penalty $\beta\big(\eta^{-1}_{\rm main}(F)-1\big)$ (i.e.,  the last term in (\ref{eq:m2_bounded})) for $\beta=1.5,2.5,3.5$.}
\label{fig:eta-main-penalty-pair}
\end{figure}
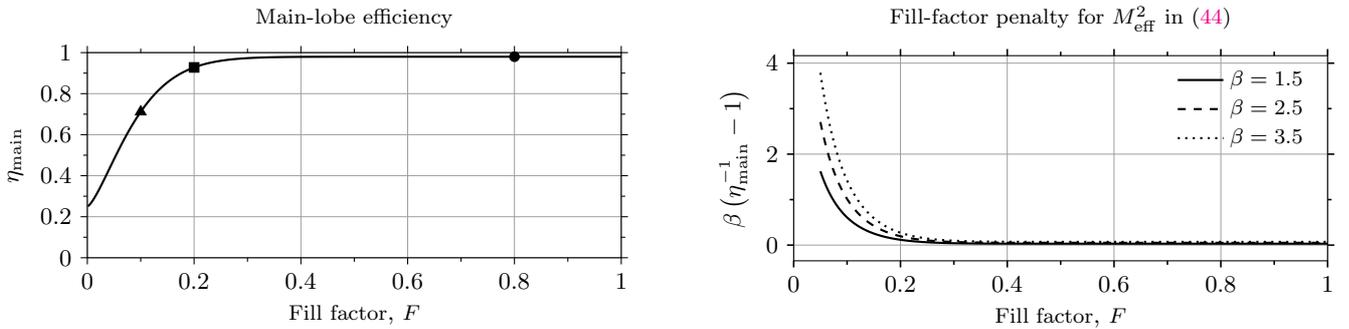

Figure~\ref{fig:eta-main-penalty-pair} shows $\eta_{\mathrm{main}}(F)$ from \eqref{eq:eta_main_bounded} with parameters in Table~\ref{tab:pa-cal} (left), and the corresponding penalty term $\beta(\eta^{-1}_{\mathrm{main}}(F)-1)$ from \eqref{eq:m2_bounded} (right). Markers indicate the operating fills present in Table~\ref{tab:pa-cal}. The plots show that reduced fill-factor $F$ lowers $\eta_{\mathrm{main}}$ and increases the $M^2$ penalty, enlarging the far-field spot.

\begin{table}[h!]
\centering
\caption{Example calibrated parameters for the bounded main-lobe efficiency model in \eqref{eq:eta_main_bounded} and the effective beam-quality factor in \eqref{eq:m2_bounded}, based on 2-D wave--optics simulations for representative phased-array layouts.}
\label{tab:pa-cal}
\begin{tabular}{lccccc}
\hline
Layout & $\eta_{0}$ & $\eta_{\infty}$ & $F_{c}$ & $p$ & $\beta$ \\
\hline\hline
Dense hex ($F\approx0.80$)       & 0.25 & 0.98 & 0.10 & 1.4 & 1.5 \\
Optimized aperiodic ($F\approx0.20$) & 0.25 & 0.98 & 0.10 & 1.4 & 2.5 \\
Sparse-corrected ($F\approx0.10$)& 0.25 & 0.98 & 0.10 & 1.4 & 3.5 \\
\hline
\end{tabular}
\end{table}

\begin{itemize}
\item \textit{Wavefront errors:} 
  $\sigma_{\phi}^2$ (in radians$^2$) is the variance of phase noise induced by 
  wavefront distortions and mechanical misalignments. One typically measures wavefront RMS errors $e$ (nm) at a laser wavelength $\lambda$ (nm), then convert 
  to $\sigma_{\phi} = (2\pi / \lambda)\, e$. 
  It is often useful to also express wavefront quality in terms of the \emph{Strehl ratio} 
  \begin{equation}
    S \simeq \exp(-\sigma_{\phi}^2),
    \label{eq:strehl}
  \end{equation}
  which quantifies the normalized on-axis irradiance relative to the diffraction-limited case. 
  For a diffraction-limited Gaussian beam, empirical correlations indicate that 
  $S \gtrsim 0.6$ (i.e., $\sigma_{\phi} \lesssim 0.64$~rad, or $\approx 100$~nm RMS 
  at $\lambda = 1.064\,\mu\mathrm{m}$) typically keeps $M_{\mathrm{eff}}^2 \lesssim 2$ 
  for the array layouts considered here.
  For example, at $\lambda=1000\,\mathrm{nm}$:
  \begin{itemize}
    \item A 20\,nm wavefront error implies $\sigma_{\phi}\approx0.126$\,rad, 
    $S \approx 0.984$, and contributes $\sigma_{\phi}^2/(2\pi)^2 \approx 4\times10^{-4}$ 
    to $M_{\mathrm{eff}}^2$.
    \item A 50\,nm wavefront error yields $\sigma_{\phi}\approx0.314$\,rad, 
    $S \approx 0.906$, and contributes $\sim2.5\times10^{-3}$ to $M_{\mathrm{eff}}^2$.
  \end{itemize}

\item \textit{Pointing misalignment:}
  $\sigma_{\text{pos}}^2$ ($\mathrm{m}^2$) captures pointing or alignment variance 
  relative to an effective aperture diameter $D_{\text{eff}}$. Lunar-orbit 
  spacecraft often experience micro-vibrations from reaction wheels or thermal 
  flex, leading to 0.1--0.5\,mm misalignments. For example, a 0.2\,mm offset 
  $(2\times10^{-4}\,\mathrm{m})$ on a 2\,m array $(D_{\text{eff}}^2=4\,\mathrm{m}^2)$ 
  implies $\sigma_{\text{pos}}^2=4\times10^{-8}\,\mathrm{m}^2$ and
  $\sigma_{\text{pos}}^2/D_{\text{eff}}^2=1\times10^{-8}$.

\item \textit{Fill factor:}
  $F$ (ranging 0--1) represents the fraction of the phased-array aperture that 
  actively transmits. Arrays with $F<0.10$ can suffer strong sidelobes unless 
  aided by advanced phase correction or additional sub-apertures. In practice, 
  $F=0.15\text{--}0.25$ often balances mass savings with acceptable beam coherence under 
  lunar-orbit constraints, and its impact is represented through $\eta_{\mathrm{main}}(F)$ in Eq.~\eqref{eq:m2_bounded}.
\end{itemize}

We instantiate \eqref{eq:eta_main_bounded} with wave--optics calibration; the parameter sets for the layouts considered are listed in Table~\ref{tab:pa-cal}.
Using the representative values $(\eta_0,\eta_\infty,F_c,p)=(0.25,\,0.98,\,0.10,\,1.4)$,
the main--lobe allocation evaluates to
\[
\eta_{\rm main}(F{=}0.80)\approx 0.98,\quad
\eta_{\rm main}(0.20)\approx 0.93,\quad
\eta_{\rm main}(0.10)\approx 0.71,\quad
\eta_{\rm main}(0.05)\approx 0.48,
\]
illustrating the degradation as fill decreases from dense hex through optimized aperiodic to sparse‐corrected layouts. Combined with the layout--dependent \(\beta\) in \eqref{eq:m2_bounded} (see Table~\ref{tab:pa-cal}), the corresponding main--lobe--induced beam-quality inflation
\(\Delta M^2_{\rm main}=\beta\bigl(\eta_{\rm main}^{-1}-1\bigr)\)
is, for example:
\[
\begin{aligned}
F=0.80:&\quad \Delta M^2_{\rm main}\approx\{0.03,\,0.05,\,0.07\}\ \text{for}\ \beta=\{1.5,\,2.5,\,3.5\};\\
F=0.20:&\quad \Delta M^2_{\rm main}\approx\{0.12,\,0.19,\,0.27\}\ \text{for}\ \beta=\{1.5,\,2.5,\,3.5\};\\
F=0.10:&\quad \Delta M^2_{\rm main}\approx\{0.61,\,1.01,\,1.42\}\ \text{for}\ \beta=\{1.5,\,2.5,\,3.5\}.
\end{aligned}
\]
These numbers make explicit how sparse fill (\(F\!\downarrow\)) shifts energy out of the main lobe and, via \eqref{eq:m2_bounded}, inflates \(M_{\rm eff}^2\).

Under typical lunar mission conditions:
\begin{itemize}
  \item Keeping wavefront RMS below 20--50\,nm at $\lambda=1\,\mu\mathrm{m}$ restricts 
        $\sigma_{\phi}^2/(2\pi)^2$ to $\lesssim 10^{-3}$.
  \item Limiting misalignment to $\lesssim 0.2$\,mm on a 1--2\,m array keeps 
        $\sigma_{\text{pos}}^2 / D_{\text{eff}}^2$ around $10^{-8}\text{--}10^{-7}$.
\end{itemize}

Following \eqref{eq:m2_bounded}, residual phase RMS \(\sigma_\phi\) (in radians at the operating wavelength) contributes additively as
\(\Delta M^2_{\phi}=\sigma_\phi^2/(2\pi)^2\).
For reference, \(\sigma_\phi=2\pi/20\) (i.e., \(\lambda/20\) WFE) yields \(\Delta M^2_{\phi}\simeq 2.5\times10^{-3}\),
while \(\sigma_\phi=2\pi/10\) (\(\lambda/10\)) yields \(\Delta M^2_{\phi}\simeq 10^{-2}\).
This is consistent with the Strehl approximation \(S\approx \exp(-\sigma_\phi^2)\);
to avoid double-counting, we model geometric main-lobe allocation with \(\eta_{\rm main}(F)\) and place phase-control penalties in \(M_{\rm eff}^2\) via \(\Delta M^2_{\phi}\) (and sub-aperture misalignment via \(\Delta M^2_{\rm pos}=\sigma_{\rm pos}^2/D_{\rm eff}^2\)).

For completeness, we report three metrics central to link performance:  
(i) the effective beam‐quality factor $M_{\mathrm{eff}}^2$ \eqref{eq:m2_bounded}, which captures divergence growth from wavefront errors, pointing misalignment, and fill‐factor effects;  
(ii) the Strehl ratio $S$ \eqref{eq:strehl}, the normalized on‐axis irradiance relative to the diffraction‐limited case; and  
(iii) the Gaussian capture fraction $\eta_{\rm cap}(r_{\mathrm{rx}},d)$, defined as the fraction of total transmitted optical power intercepted by a circular receiver of radius $r_{\mathrm{rx}}$ at range $d$.  
With the $1/e^2$ beam radius $w=\theta_{\mathrm{tx}}\,d$ from \eqref{eq:slant-range},  
\begin{equation}
\label{eq:cap_frac}
\eta_{\rm cap}(r_{\mathrm{rx}},d)
\;\equiv\;
1-\exp\!\Big[-\,2\Big(\frac{r_{\mathrm{rx}}}{w}\Big)^2\Big]
=
1-\exp\!\Big[-\,\frac{2\,r_{\mathrm{rx}}^2}{\theta_{\mathrm{tx}}^{\,2} d^{2}}\Big],
\end{equation}
where $r_{\mathrm{rx}}=\sqrt{A_{\mathrm{rx}}/\pi}$ for a receiver of projected area $A_{\mathrm{rx}}$.  
Because $\eta_{\rm cap}$ directly sets the fraction of transmitted power intercepted by the receiver, it provides the most immediate link between optical beam properties and the received‐power term in the link budget. By jointly managing sub‐aperture quality $M_{\text{subap}}^2$, wavefront stability, pointing control, and fill‐factor design---and calibrating the parameters in the bounded $\eta_{\mathrm{main}}(F)$ model of \eqref{eq:eta_main_bounded}---one can maintain $M_{\mathrm{eff}}^2 \approx 2\text{--}5$, supporting near‐diffraction‐limited performance for high‐power lunar laser power beaming.

\subsubsection{Comparison of Beam Divergence and Spot Size Across Architectures}

Table~\ref{tab:beam_divergence_comparison} summarizes the divergence characteristics for different phased-array configurations. The beam-quality factor ranges $M_{\text{eff}}^2$ are obtained from \eqref{eq:m2_bounded} using the representative parameter sets in Table~\ref{tab:pa-cal} for each architecture, with $D_{\text{eff}}$ ranges reflecting typical design scales. Divergence, spot radius, and spot area follow directly from $\theta_{\mathrm{actual}} = M_{\mathrm{eff}}^{2}\,2\lambda/(\pi D_{\mathrm{eff}})$ and $w=\theta_{\mathrm{actual}}d$ for $d=1000$~km.

\begin{table}[h!]
\centering
\renewcommand{\arraystretch}{1.0}
\caption{Comparison of beam divergence and spot size for different phased-array architectures. Assumptions: laser wavelength $\lambda = 1.064\,\mu$m, slant range $d = 1000\,\mathrm{km}$, and $M_{\mathrm{eff}}^2$ calculated from \eqref{eq:m2_bounded} with layout-specific parameters from Table~\ref{tab:pa-cal}.}
\label{tab:beam_divergence_comparison}
\begin{tabular}{lcccc}
\hline
Metric & Dense Array & Optimized Aperiodic & Sparse Corrected & Sparse Array \\
\hline\hline
Effective aperture diameter, $D_{\text{eff}}$ (m) & 1--2 & 3--10 & 5--15 & 5--20 \\
Beam quality, $M_{\text{eff}}^2$ (from \eqref{eq:m2_bounded}) & 1.3--1.5 & 1.8--2.5 & 2.8--3.5 & 5--8 \\
Beam divergence $\theta_{\text{actual}}$ ($\mu$rad) & 0.44--0.95 & 0.38--1.13 & 0.40--1.36 & 0.53--2.71 \\
Spot radius at $d = 1000$ km, $w$ (m)  & 0.44--0.95 & 0.38--1.13 & 0.40--1.36 & 0.53--2.71 \\
Spot area at $d = 1000$ km, $A_{\text{spot}}$ (m$^2$) & 0.61--2.85 & 0.45--4.01 & 0.50--5.81 & 0.88--23.07 \\
\hline
\end{tabular}
\end{table}

\subsubsection{Optimization Strategies for Beam Quality}

Several strategies can mitigate degradation in \(M_{\text{eff}}^2\). These strategies act directly on the terms in \eqref{eq:m2_bounded}, lowering wavefront error, positional misalignment, or improving $\eta_{\mathrm{main}}(F)$:
(1)  \textit{Real-time phase correction}: Adaptive optics and wavefront sensors actively adjust phase errors across sub-apertures.
(2) \textit{Active structural and thermal management}: Using low-expansion materials (e.g., Zerodur, SiC) and active thermal control mitigates expansion-induced phase distortions.
(3) \textit{Enhanced pointing control}: High-precision attitude determination and control systems reduce dynamic misalignment errors.
(4) \textit{Array optimization}: Increasing fill factor \(F\) and using optimized element placement (e.g., aperiodic layouts) reduce sidelobe power and improve beam quality.

\subsection{Different Phased Array Architectures}

Optical phased arrays provide a scalable approach to reducing beam divergence by coherently combining multiple sub-apertures, thereby synthesizing a larger effective aperture. The selection of a phased array architecture significantly impacts beam quality, divergence, sidelobe suppression, and overall power delivery efficiency. 

A monolithic aperture could, in principle, achieve diffraction-limited performance, but mass constraints and structural limitations make such designs impractical for spacecraft. Instead, phased arrays enable large effective apertures while maintaining modularity and redundancy. 

Here we examine four primary phased array architectures: i) dense (filled) arrays, ii) sparse arrays, iii) sparse fill-factor-corrected arrays, and iv) optimized aperiodic arrays. Each configuration balances trade-offs in aperture size, fill factor, beam coherence, and system complexity. The optimal architecture depends on mission-specific constraints such as spacecraft mass, pointing accuracy, available power, and beam quality requirements.

\subsubsection{Dense (Filled) Arrays}

Dense (filled) arrays seek to maximize the fill factor $F$, often achieving $F \approx 50\%\text{--}90\%$, so that nearly the entire geometric aperture actively participates in beam formation. A highly effective layout is a \emph{hexagonal (close‐packed)} arrangement of sub‐apertures, where each new ``ring'' encloses all existing sub‐apertures, thereby preserving a uniform coverage of the aperture plane.

We begin by discussing the hexagonal geometry and array diameter:
{}
\begin{enumerate}
\item \textit{Ring Indexing and Sub‐Aperture Count:}  
   Let the central sub‐aperture be ring $k=0$. For $k\ge1$, imagine a regular hexagon of side 
   length $k$ (in units of center‐to‐center spacing). The number of newly added sub‐apertures 
   on ring $k$ is:
   \begin{equation}
      N_{\mathrm{ring}}(k) 
      \;=\;
      \begin{cases}
         1, & k=0,\\
         6\,k, & k \ge 1,
      \end{cases}
      \label{eq:dense_ring_count}
   \end{equation}
   and the cumulative total of sub‐apertures from ring $0$ up to ring $k$ is
   \begin{equation}
      N_{\mathrm{hex}}(k)
      \;=\;
      1
      \;+\;
      \sum_{m=1}^{k}
      \bigl(6\,m\bigr)
      \;=\;
      1
      +
      3\,k\,(k+1),
      \label{eq:dense_hex_total}
   \end{equation}
   where the initial ``1'' accounts for the single center element at $k=0$.

\item \textit{Array Diameter:}
   If $d_{\mathrm{spacing}}$ is the center-to-center distance between adjacent sub‐apertures, 
   then the effective diameter across opposite edges is
   \begin{equation}
      D_{\mathrm{eff,dense}} 
      \;\approx\;
      (2k+1)\,d_{\mathrm{spacing}},
      \label{eq:dense_eff_diameter}
   \end{equation}
   which can be viewed as $k$ rings outward in each radial direction from the center, plus 
   the center element itself.
\end{enumerate}

Table~\ref{tab:dense_hex_table} illustrates how each ring contributes new sub‐apertures and 
how $D_{\mathrm{eff,dense}}$ grows with ring index $k$.

\begin{table}[ht]
\centering
\caption{Hexagonal dense‐array parameters for ring index $k$. 
$N_{\mathrm{ring}}(k)$ [Eq.~(\ref{eq:dense_ring_count})] is the number of new sub‐apertures
that appear on ring $k$, $N_{\mathrm{hex}}(k)$ [Eq.~(\ref{eq:dense_hex_total})] is
the cumulative total from ring $0$ through $k$, and 
$D_{\mathrm{eff,dense}}\approx(2k+1)\,d_{\mathrm{spacing}}$
[Eq.~(\ref{eq:dense_eff_diameter})] is the approximate diameter.}
\label{tab:dense_hex_table}
\begin{tabular}{r c c c}
\hline
\multicolumn{1}{c}{$k$}
& $N_{\mathrm{ring}}(k)$
& $N_{\mathrm{hex}}(k)$
& $D_{\mathrm{eff,dense}} \approx (2k+1)\,d_{\mathrm{spacing}}$ \\
\hline\hline
0 & 1  & 1   & $1 \times d_{\mathrm{spacing}}$ \\
1 & 6  & 7   & $3 \times d_{\mathrm{spacing}}$ \\
2 & 12 & 19  & $5 \times d_{\mathrm{spacing}}$ \\
3 & 18 & 37  & $7 \times d_{\mathrm{spacing}}$ \\
4 & 24 & 61  & $9 \times d_{\mathrm{spacing}}$ \\
5 & 30 & 91  & $11 \times d_{\mathrm{spacing}}$ \\
6 & 36 & 127 & $13 \times d_{\mathrm{spacing}}$ \\
7 & 42 & 169 & $15 \times d_{\mathrm{spacing}}$ \\
8 & 48 & 217 & $17 \times d_{\mathrm{spacing}}$ \\
9 & 54 & 271 & $19 \times d_{\mathrm{spacing}}$ \\
10 & 60 & 331 & $21 \times d_{\mathrm{spacing}}$ \\
\hline
\end{tabular}
\end{table}

Often, each sub‐aperture has diameter $d_{\mathrm{aperture}}$, with $d_{\mathrm{spacing}} \approx d_{\mathrm{aperture}} + \delta$, where $\delta$ is a small structural margin.  Table~\ref{tab:dense_hex_examples} provides sample arrays for $d_{\mathrm{aperture}}=0.10$\,m and $d_{\mathrm{spacing}}=0.10$\,m, targeting overall diameters near 1--2\,m.

\begin{table}[ht]
\centering
\caption{Examples of dense hexagonal arrays approximating 1--2\,m diameters. Assumes $d_{\mathrm{aperture}}=0.10$\,m and $d_{\mathrm{spacing}}=0.10$\,m. The fill factor $F$ values shown here correspond to the input for $\eta_{\mathrm{main}}(F)$ in \eqref{eq:eta_main_bounded}.}
\label{tab:dense_hex_examples}
\begin{tabular}{lccccc}
\hline
Target Diameter & $k$ & $D_{\mathrm{eff,dense}}$(m)
& $N_{\mathrm{hex}}(k)$ 
& $F$ (for $\eta_{\mathrm{main}}$) 
& Notes \\
\hline\hline
$\sim1.0$\,m & 3 & $7\times0.10=0.70$ & 37 & 0.77 & about 0.7\,m total \\[0pt]
             & 4 & $9\times0.10=0.90$ & 61 & 0.78 & near 1\,m diameter \\
\hline
$\sim2.0$\,m & 9  & $19\times0.10=1.90$ & 271 & 0.79 & nearly 2\,m total \\
             & 10 & $21\times0.10=2.10$ & 331 & 0.80 & slightly above 2\,m \\
\hline
\end{tabular}
\end{table}

Arrays beyond 2\,m are achievable but significantly increase mass, deployment complexity, and alignment demands. If the mission architecture tolerates these challenges, larger arrays can deliver still narrower beam divergence at the lunar surface. For dense hexagonal layouts, the typical fill factor is $F\approx0.75$--$0.80$, which feeds directly into the main‐lobe efficiency model $\eta_{\mathrm{main}}(F)$ in \eqref{eq:eta_main_bounded} when computing the effective beam‐quality factor $M_{\mathrm{eff}}^2$ via \eqref{eq:m2_bounded}.

Realistic beam divergence and $M_{\mathrm{eff}}^2$ will be affected by the array design choice. When each sub‐aperture is nearly diffraction‐limited ($M_{\mathrm{subap}}^2\approx1$),
the entire array can still exhibit a higher $M_{\mathrm{eff}}^2$ from finite fill factor,
alignment errors, or wavefront distortions. In an ideal case,
\begin{equation}
   \theta_{\mathrm{dense}}
   \approx
   M_{\mathrm{eff}}^2
   \;\frac{2 \lambda}{\pi D_{\mathrm{eff,dense}}},
   \label{eq:dense_divergence_ideal}
\end{equation}
where $D_{\mathrm{eff,dense}}$ is from \eqref{eq:dense_eff_diameter}.
In our unified model (\ref{eq:m2_bounded}), $M_{\mathrm{eff}}^2$ for a dense array is given by
\begin{equation}
   M_{\mathrm{eff}}^{2} = M_{\mathrm{subap}}^{2}
   + \frac{\sigma_{\phi}^{2}}{(2\pi)^{2}}
   + \frac{\sigma_{\mathrm{pos}}^{2}}{D_{\mathrm{eff}}^{2}}
   + \beta\!\left(\frac{1}{\eta_{\mathrm{main}}(F)} - 1\right),
   \label{eq:M2_eff_formula_dense_unified}
\end{equation}
where $\eta_{\mathrm{main}}(F)$ is from Eq.~\eqref{eq:eta_main_bounded} and the
dense‐array parameters $\{\eta_0,\eta_\infty,F_c,p,\beta\}$ come from Table~\ref{tab:pa-cal}.

The terms are interpreted as follows:
\begin{itemize}
\item $M_{\mathrm{subap}}^2$ is the intrinsic beam quality of each sub‐aperture (typically $\approx 1$ for diffraction‐limited elements).
\item $\sigma_{\phi}^2/(2\pi)^2$ is the variance in phase (in radians$^2$) due to optical/thermal distortions.  
      For example, $\Delta\ell\approx 20\ \mathrm{nm}$ at $\lambda=1\ \mu\mathrm{m}$ yields  
      $\sigma_{\phi}\approx 0.126$ and $\sigma_{\phi}^2/(2\pi)^2\approx 4\times10^{-4}$.
\item $\sigma_{\mathrm{pos}}^2/D_{\mathrm{eff}}^2$ quantifies RMS misalignment of sub‐apertures relative to the array diameter.
\item The last term, with $\eta_{\mathrm{main}}(F)$, captures the penalty from finite fill factor and sidelobes for the given geometry.
\end{itemize}

Now we will estimating the fill factor $F$. For circular sub‐apertures of diameter $d_{\mathrm{aperture}}$, arranged in a
hexagonal or roughly circular boundary of diameter $D_{\mathrm{eff,dense}}$, a
first‐order fill factor estimate is:
\begin{equation}
   F
   \;\approx\;
   \frac{
     \pi\,\bigl(\tfrac12{d_{\mathrm{aperture}}}\bigr)^2
     \;N_{\mathrm{hex}}
   }{
     \pi\,\bigl(\tfrac12{D_{\mathrm{eff,dense}}}\bigr)^2
   }
   \;=\;
   \frac{
     N_{\mathrm{hex}}
     \,\bigl(d_{\mathrm{aperture}}\bigr)^2
   }{
     \bigl(D_{\mathrm{eff,dense}}\bigr)^2
   },
   \label{eq:fill_factor_example}
\end{equation}
neglecting structural margins. Real designs have frames or partial ring edges that reduce $F$ slightly; dense arrays typically reach $F\approx 0.75\text{--}0.85$.

Table~\ref{tab:dense_m2_examples} shows representative calculations for a 1\,m‐diameter dense array using \eqref{eq:M2_eff_formula_dense_unified} with $M_{\mathrm{subap}}^2=1$ and
the dense‐array parameters from Table~\ref{tab:pa-cal}.
The fill‐factor term is computed  $\beta\big(\eta^{-1}_{\mathrm{main}}(F)-1\big)$.

\begin{table}[ht]
\centering
\caption{Example $M_{\mathrm{eff}}^2$ estimates for a 1\,m‐diameter dense array,
using Eq.~\eqref{eq:M2_eff_formula_dense_unified} with $M_{\mathrm{subap}}^2=1$.
The fill‐factor penalty is computed from $\beta[1/\eta_{\mathrm{main}}(F)-1]$,
where $\eta_{\mathrm{main}}(F)$ is from Eq.~\eqref{eq:eta_main_bounded} with dense‐array parameters.}
\label{tab:dense_m2_examples}
\begin{tabular}{lcccc}
\hline
Case & $\sigma_{\phi}^2/(2\pi)^2$ & $\sigma_{\mathrm{pos}}^2/D_{\mathrm{eff}}^2$
& $F$ & $M_{\mathrm{eff}}^2$ \\
\hline\hline
Ideal & 0 & 0 & 0.80 & 1.35 \\
A     & 0.0010 & $5\times 10^{-8}$ & 0.80 & 1.36 \\
B     & 0.0025 & $2\times 10^{-7}$ & 0.75 & 1.40 \\
C     & 0.0005 & 0 & 0.85 & 1.33 \\
\hline
\end{tabular}
\end{table}

\begin{itemize}
\item \textit{Case A}: Moderate wavefront error and negligible misalignment; $F=0.80$ yields $M_{\mathrm{eff}}^2\approx 1.36$.
\item \textit{Case B}: Larger wavefront error and lower fill factor ($F=0.75$) increase $M_{\mathrm{eff}}^2$ to $\approx 1.40$.
\item \textit{Case C}: Smaller wavefront error and high fill factor ($F=0.85$) give $M_{\mathrm{eff}}^2\approx 1.33$.
\end{itemize}

These scenarios show that, for dense arrays, $M_{\mathrm{eff}}^2$ remains close to $\approx 1.3$--$1.4$ when mechanical and thermal errors are well‐controlled and $F\gtrsim 0.75$. Lower fill factors or degraded stability drive $M_{\mathrm{eff}}^2$ higher, increasing divergence.

As a result, we arrive to the following conclusions on dense Arrays:
\begin{enumerate}
\item \emph{Narrow Divergence:} Dense, close‐packed arrays can approach near‐diffraction‐limited performance with $M_{\mathrm{eff}}^2$ in the low‐1.x range.
\item \emph{Engineering Trade‐Offs:} Each additional ring increases mass, thermal complexity, and alignment demands; arrays beyond $\sim$2\,m require careful structural design.
\item \emph{Fill Factor Importance:} High $F$ values feed into $\eta_{\mathrm{main}}(F)$, directly impacting $M_{\mathrm{eff}}^2$; keeping $F\gtrsim0.75$ is critical.
\end{enumerate}

In summary, a dense hexagonal array can achieve near‐ideal beam quality up to practical diameters of 1--2\,m, subject to overall mass and thermal constraints. For missions that demand minimal divergence and can support the structural overhead, dense arrays are a high‐performance solution for cislunar laser power beaming.

\subsubsection{Sparse Arrays: Large Synthetic Apertures with Extremely Low Fill Factor}
\label{sec:sparse_arrays}

Sparse arrays place sub-apertures at relatively large center-to-center separations, producing an effectively large diameter \(D_{\mathrm{eff,sparse}}\) while drastically reducing the fill factor (\(F \lesssim 5\%\)). Although such layouts can reduce the number of sub-apertures (and hence some structural mass), they typically exhibit significant sidelobes, partial coherence losses,  large beam-quality factors, thereby demanding advanced phasing techniques to achieve useful main-lobe efficiency.

Assume \(N\) sub-apertures are distributed in a sparse configuration with
center-to-center spacing \(d_{\mathrm{spacing}} \gg d_{\mathrm{aperture}}\). For a
simple one-dimensional arrangement, the synthetic aperture diameter is
\begin{equation}
   D_{\mathrm{eff,sparse}} \;\approx\; (N - 1)\,d_{\mathrm{spacing}},
   \label{eq:sparse_deff_approx}
\end{equation}
while a two-dimensional array (e.g.\ an \(N\times N\) grid) could produce
\(\,D_{\mathrm{eff,sparse}} \approx (N-1)\,d_{\mathrm{spacing}}\) in each dimension.
The fill factor \(F\), defined as the fraction of the notional aperture area actually covered by sub-apertures, generally satisfies $ F \approx ({d_{\mathrm{aperture}}}/{d_{\mathrm{spacing}}})^2   \ll 1.$ For example, if \(d_{\mathrm{aperture}}=0.05\)\,m and \(d_{\mathrm{spacing}}=1.0\)\,m,
then \(F=(0.05/1.0)^2=0.25\%\). Such a low fill factor often induces substantial sidelobes.

While enlarging \(D_{\mathrm{eff,sparse}}\) decreases the diffraction-limited divergence, sparse coverage severely degrades beam coherence. In the unified model, the practical half-angle divergence is
\begin{equation}
   \theta_{\mathrm{sparse}}
   \;\approx\;
   M_{\mathrm{eff}}^2
   \,\frac{2\lambda}{\pi D_{\mathrm{eff,sparse}}},
   \label{eq:sparse_theta_approx}
\end{equation}
where \(M_{\mathrm{eff}}^2\) is given by \eqref{eq:m2_bounded} with the sparse-array parameters \(\{\eta_0,\eta_\infty,F_c,p,\beta\}\) in Table~\ref{tab:pa-cal}. For sparse arrays, \(\beta\) is large, so even modest decreases in \(\eta_{\mathrm{main}}(F)\) from \eqref{eq:eta_main_bounded} can raise \(M_{\mathrm{eff}}^2\) by an order of magnitude exceeding 5--30 due to:
\begin{enumerate}
\item \textit{Wide sub-aperture spacing:}
   Broad gaps produce strong sidelobes and partial coherence unless finely tuned    wavefront corrections are applied.
\item \textit{Mechanical alignment complications:}
   Widely spaced elements can shift relative to one another under spacecraft  vibrations or thermal distortions, introducing large phase errors.
\item \textit{Thermal gradients:}
   Each distant sub-aperture may experience different illumination or cooling,  complicating the array’s overall wavefront matching.
\end{enumerate}
If $F\ll1$ without rigorous metrology and phase adjustment, \(M_{\mathrm{eff}}^2\) is likely to grow large, pushing most of the transmitted power outside the main lobe.

The main-lobe efficiency for a sparse array is computed from the bounded model (\ref{eq:eta_main_bounded}),
\begin{equation}
   \eta_{\mathrm{main}}(F) =
   \eta_{\infty} - (\eta_{\infty} - \eta_{0})
   \exp\!\Big[-\Big(\frac{F}{F_c}\Big)^{p}\Big],
   \label{eq:sparse_eta_main_bounded2a}
\end{equation}
with \(\{\eta_0,\eta_\infty,F_c,p\}\) chosen for the sparse layout. This form is physically bounded and can be calibrated to diffraction simulations of specific
array geometries. In practice, very low \(F\) for sparse arrays leads to low \(\eta_{\mathrm{main}}\) unless elaborate beam-shaping or phase correction is applied.

Because a sparse array leaves most of the nominal aperture empty, strong sidelobes and multi-lobed far-field patterns are common without advanced beam-shaping. In the unified model, the high \(\beta\) values for sparse arrays (Table~\ref{tab:pa-cal}) mean that low \(F\) drives the last term in Eq.~\eqref{eq:m2_bounded} to dominate,
producing \(M_{\mathrm{eff}}^2\) in the 10--30 range unless real-time phase correction and wavefront control are applied.

Table~\ref{tab:sparse_examples} shows how $M_{\mathrm{eff}}^2$ changes with very low
fill factors for the sparse layout, computed from Eq.~\eqref{eq:m2_bounded} with 
$M_{\mathrm{subap}}^2=1$, $\sigma_{\phi}^2/(2\pi)^2=10^{-3}$, 
$\sigma_{\mathrm{pos}}^2/D_{\mathrm{eff}}^2=10^{-7}$, and the sparse-array parameters from
Table~\ref{tab:pa-cal}. The fill-factor penalty uses $\beta (\eta^{-1}_{\mathrm{main}}(F)-1)$.

\begin{table}[ht]
\centering
\caption{Example $M_{\mathrm{eff}}^2$ for sparse arrays at very low fill factors, using sparse-array parameters from Table~\ref{tab:pa-cal}.}
\label{tab:sparse_examples}
\begin{tabular}{cccc}
\hline
$F$ & $\eta_{\mathrm{main}}(F)$ & Penalty term $\beta(\eta^{-1}_{\mathrm{main}} - 1)$ & $M_{\mathrm{eff}}^2$ \\
\hline\hline
0.01 & 0.18 & 22.78 & 23.78 \\
0.02 & 0.27 & 13.56 & 14.56 \\
0.05 & 0.41 & 8.59 & 9.59 \\
\hline
\end{tabular}
\end{table}

These values illustrate the steep penalty at low $F$: even at $F=0.05$, $M_{\mathrm{eff}}^2$ remains near~9.6 for the assumed wavefront and pointing conditions, showing why sparse arrays require either significant fill-factor correction or highly effective sidelobe suppression to achieve good on-target performance.

Below we present some practical considerations for sparse arrays:
\begin{itemize}
\item \textit{Mass and Deployment Advantages:}
   A sparse array can consist of fewer sub-apertures, simplifying some mechanical  and stowage aspects. However, the electronics and optical metrology needed to synchronize widely spaced elements can negate mass savings.
\item \textit{High $M_{\mathrm{eff}}^2$ and Sidelobes:}
   Extremely low fill factors increase $M_{\mathrm{eff}}^2$ dramatically, diverting
   much of the transmitted power into sidelobes unless advanced wavefront control
   is used.
\item \textit{Pointing and Alignment Complexity:}
   Large inter-element separations magnify any alignment errors or s/c jitter. Precision attitude control and sub-wavelength metrology are needed to preserve coherence in the main lobe.
\item \textit{Suitability for Long-Range Beaming:}
   A large synthetic diameter can, in principle, reduce nominal divergence, but only if the array is phased with high accuracy at high update rates. This is feasible only if the mission architecture supports intensive real-time control.
\end{itemize}

Sparse arrays can achieve very large \(D_{\mathrm{eff,sparse}}\) with relatively few sub-apertures, potentially reducing some mass or stowage constraints. However, the combination of low fill factor \((F \lesssim 5\%)\) and large mechanical separation greatly inflates \(M_{\mathrm{eff}}^2\) and sidelobe levels, often compromising on-target power. In the unified model, the sparse-array parameters in Table~\ref{tab:pa-cal} quantify this penalty through the
\(\beta\big(\eta^{-1}_{\mathrm{main}}(F) - 1\big)\) term. Missions considering sparse arrays must balance the potential mass advantage against the complexity of maintaining coherence in the main lobe via precise metrology and advanced beam-steering systems.

\subsubsection{Sparse (Fill-Factor-Corrected) Arrays: Moderate Fill Factor for Improved Coherence}
\label{sec:sparse_fill_corrected}

Sparse fill-factor-corrected arrays augment an otherwise sparse layout with additional sub-apertures, increasing the fill factor to $F \approx 5\%\text{--}20\%$ and thereby improving beam quality relative to a purely sparse configuration. This approach aims to balance mass savings (compared to a fully dense array) against the need for acceptable main-lobe efficiency and a manageable $M_{\mathrm{eff}}^2$.

In the unified framework, the beam-quality factor for a sparse fill-factor-corrected array is given by (\ref{eq:m2_bounded}) as below
\begin{equation}
    M_{\mathrm{eff}}^{2} = M_{\mathrm{subap}}^{2}
    + \frac{\sigma_{\phi}^{2}}{(2\pi)^{2}}
    + \frac{\sigma_{\mathrm{pos}}^{2}}{D_{\mathrm{eff}}^{2}}
    + \beta\!\left(\frac{1}{\eta_{\mathrm{main}}(F)} - 1\right),
    \label{eq:m2_sparse_fillcorr_unified}
\end{equation}
where $\eta_{\mathrm{main}}(F)$ is given by \eqref{eq:eta_main_bounded} and the parameters
$\{\eta_0,\eta_\infty,F_c,p,\beta\}$ correspond to the sparse-corrected layout in Table~\ref{tab:pa-cal}.
Compared to purely sparse arrays, the higher $F$ here raises $\eta_{\mathrm{main}}(F)$ significantly,
reducing the fill-factor penalty term $\beta\big(\eta^{-1}_{\mathrm{main}}(F) - 1\big)$ and thereby lowering $M_{\mathrm{eff}}^2$ into the 2--5 range for $F \approx 0.10$--0.20.

Once $M_{\mathrm{eff}}^2$ is computed from \eqref{eq:m2_sparse_fillcorr_unified},
the half-angle beam divergence is
\begin{equation}
    \theta_{\mathrm{sparse+fill}}
    \approx
    M_{\mathrm{eff}}^2
    \frac{2 \lambda}{\pi D_{\mathrm{eff,sparse+fill}}},
    \label{eq:theta_sparse_fillcorr}
\end{equation}
where $D_{\mathrm{eff,sparse+fill}}$ is set by the outer dimensions of the array and the chosen
sub-aperture spacing. Higher $\eta_{\mathrm{main}}(F)$ directly improves $M_{\mathrm{eff}}^2$ and thus
reduces $\theta_{\mathrm{sparse+fill}}$.

Key design considerations include the following areas:
\begin{itemize}
\item \textit{Reduced Sidelobes Compared to Purely Sparse Layouts:}
  Adding sub-apertures in selected “gap” regions boosts $\eta_{\mathrm{main}}(F)$ from a few percent to  the 5--20\% range, which can substantially cut sidelobe power. Even modest fill-factor increases (e.g.\ from 5\% to 10\%) may prevent $M_{\mathrm{eff}}^2$ from ballooning above 10--20.
  \item \textit{Maintaining Large Effective Diameter:}
      Although fill-factor correction raises $F$, the array still remains far from a fully dense design, preserving much of the mass advantage. The larger synthetic diameter still potentially yields a lower diffraction‐limited divergence, provided sub-aperture phasing is well controlled.
\item \textit{Phase Synchronization Complexity:}
 More sub-apertures and partial fill in non-uniform patterns demand intricate placement algorithms and real-time wavefront control  to maintain high $\eta_{\mathrm{main}}(F)$. Aligning these sub-apertures can be challenging, requiring additional sensors or actuators to maintain coherence across the array.       
\item \textit{Thermal and Mechanical Stability:}
      Uneven distribution of added elements can cause local phase errors if not carefully managed
      with thermal control and structural precision.
\end{itemize}

Sparse fill-factor-corrected arrays retain many benefits of sparse arrays---lower total sub-aperture count and potentially lighter structures---while achieving markedly improved beam quality. With $F=0.10$--0.20 and appropriate sparse-corrected parameters in
Table~\ref{tab:pa-cal}, $\eta_{\mathrm{main}}(F)$ remains high enough to keep $M_{\mathrm{eff}}^2$ in the 2--5 range, limiting sidelobes to a manageable fraction of the total power.

Table~\ref{tab:sparse_fillcorr_examples} illustrates how $M_{\mathrm{eff}}^2$ varies with fill factor $F$ for the sparse‐corrected layout, using \eqref{eq:m2_sparse_fillcorr_unified} 
with $M_{\mathrm{subap}}^2=1$, $\sigma_{\phi}^2/(2\pi)^2=10^{-3}$, and 
$\sigma_{\mathrm{pos}}^2/D_{\mathrm{eff}}^2=10^{-7}$. 
The fill‐factor penalty is computed from $\beta(\eta^{-1}_{\mathrm{main}}(F)-1)$ with $\{\eta_0,\eta_\infty,F_c,p,\beta\}$ from Table~\ref{tab:pa-cal}. These values show how increasing $F$ from 0.10 to 0.20 can nearly halve the fill‐factor penalty term, reducing $M_{\mathrm{eff}}^2$ from above~4 to near~2.3 
for the assumed wavefront and alignment conditions.

\begin{table}[ht]
\centering
\caption{Example $M_{\mathrm{eff}}^2$ estimates for a sparse fill‐factor‐corrected array. 
Parameters are from the sparse‐corrected row of Table~\ref{tab:pa-cal}.}
\label{tab:sparse_fillcorr_examples}
\begin{tabular}{cccc}
\hline
$F$ & $\eta_{\mathrm{main}}(F)$ & Penalty term $\beta(\eta^{-1}_{\mathrm{main}} - 1)$ & $M_{\mathrm{eff}}^2$ \\
\hline\hline
0.10 & 0.53 & 3.11 & 4.11 \\
0.15 & 0.65 & 1.88 & 2.88 \\
0.20 & 0.73 & 1.29 & 2.29 \\
\hline
\end{tabular}
\end{table}

By selectively adding sub-apertures to raise $F$ beyond the few-percent levels of purely sparse layouts, one can significantly reduce the $\beta\,[\eta^{-1}_{\mathrm{main}}(F) - 1]$ penalty and thus $M_{\mathrm{eff}}^2$. The main trade-off is increased mechanical complexity and real-time phasing overhead. Despite these added demands, fill-factor-corrected arrays often strike a practical balance between aperture size, mass constraints, and beam coherence, making them attractive when dense arrays are infeasible but purely sparse
layouts suffer unacceptable sidelobe penalties.

\subsubsection{Optimized Aperiodic Arrays: Computationally Enhanced Layouts}
\label{sec:aperiodic_arrays}

Whereas sparse or fill-factor-corrected designs mitigate some sidelobe power, they do not fully eliminate the diffraction artifacts introduced by partial filling. In contrast, \emph{optimized aperiodic arrays} employ computationally intensive techniques (e.g., genetic algorithms, simulated annealing, iterative Fourier transforms, or discrete structures such as Golomb rulers) to determine precise sub-aperture placements to maximize main-lobe efficiency and suppress sidelobes.

In the unified framework (\ref{eq:m2_bounded}), the beam-quality factor for an optimized aperiodic array is
\begin{equation}
    M_{\mathrm{eff}}^{2} = M_{\mathrm{subap}}^{2}
    + \frac{\sigma_{\phi}^{2}}{(2\pi)^{2}}
    + \frac{\sigma_{\mathrm{pos}}^{2}}{D_{\mathrm{eff}}^{2}}
    + \beta\!\left(\frac{1}{\eta_{\mathrm{main}}(F)} - 1\right),
    \label{eq:m2_aperiodic_unified}
\end{equation}
where $\eta_{\mathrm{main}}(F)$ is from \eqref{eq:eta_main_bounded} and the parameters
$\{\eta_0,\eta_\infty,F_c,p,\beta\}$ correspond to the optimized aperiodic layout in 
Table~\ref{tab:pa-cal}. Optimization raises $\eta_{\mathrm{main}}(F)$ for a given $F$ 
by strategically distributing sub-apertures to minimize sidelobe levels, thus reducing
the penalty term $\beta(1/\eta_{\mathrm{main}}(F) - 1)$ and keeping 
$M_{\mathrm{eff}}^2$ in the $\sim$1.5--3 range for $F \approx 0.15$--0.25.

The half-angle beam divergence for an optimized aperiodic array is
\begin{equation}
    \theta_{\mathrm{opt}}
    \approx
    M_{\mathrm{eff}}^2
    \frac{2\lambda}{\pi D_{\mathrm{eff,opt}}},
    \label{eq:theta_aperiodic}
\end{equation}
where $D_{\mathrm{eff,opt}}$ is the outer boundary of the array geometry. Higher $\eta_{\mathrm{main}}(F)$ in these layouts reduces $M_{\mathrm{eff}}^2$, yielding narrower divergence compared to sparse or sparse-fill-corrected arrays of similar diameter.

Table~\ref{tab:aperiodic_examples} shows example $M_{\mathrm{eff}}^2$ results for selected $F$ values, assuming $M_{\mathrm{subap}}^2=1$, $\sigma_{\phi}^2/(2\pi)^2=10^{-3}$, $\sigma_{\mathrm{pos}}^2/D_{\mathrm{eff}}^2=10^{-7}$, and the optimized aperiodic parameters from Table~\ref{tab:pa-cal}. These values demonstrate that optimized aperiodic layouts can achieve $M_{\mathrm{eff}}^2$ values close to dense arrays while operating at significantly lower fill factors.

\begin{table}[ht]
\centering
\caption{Example $M_{\mathrm{eff}}^2$ for optimized aperiodic arrays.}
\label{tab:aperiodic_examples}
\begin{tabular}{cccc}
\hline
$F$ & $\eta_{\mathrm{main}}(F)$ & Penalty term $\beta\big(1/\eta_{\mathrm{main}} - 1\big)$ & $M_{\mathrm{eff}}^2$ \\
\hline\hline
0.15 & 0.71 & 1.01 & 2.01 \\
0.20 & 0.78 & 0.64 & 1.64 \\
0.25 & 0.83 & 0.43 & 1.43 \\
\hline
\end{tabular}
\end{table}

As far as the relevant design methods are concerned, techniques such as genetic algorithms, simulated annealing, or iterative FFT-based optimizations iteratively adjust sub-aperture locations to minimize sidelobes or maximize main-lobe efficiency. Cost functions may penalize peak sidelobe levels, target a specific
beam pattern, or balance multiple criteria. For arrays with tens or hundreds of sub-apertures, these methods can require substantial offline computation; the final layout is then implemented in hardware.

One also needs to address implementation complexity and real-time phasing. Although optimized aperiodic arrays achieve better sidelobe suppression for a given fill factor, they also introduce design and operational challenges:
\begin{itemize}
\item \textit{High Computational Overhead:}
   Generating an aperiodic layout often involves iterative numerical methods and multi-parameter optimizations. Large arrays (e.g.\ 100--500 sub-apertures) can require substantial computation time even before launch.
\item \textit{Precise Metrology and Real-Time Phase Control:}
   The sub-apertures must be phased accurately according to the optimized layout. Any mechanical offset or thermal drift can degrade $M_{\mathrm{opt}}^2$. Maintaining the array’s intended wavefront quality in lunar orbit necessitates active wavefront sensing, feedback loops, and possibly micro-actuated optical elements.
\item \textit{Structural and Thermal Constraints:}
   Irregular sub-aperture placements can complicate mechanical design (e.g.\ foldable booms, partial frames). Each sub-aperture may also experience different thermal environments, requiring local or zonal thermal management.
\end{itemize}

For lunar power beaming, aperiodic arrays can deliver a high main-lobe fraction at a moderate fill factor, balancing some mass savings against dense-array complexity. The improvement over simpler sparse or fill-factor-corrected designs can be substantial, sometimes halving the sidelobe fraction. However, mission planners must weigh the increased on-board computational load (if in-flight reconfiguration is planned), the detailed mechanical design for irregular placements, and the demanding real-time phasing and thermal management required.

\subsubsection{Comparative Performance among Phased-Array Architectures}
\label{sec:phased_array_performance}

Table~\ref{tab:phased_array_perf} compares four common phased-array architectures in lunar laser power beaming by listing typical ranges for effective aperture diameter, beam-quality factor \((M_{\mathrm{eff}}^2)\), resulting beam divergence, and sidelobe power fraction. Values are derived from \eqref{eq:m2_bounded} using the parameter sets in Table~\ref{tab:pa-cal} and assuming $\lambda=1.064\,\mu$m and $d=1000$\,km.
Numbers are indicative design ranges, not absolute limits.

\begin{table}[ht]
\centering
\renewcommand{\arraystretch}{1.0}
\caption{Representative performance metrics of phased-array architectures for lunar laser power beaming. Divergence is computed as $\theta_{\mathrm{actual}} = M_{\mathrm{eff}}^2\, 2\lambda / (\pi D_{\mathrm{eff}})$. 
Sidelobe fraction is $100\times(1-\eta_{\mathrm{main}})$.}
\label{tab:phased_array_perf}
\begin{tabular}{lcccc}
\hline
Metric 
& Dense Array 
& Sparse Array 
& Sparse Corrected 
& Optimized Aperiodic \\
\hline\hline
Effective aperture diameter, 
\(D_{\mathrm{eff}}\) (m)
& 1--2
& 5--20
& 5--15
& 3--10
\\

Beam quality, 
\((M_{\mathrm{eff}}^2)\)
& 1.3--1.5
& 9.5--23.8
& 2.3--4.1
& 1.4--2.0
\\

Beam divergence,
\(\theta_{\mathrm{actual}}\) 
($\mu$rad)
& 0.43--0.95
& 0.34--2.56
& 0.31--1.08
& 0.27--0.72
\\

Sidelobe power fraction (\%)
& $\lesssim$5
& 30--50
& 20--30
& $\lesssim$15
\\
\hline
\end{tabular}
\end{table}

Architectural notes:
\begin{itemize}
\item \textit{Dense array:} Near-diffraction-limited beam quality 
(\(M_{\mathrm{eff}}^2\approx1.3\text{--}1.5\)), sidelobes typically under 5\%. Diameter often limited to 1--2\,m by mass and thermal constraints.

\item \textit{Sparse array:} Larger diameter (5--20\,m) from fewer sub-apertures, but \(M_{\mathrm{eff}}^2\) in the 10--20+ range and sidelobes of 30--50\%. Requires robust real-time phasing to avoid main-lobe losses.

\item \textit{Sparse corrected:} Adds sub-apertures to improve coherence, reducing sidelobes to about 20--30\% and capping \(M_{\mathrm{eff}}^2\) near 2--4. 
Balances mass savings with beam quality improvements.

\item \textit{Optimized aperiodic:} Computationally optimized placement yields  \(M_{\mathrm{eff}}^2\approx1.4\text{--}2.0\) and sidelobes under 15\%, 
while keeping \(F\) moderate. Requires high-precision metrology and real-time control.
\end{itemize}

\subsection{Comparison of Phased Array Architectures}
\label{sec:phased_array_architectures}

The performance characteristics of different phased array architectures are summarized in Table~\ref{tab:phased_array_performance}, which outlines key trade-offs in beam quality, divergence, sidelobe levels, and system complexity. This comparison is essential for selecting the optimal architecture based on mission constraints such as mass limitations, pointing accuracy, available power, and beam quality requirements.

\begin{table}[h!]
\centering
\renewcommand{\arraystretch}{1.0}
\caption{Representative performance metrics of phased-array architectures for lunar laser power beaming. Beam-quality factors and divergences are calculated from the unified model~(\ref{eq:m2_bounded}) using parameter sets from Table~\ref{tab:pa-cal}, assuming transmitted optical power $P_{\text{tx}}=1000$\,W, laser wavelength $\lambda=1.064\,\mu$m, and slant range $d=1000$\,km. Beam divergence values include practical degradation from $M_{\mathrm{eff}}^2$. {Note:} The divergence, spot size, and flux density ranges are \emph{representative mid-case values} from Eqs.~(\ref{eq:theta_actual})--(\ref{eq:power_flux}), computed for typical $M_{\mathrm{eff}}^2$ and $D_{\mathrm{eff}}$ within each architecture’s range. They do not represent the full extrema implied by the $D_{\mathrm{eff}}$ and $M_{\mathrm{eff}}^2$ bands in the left columns.}
\label{tab:phased_array_performance}
\begin{tabular}{lcccc}
\hline
Metric & Dense Array & Sparse Array & Sparse Corrected & Optimized Aperiodic \\
\hline\hline
Effective aperture $D_{\mathrm{eff}}$ (m) & 1--2 & 5--20 & 5--15 & 3--10 \\
Fill factor $F$ & 50--90\% & $<$5\% & 5--20\% & 10--30\% \\
Beam quality $M_{\mathrm{eff}}^2$ & 1.3--1.5 & 9.6--23.8 & 2.3--4.1 & 1.4--2.0 \\
Beam divergence $\theta_{\mathrm{actual}}$ ($\mu$rad) & 0.43--0.95 & 0.34--2.56 & 0.31--1.08 & 0.27--0.72 \\
Spot radius $w$ (m) & 0.43--0.95 & 0.34--2.56 & 0.31--1.08 & 0.27--0.72 \\
Spot area $A_{\mathrm{spot}}$ (m$^2$) & 0.58--2.84 & 0.36--20.6 & 0.30--3.67 & 0.23--1.63 \\
Ideal power density $I_{\mathrm{ideal}}$ (W/m$^2$) & 352--1724 & 48--2778 & 272--3333 & 613--4348 \\
Sidelobe power fraction (\%) & $\lesssim$5 & 30--50 & 20--30 & $\lesssim$15 \\
Complexity$^\dagger$ & High & Moderate & Moderate & Moderate--High \\
\hline
\end{tabular}
\begin{flushleft}
\footnotesize{$^\dagger$``Complexity'' reflects qualitative system integration difficulty, including metrology, structural stability, thermal control, and real-time phasing requirements.}
\end{flushleft}
\end{table}

Sparse arrays achieve large effective apertures but suffer from high sidelobe losses, which significantly reduce beam efficiency unless advanced phase correction techniques are applied. Dense arrays offer superior beam quality and minimal sidelobes but impose strict structural stability and mass constraints, making them more suitable for spacecraft with high-power budgets. Sparse fill-factor-corrected arrays improve beam quality by increasing the aperture fill factor, offering a compromise between beam efficiency and system complexity. Optimized aperiodic arrays refine aperture placement to suppress sidelobes while maintaining a large effective aperture, making them a promising candidate for adaptive power beaming systems.

The choice of an array architecture depends on mission requirements, including spacecraft mass constraints, available power, and beam-steering precision. Dense arrays are suitable for applications prioritizing beam quality over mass efficiency, while sparse and optimized aperiodic arrays are preferable for missions requiring large apertures with manageable complexity. Sparse corrected arrays provide a balance between aperture efficiency and sidelobe suppression, making them a viable option where phase control limitations exist.

\subsection{Technical Realization of Phased Arrays on Spacecraft}
\label{subsec:implementation}

Deploying a large-aperture phased array on a lunar-orbiting spacecraft entails meeting stringent mechanical, thermal, and computational requirements while preserving phase coherence across potentially hundreds of sub-apertures. Although the four primary architectures (dense, sparse, fill-factor-corrected, and optimized aperiodic) differ in fill factor \(F\) and beam quality \(M_{\text{eff}}^2\), each must address similar challenges in structural deployment, thermal stabilization, sub-aperture metrology, and real-time wavefront control. Here \(M_{\mathrm{eff}}^2\) and \(F\) are defined in the unified beam-quality model of \eqref{eq:m2_bounded}, with architecture-specific parameters in Table~\ref{tab:pa-cal}. The following subsections outline the engineering approaches required to implement these arrays effectively.

\begin{itemize}
\item{\it Mechanical Structure and Deployment:}
Phased arrays designed for orbital laser transmission frequently exceed the dimensions of standard launch fairings, necessitating segmented or foldable assemblies. Dense arrays (with fill factors above 50--90\%) generally employ heavier, high-stiffness frameworks to maintain inter-segment alignment to within \(\pm10\)--100\,\(\mu\)m. Sparse or aperiodic arrays, with fill factors under 30\%, can use lighter truss or boom structures that cut overall mass by up to 50\%. However, they demand more extensive metrology to track each sub-aperture’s position to sub-wavelength tolerances. During operation, reaction wheels and thruster firings induce micro-vibrations in the 1--10\,Hz range, risking loss of phase coherence unless passive or active isolators reduce relative motion below tens of nanometers.

\item{\it Thermal and Radiation Management:}
A spacecraft in lunar orbit experiences temperature differentials of over 100\,K during its orbital cycle, and accumulative radiation doses can reach 10--100\,krad over multi-year missions. Structural materials with low coefficients of thermal expansion (e.g., Zerodur at \(\text{CTE}\approx10^{-8}\,\text{K}^{-1}\) or SiC at \(\text{CTE}\approx2\times10^{-6}\,\text{K}^{-1}\)) help mitigate thermally induced distortions across large apertures. In addition, localized heaters or zonal cooling loops can even out temperature gradients essential for maintaining \(M_{\text{eff}}^2<5\). Meanwhile, electronic modules (laser diodes, phase modulators, wavefront sensors) must be rad-hard or shielded against cislunar radiation---particularly for systems expected to deliver 100s to 1000s of hours of continuous operation.

\item{\it Sub-Aperture Alignment and Wavefront Control:}
Each sub-aperture in the array must remain phase-locked to a reference oscillator, typically within \(\pm50\)--100\,nm path length error for a 1.064\,\(\mu\)m wavelength (\(\lambda/20\) criterion). Real-time wavefront sensing at 1--10\,kHz update rates detects misalignments caused by mechanical drift and thermal expansion, while electro-optic modulators or micro-actuated mirrors perform fine phase or tip/tilt corrections. Sparse or aperiodic arrays with element spacing up to 10--20\,m may need laser interferometric metrology systems capable of measuring sub-aperture positions with \(\pm50\)--100\,nm precision. By continuously adjusting each segment’s piston and tilt, the collective aperture can be summed coherently to form a narrow beam aimed at a ground receiver.

\item{\it Mass and Power Budgets:}
Dense arrays, which minimize sidelobes but necessitate robust frames, can weigh as much as 20--40\,kg per m\(^2\) of active aperture. Sparse arrays, while reducing mass to 10--25\,kg/m\(^2\), rely on substantial metrology hardware and more sophisticated phase correction software to fill in empty regions. For the optical source stage alone, wall-plug efficiency typically ranges from 30--50\%. At these values, producing 1\,kW of optical output requires 2--3\,kW of electrical input, along with 1--2\,kW of thermal dissipation through radiators or active cooling loops. If solar power is intermittent (e.g., in polar orbits with partial eclipses), large on-board batteries or supercapacitors buffer laser operation. Balancing these trade-offs---mass, sidelobe power, beam quality, and thermal overhead---drives architectural decisions.

\item{\it Real-Time Computation and Fault Tolerance:}
Operating an array of \(N=100\)--500 sub-apertures at 1--2\,kHz control loops results in \(10^5\)--\(10^6\) actuator commands per second, requiring on-board processors (FPGA or GPU) with 0.1--1\,TFLOPS capacity for steady-state phasing and beam steering. Optimized aperiodic arrays may push computational loads higher if adaptive sidelobe minimization or in-flight layout optimization is implemented. Sub-aperture failures or sensor dropouts must be handled gracefully by the beamforming software, which down-weights or excludes failing segments from the coherence sum. Maintaining adequate wavefront coherence despite a \(\sim\)5--10\% segment loss ensures partial functionality rather than a total system failure.
\end{itemize}

\begin{table}[h!]
\centering
\renewcommand{\arraystretch}{1.0}
\caption{Representative engineering parameters for different phased-array architectures on a spacecraft. Values are typical design ranges for 1--10\,m effective apertures; ``mass per m$^2$'' refers to structural and optical support mass, excluding lasers, power generation and storage, and thermal control systems.}
\label{tab:phased_array_eng}
\begin{tabular}{lcccc}
\hline
Parameter & Dense (Filled) & Sparse & Sparse Corrected & Aperiodic \\
\hline\hline
Nominal fill factor $F$ (\%) & 50--90 & $<$5 & 5--20 & 10--30 \\
Mass per m$^2$ (kg/m$^2$) & 20--40 & 10--25 & 15--30 & 15--35 \\
Phase alignment tolerance (nm) & 50--100 & 50--80 & 50--80 & 50--80 \\
Wavefront sensors needed & 1--2/element & 1--2/element & 1--2/element & 1--2/element \\
Computational load (TFLOPS) & 0.1--0.5 & 0.1--1.0 & 0.2--1.0 & 0.3--2.0 \\
Thermal requirements & Strict & Moderate & Moderate & Moderate \\
Structure complexity & High & Moderate & Moderate & Moderate--High \\
\hline
\end{tabular}
\end{table}

Table~\ref{tab:phased_array_eng} summarizes key engineering parameters for each array architecture, highlighting mass per unit aperture, fill factor, alignment tolerances, and computational demand. Although dense arrays feature simpler wavefront corrections and minimal sidelobes at the expense of higher structural mass, sparser or aperiodic designs reduce mechanical loads but require advanced metrology and more complex real-time control. Successful phased-array implementations integrate robust thermal management, accurate sub-aperture metrology (to \(\pm50\)--100\,nm), and efficient (kHz-level) phase correction loops, enabling large effective apertures in lunar orbit that deliver the high flux densities necessary for reliable power beaming to the lunar surface.

A $10\,$m aperiodic phased-array transmitter with an areal density of $15\text{--}35~\mathrm{kg\,m^{-2}}$ has a structural mass of $\sim 1.2\text{--}2.8~\mathrm{t}$, excluding the laser sources, radiators, and energy storage subsystems. For a transmitted optical power of $P_{\rm tx}=100~\mathrm{kW}$ and an end-to-end efficiency\footnote{Includes optical source wall-plug efficiency, beam transport, pointing, and receiver conversion losses.} of $\eta_{\rm chain}=0.35$ in favorable link geometry, the corresponding daily delivered electrical energy to the surface is  $\approx 30\text{--}50~\mathrm{kWh}$. The ratio of array structural mass to delivered energy is therefore
\[
\frac{M_{\rm struct}}{E_{\rm day}}
\; \approx \; 24\text{--}93~\mathrm{g\,(Wh/day)^{-1}}
\quad\text{or}\quad
24\text{--}93~\mathrm{kg\,(kWh/day)^{-1}},
\]
where $M_{\rm struct}$ refers only to the phased-array structural mass. In terms of continuous output, these values correspond to $\sim 580\text{--}2240~\mathrm{kg}$ per continuous kilowatt (averaged over 24 hours).

For the case of a $D_{\mathrm{eff}} = 10$\,m optimized aperiodic array, $P_{\mathrm{tx}} = 100$\,kW, and $\eta_{\mathrm{chain}} = 0.35$, the continuous‐view, full‐capture limit would yield
$
E_{\mathrm{day,max}} = P_{\mathrm{tx}}\,\eta_{\mathrm{chain}} \times 24\,\mathrm{h} \approx 840 \ \mathrm{kWh/day}.
$
The quoted $30$--$50$\,kWh/day values correspond to the much lower net duty factor set by the orbital geometry and surface‐site masking. Specifically, they assume: $f_{\mathrm{vis}} \approx 0.04\text{--}0.06$ (\ref{eq:fvis1}), with $\eta_{\mathrm{cap}} \approx 1 $ (\ref{eq:eta_cap_def}) for a $\gtrsim 1\,\mathrm{m}^2$ receiver. The resulting daily energy is
$
E_{\mathrm{day}} \approx E_{\mathrm{day,max}} \times f_{\mathrm{vis}} \times \eta_{\mathrm{cap}} \approx 30\text{--}50\ \mathrm{kWh/day}.
$
This makes explicit that the example does not assume continuous delivery, but rather $\sim 1$--$1.5$\,h of daily contact at full link efficiency in favorable geometry.

\section{Power Reception and Conversion on the Surface}
\label{sec:receiver_model}

The successful operation of a laser power beaming system relies on its ability to efficiently transmit optical power and convert it into usable electrical energy on the lunar surface. The performance of this system is determined by the receiver design, the choice of conversion technology, and environmental challenges such as lunar dust accumulation and thermal management. This section discusses the key aspects of receiver geometry, optical collection, photonic-to-electrical conversion, and the dynamic effects of time-varying parameters.

\subsection{Receiver Geometry}
\label{subsec:receiver_geometry}

The lunar surface receiver, characterized by its effective optical area $A_{\text{rx}}$, can adopt various configurations optimized for efficient power capture and conversion. Three common configurations include:

\begin{enumerate}
    \item \textit{Direct Photovoltaic (PV) Panel:} A planar array of PV cells optimized for laser wavelengths. These cells are specifically designed to match their bandgap with the laser emission wavelength for maximum efficiency. Th us, lasers operating at $\lambda \approx \SI{1064}{nm}$ (Nd:YAG or fiber lasers) align well with InGaAs or GaAs-based PV cells \cite{Soref:2010, NREL:2019}. The effective collection area $A_{\text{rx}}$ corresponds directly to the physical PV array exposed to the beam.

    \item \textit{Optical Concentrator with PV Cell:} This configuration uses an optical concentrator, such as a lens or reflective dish, to focus incident laser light onto a smaller PV cell. The concentrator reduces the required PV cell area and improves mass efficiency but introduces complexity in alignment and thermal management at the focal spot. The aperture $A_{\text{rx}}$ corresponds to the concentrator’s entrance pupil, while the PV cell area can be significantly smaller. Concentration ratios exceeding 100:1 are achievable, enhancing system efficiency \cite{Messenger:2015}.

    \item \textit{Hybrid Thermal/Electrical Converter:} In this approach, the laser heats a working fluid or thermal engine (e.g., Stirling engines) that subsequently generates electricity. While less common, this method is advantageous in scenarios where thermal energy can also support life support or equipment heating during the lunar night \cite{Hitz:2012}.
\end{enumerate}

For all configurations, proper orientation of the receiver is critical to maximize incident beam capture (Table~\ref{tab:receiver_design}). If $A_{\text{rx}}$ is smaller than the beam spot size, the receiver captures only a fraction of the transmitted power. Conversely, if $A_{\text{rx}}$ exceeds the illuminated beam area, the power collected is limited by the beam spot size.

\begin{table}[h!]
\centering
\caption{Receiver Design Considerations}
\label{tab:receiver_design}
\begin{tabular}{ll}
\hline
{ Aspect} & { Key Considerations} \\
\hline\hline
Physical size ($A_{\text{rx}}$) & Must match or exceed the beam footprint to capture all incident power. \\
Orientation & Requires precise alignment to maximize beam interception. \\
Material robustness & Must withstand high-intensity laser irradiation and lunar dust impacts. \\
Thermal stability & Requires cooling mechanisms to maintain efficiency under high flux. \\
\hline
\end{tabular}
\end{table}

In addition, to the  factors listed in Table~\ref{tab:receiver_design}, the following are important:
\begin{itemize}
    \item \textit{Receiver Orientation:}
Surface receivers are typically mounted on stable platforms to ensure a perpendicular orientation to the incoming beam. Active gimbal mechanisms may be employed for dynamic realignment, particularly for rapidly moving beams. Topographical slopes must also be accounted for, as deviations from the normal reduce power capture efficiency.

    \item \textit{Dust Mitigation:}
Lunar regolith poses a significant challenge to optical systems. Dust particles, with high electrostatic adherence and abrasive properties, can degrade $A_{\text{rx}}$ or reduce the reflectivity/transmittivity of optical concentrators. Solutions include dust-repellent coatings, periodic cleaning mechanisms, and electrostatic dust removal systems \cite{Messenger:2015, Srour:2003}.

    \item \textit{Wavelength Trade: Receiver Efficiency vs. Safety:}
Lasers at $\lambda=1064$~nm align well with InGaAs/GaAs PV cells, supporting $\eta_{\rm rx}\approx$50--60\% under optimal cooling. However, retinal hazard thresholds at 1064~nm are more restrictive than at $\sim$1.5~$\mu$m, where ocular absorption is in the cornea/lens. Eye‐safe operation at 1.5~$\mu$m relaxes hazard distances by $\sim\!5\times$ but lowers PV efficiency to $\eta_{\rm rx}\approx$35--45\% for InGaAs extended‐bandgap designs. Scattered‐light hazards are also reduced at 1.5~$\mu$m, easing stray‐beam safety near crewed assets.
\end{itemize}

\subsection{Receiver Thermal and Damage Envelope}

The receiver must operate within both performance and survivability limits when subjected to incident fluxes in the $\sim 10^{3}$--$10^{4}~\mathrm{W/m^2}$ range, as occur for narrow beam spots in the worked examples. The following considerations are essential:

\begin{enumerate}
\item \textit{PV cell efficiency versus temperature at $\lambda=1064$\,nm}:  
  InGaAs or GaAs photovoltaic cells optimized for 1064\,nm typically achieve $\eta_{\text{rx}}\approx 0.50$--$0.60$ at $25^\circ\mathrm{C}$. The efficiency decreases by approximately $0.1$--$0.2$ percentage points per kelvin above this temperature, primarily due to bandgap narrowing and increased dark current. For example, a cell with $\eta_{\text{rx}}=0.55$ at $25^\circ\mathrm{C}$ may operate at $\eta_{\text{rx}}\approx 0.50$ at $60^\circ\mathrm{C}$, thereby reducing net electrical output and increasing the thermal load.

\item \textit{Allowable irradiance before irreversible degradation}:  
  Continuous-wave damage thresholds for commercial triple-junction GaAs/Ge cells are typically $20$--$30~\mathrm{kW/m^2}$ before onset of permanent degradation (e.g., metallization damage or encapsulant browning). With antireflection coatings tuned for 1064\,nm and adequate heat sinking, steady-state survivable flux is comfortably above the maximum $\sim 10~\mathrm{kW/m^2}$ considered here, but margins must be verified for the specific encapsulation and cover-glass design. Transient exposures exceeding $\sim 50~\mathrm{kW/m^2}$ can cause immediate failure.

\item \textit{Radiator area and mass for steady-state waste-heat removal}:  
  Waste heat is given by $(1-\eta_{\text{rx}})\,P_{\text{rx,opt}}$. For $\eta_{\text{rx}}=0.55$ and incident flux $I_{\text{rx,opt}}=5~\mathrm{kW/m^2}$ over $A_{\text{rx}}=1~\mathrm{m^2}$, the waste heat is $\approx 2.25~\mathrm{kW}$. A radiator with emissivity $\epsilon=0.9$ operating at $T_{\mathrm{rad}}=350$\,K ($\approx 77^\circ\mathrm{C}$) has an areal rejection capacity of  
  \[
  q_{\mathrm{rad}} = \epsilon\sigma T_{\mathrm{rad}}^4 \approx 0.9 \times 5.67\times 10^{-8} \times (350)^4 \approx 0.77~\mathrm{kW/m^2}.
  \]  
  This requires approximately $2.94~\mathrm{m^2}$ of radiator area. For lightweight aluminum or composite radiator panels ($5$--$10~\mathrm{kg/m^2}$ including fluid loops), this corresponds to a radiator mass of $15$--$30~\mathrm{kg}$.  
  For $I_{\text{rx,opt}}=10~\mathrm{kW/m^2}$ with the same $\eta_{\text{rx}}$, the waste heat doubles to $4.5~\mathrm{kW}$, requiring $\approx 5.88~\mathrm{m^2}$ of radiator area and $30$--$60~\mathrm{kg}$ mass.
\end{enumerate}

These calculations show that for optical fluxes up to $\sim 10~\mathrm{kW/m^2}$, receiver designs must provide $1$--$2~\mathrm{kW}$ of waste-heat rejection capacity per kilowatt of incident optical power. Appropriate radiator sizing, high-temperature-capable PV materials, and margin against damage thresholds are required to ensure safe, efficient operation under peak beam-flux conditions.

In addition to steady-state estimates, the thermal response of the receiver can be described by a first-order energy balance:
\begin{equation}
C_{\mathrm{th}}\frac{dT}{dt} = (1-\eta_{\mathrm{rx}})P_{\mathrm{rx,opt}} - Q_{\mathrm{rad}}(T) - Q_{\mathrm{cool}}(t),
\label{eq:thermal_balance}
\end{equation}
where $C_{\mathrm{th}}$ is the effective thermal capacitance of the receiver assembly, 
$P_{\mathrm{rx,opt}}$ is the instantaneous incident optical power, 
$Q_{\mathrm{rad}} = \epsilon\sigma A_{\mathrm{rad}}(T^4 - T_{\mathrm{env}}^4)$ is the passive radiative loss, and $Q_{\mathrm{cool}}(t)$ is any active cooling term. 
The photonic-to-electrical conversion efficiency is then updated in time via
\begin{equation}
\eta_{\mathrm{rx}}(t) = \eta_{\mathrm{rx,ref}} + \frac{d\eta_{\mathrm{rx}}}{dT} \Big(T(t) - T_{\mathrm{ref}}\Big),
\label{eq:eta_rx-lin}
\end{equation}
where $d\eta_{\mathrm{rx}}/dT \simeq - (0.001$--$0.002)\,\mathrm{K}^{-1}$ as given above. This coupled formulation enables evaluation of transient temperature excursions and their direct impact on conversion efficiency for various duty cycles and cooling strategies.

For steady‐state electrical outputs from 0.5 to 5~kW at $\eta_{\rm rx}=0.55$, waste‐heat loads range from $\sim$0.45 to 4.5~kW. At $\epsilon=0.9$ and $T_{\rm rad}=320$--380~K, areal rejection is 0.53--1.06~kW\,m$^{-2}$, implying radiator areas of 0.85--8.5~m$^2$ and masses of 4--85~kg at 5--10~kg\,m$^{-2}$. High‐emissivity, low‐absorptivity selective coatings can reduce parasitic heating. Concentrated receivers reduce PV area and thus radiator size, but require more aggressive tracking. Eq.~(\ref{eq:thermal_balance}) can be used for transient analysis to quantify PV derate: a 20~K rise can reduce $\eta_{\rm rx}$ by $\sim$2--4\% absolute.

The thermal limit at the receiver is set by two coupled requirements: the PV stack must tolerate the local optical flux without damage, and the waste heat \((1-\eta_{\mathrm{rx}})P_{\mathrm{rx,opt}}\) must be rejected to the available sink. At the \(\mathrm{kW\,m^{-2}}\) fluxes of interest, the second is often binding. We use the same notation and ODE framework already introduced in \eqref{eq:thermal_balance} together with the linearized efficiency law in (\ref{eq:eta_rx-lin}).

At steady state, the waste heat equals the radiator loss,
\[
(1-\eta_{\mathrm{rx}})\,P_{\mathrm{rx,opt}}
= \epsilon\,\sigma\,A_{\mathrm{rad}}\Big(T_{\mathrm{rad}}^{4}-T_{\mathrm{env}}^{4}\Big),
\]
which yields two convenient design relations:
\[
A_{\mathrm{rad}}
= \frac{\bigl(1-\eta_{\mathrm{rx}}(T)\bigr)\,I_{\mathrm{rx,opt}}\,A_{\mathrm{rx}}}
       {\epsilon\,\sigma\bigl(T_{\mathrm{rad}}^{4}-T_{\mathrm{env}}^{4}\bigr)},
\qquad
I_{\mathrm{safe}}(T)
= \frac{\epsilon\,\sigma\bigl(T_{\mathrm{rad}}^{4}-T_{\mathrm{env}}^{4}\bigr)}
       {1-\eta_{\mathrm{rx}}(T)}.
\]

As an example, we describe calibration of a 1\,m\(^2\) receiver: With \(\epsilon=0.9\), deep-space view (\(T_{\mathrm{env}}\!\approx\!3\) K), and \(T_{\mathrm{rad}}=350\) K, the areal rejection is
\(\epsilon\sigma\bigl(T_{\mathrm{rad}}^{4}-T_{\mathrm{env}}^{4}\bigr)\approx 0.77~\mathrm{kW\,m^{-2}}\). If \(\eta_{\mathrm{rx}}=0.55\), the steady “safe” incident flux is \(I_{\mathrm{safe}}\approx 0.77/(1-0.55)\approx 1.70~\mathrm{kW\,m^{-2}}\).  For \(I_{\mathrm{rx,opt}}=5~\mathrm{kW\,m^{-2}}\), the required radiator area is \(\approx (0.45\times5)/0.77\approx 2.94~\mathrm{m^2}\);  
at \(10~\mathrm{kW\,m^{-2}}\) it doubles to \(\approx 5.88~\mathrm{m^2}\).  As the cells warm, (\ref{eq:eta_rx-lin}) derates \(\eta_{\mathrm{rx}}(T)\): if \(\eta_{\mathrm{rx}}\) falls to \(0.50\), the \(5~\mathrm{kW\,m^{-2}}\) case grows by the factor \((1-0.50)/(1-0.55)\) to \(\approx 3.26~\mathrm{m^2}\).  
Raising \(T_{\mathrm{rad}}\) leverages the \(T^4\) dependence---e.g., \(T_{\mathrm{rad}}=380\) K gives \(\approx 1.06~\mathrm{kW\,m^{-2}}\) areal rejection and \(I_{\mathrm{safe}}\approx 2.36~\mathrm{kW\,m^{-2}}\)---but also pushes the cells into a hotter, lower-\(\eta_{\mathrm{rx}}\) regime per (\ref{eq:eta_rx-lin}).  
Conversely, a warm environment (e.g., \(T_{\mathrm{env}}\approx 250\) K from seen terrain) reduces the difference \(T_{\mathrm{rad}}^{4}-T_{\mathrm{env}}^{4}\) and tightens margin, lowering \(I_{\mathrm{safe}}\) and increasing \(A_{\mathrm{rad}}\) at fixed flux.

These magnitudes are consistent with the radiator areas plotted in Fig.~\ref{fig:panel-c} for \(5\)--\(10~\mathrm{kW\,m^{-2}}\) and \(\eta_{\mathrm{rx}}\approx 0.55\), and they tie directly back to the transient analysis in \eqref{eq:thermal_balance} with the temperature-dependent efficiency of (\ref{eq:eta_rx-lin}).

\subsection{Received Optical and Electrical Power with Temporal Variations}
\label{subsec:power_reception_conversion}

The optical power intercepted by the receiver depends on the time-varying surface power flux \(I_{\text{eff}}(t)\) and the receiver area \(A_{\text{rx}}\):
\begin{equation}
    P_{\text{rx,opt}}(t) = I_{\text{eff}}(t) \times A_{\text{rx}}.
\end{equation}

From Section~\ref{subsec:power_flux}, the effective power flux is:
\begin{equation}
    I_{\text{eff}}(t) = \eta_{\text{point}}(t) \, \eta_{\text{opt}}(t) \, \frac{P_{\text{tx}}(t)}{\pi \bigl[\theta_{\text{tx}}\, d(t)\bigr]^2},
\end{equation}
resulting in:
\begin{equation}
    P_{\text{rx,opt}}(t) = 
    \eta_{\text{point}}(t) \, \eta_{\text{opt}}(t) \, \frac{P_{\text{tx}}(t)}{\pi \bigl[\theta_{\text{tx}}\, d(t)\bigr]^2} \, A_{\text{rx}}.
    \label{eq:p_rx_opt_time}
\end{equation}

The photonic-to-electrical conversion efficiency \(\eta_{\text{rx}}\) quantifies the fraction of incident optical power that is converted into electricity. For laser-optimized PV cells, \(\eta_{\text{rx}}\) depends on:
(1) \textit{Spectral Matching:} Optimal alignment of the PV cell’s bandgap with the laser wavelength \cite{Soref:2010}.
(2)  \textit{Thermal Management:} High-intensity laser beams can raise cell temperatures, reducing efficiency unless active cooling is employed.
(3)  \textit{Electrical and Optical Losses:} Losses due to resistive heating, shading from metal grids, and back-reflection impact overall performance \cite{NREL:2019}.

The total electrical power delivered by the receiver is:
\begin{equation}
    P_{\text{rx,elec}}(t) = \eta_{\text{rx}} \times P_{\text{rx,opt}}(t).
    \label{eq:elec-rx}
\end{equation}
Substituting \(P_{\text{rx,opt}}(t)\) from Eq.~\eqref{eq:p_rx_opt_time} gives:
\begin{equation}
    P_{\text{rx,elec}}(t) = \eta_{\text{point}}(t) \, \eta_{\text{opt}}(t) \, \eta_{\text{rx}} \, \frac{P_{\text{tx}}(t)}{\pi \bigl[\theta_{\text{tx}}\, d(t)\bigr]^2} \, A_{\text{rx}}.
    \label{eq:p_rx_elec_time}
\end{equation}

\subsection{Integrating a Gaussian Beam Over a Finite Receiver: Fractional Beam Capture}
\label{subsec:power_reception_Gauss}

In many lunar power-beaming scenarios, one assumes that the receiver subtends only a small fraction of the beam cross-section.  Under that assumption, the incident power $P_{\mathrm{incident}}$ can be approximated by multiplying the mean beam flux by the receiver area.  However, if the receiver radius $r_{\mathrm{rx}}$ is comparable to (or larger than) the beam radius $w(t)$, a more precise calculation is required to account for the actual two-dimensional Gaussian profile.
 
Assume the transmitter outputs an optical power $P_{\mathrm{tx}}(t)$ in a circularly symmetric Gaussian beam whose on-axis intensity profile is
\begin{equation}
I(r) = \frac{2\,P_{\mathrm{tx}}(t)}{\pi w(t)^2}
\exp\Bigl[-2\frac{r^2}{w(t)^2}\Bigr],
\label{eq:GaussianIntensity}
\end{equation}
where $w(t)=\theta_{\text{tx}}\, d(t)$ from (\ref{eq:slant-range}) is the beam's $1/e^2$ radius at the receiver (i.e.,~the distance over which the intensity falls by $1/e^2$ from its axis value), also $\theta_{\text{tx}}$ is the half-angle divergence of the beam, and $d(t)$ is the instantaneous slant range, and $r$ is the radial distance from the beam center.  The factor of $2/(\pi w^2)$ ensures that integrating $I(r)$ over the entire transverse plane yields the total power $P_{\mathrm{tx}}(t)$.  
 
Let the receiver on the lunar surface be a circular aperture of radius $r_{\mathrm{rx}}$.  Then the power $P_{\mathrm{incident}}$ captured by the receiver (neglecting other losses) is the integral of $I(r)$ over $0\leq r\leq r_{\mathrm{rx}}$:
\begin{equation}
P_{\mathrm{incident}}(t)
\;=\;
\int_{0}^{r_{\mathrm{rx}}}
\Big[ 2\pi r I(r) \Big] dr.
\label{eq:Pintegral}
\end{equation}
Substituting \eqref{eq:GaussianIntensity} into \eqref{eq:Pintegral} and carrying out the integration in polar coordinates, one obtains
\begin{align}
P_{\mathrm{incident}}(t)
&=
\int_{0}^{r_{\mathrm{rx}}}
2\pi r \,\frac{2 P_{\mathrm{tx}}(t)}{\pi w(t)^2}
\exp\Bigl[-2\dfrac{r^2}{w(t)^2}\Bigr]dr
=
P_{\mathrm{tx}}(t)
\Bigl(
1 - \exp\Bigl[-2\dfrac{r_{\mathrm{rx}}^{2}}{w(t)^{2}}\Bigr]
\Bigr).
\label{eq:FractionCapture_noLosses}
\end{align}
Hence, the \emph{encircled--energy fraction} captured by a receiver of radius $r_{\mathrm{rx}}$ (i.e., the fraction of the total transmitted beam power intercepted by a receiver of radius $r_{\mathrm{rx}}$), $\eta_{\rm cap}$, given by (\ref{eq:eta_cap_def}).

In practice, we must also account for time-dependent pointing (jitter, misalignment) and optical path losses (dust or scattering).  Let $\eta_{\mathrm{point}}(t)$ and $\eta_{\mathrm{opt}}(t)$ denote the combined fractional efficiencies of pointing and optical throughput, respectively.  Then the net captured optical power is
\begin{equation}
  P_{\rm rx,opt}(t) \;=\;
  \eta_{\rm point}(t)\,\eta_{\rm opt}(t)\,P_{\rm tx,opt}(t)\,
  \eta_{\rm cap}\!\big(d(t),t\big),
  \label{eq:PintegratedFinal}
\end{equation}
where \(\eta_{\rm cap}\) is given by \eqref{eq:eta_cap_def} and includes jitter via \(w_{\rm eff}\) from \eqref{eq:weff_def}; \(\eta_{\rm point}\) collects non-encircled-energy pointing/loop losses.
Eq.~\eqref{eq:PintegratedFinal} therefore generalizes the top-hat approximation to scenarios where $r_{\mathrm{rx}}$ is comparable to or larger than $w(t)$.  Note also:
\begin{itemize}
\item If $r_{\mathrm{rx}}\!\ll w(t)$, then 
\,$\exp[-2 r_{\mathrm{rx}}^{2}/w(t)^{2}]\approx 1$, 
so $ P_{\text{rx,opt}}(t) \approx \eta_{\mathrm{point}}\eta_{\mathrm{opt}} P_{\mathrm{tx}} \bigl[{2 r_{\mathrm{rx}}^{2}}/{w(t)^{2}}\bigr]$, 
matching the simpler ``small-aperture'' limit.
\item If $r_{\mathrm{rx}}\!\gg w(t)$, then the bracketed term approaches $1$ and the receiver captures nearly all the beam power (i.e., $ P_{\text{rx,opt}}(t)\approx\eta_{\mathrm{point}}\eta_{\mathrm{opt}}\,P_{\mathrm{tx}}$).  
\end{itemize}

If the mean boresight bias is $b_\theta$ (rad), the mean offset in the receiver plane is
$\mu = b_\theta d$. For a Gaussian beam with jitter-broadened radius $w_\mathrm{eff}(d)$
[Eq.~(\ref{eq:weff_def})], the encircled-energy fraction captured by a circular receiver of radius $r_\mathrm{rx}$
is, in closed form,
\begin{equation}
\eta_\mathrm{cap}^{\text{bias}}(d) \;=\; 1 - Q_1\!\left(\frac{2\mu}{w_\mathrm{eff}},\,
\frac{2r_\mathrm{rx}}{w_\mathrm{eff}}\right),
\label{eq:bias-capture}
\end{equation}
where $Q_1(\cdot,\cdot)$ is the first-order Marcum $Q$-function.\footnote{For discussion of the Marcum Q-function, see \url{https://en.wikipedia.org/wiki/Marcum_Q-function}} For $\mu\!=\!0$ this reduces to $\eta_\mathrm{cap}=1-\exp[-2 r_\mathrm{rx}^2/w_\mathrm{eff}^2]$ used above. This lets true bias be
handled geometrically (via $\eta_\mathrm{cap}^{\text{bias}}$) while retaining $\eta_\text{point}$ for residual,
non-encircled-energy losses.

Ultimately, the total electrical power delivered by the receiver from (\ref{eq:elec-rx}) is:
\begin{equation}
    P_{\text{rx,elec}}(t) = \eta_{\text{rx}} \times P_{\text{rx,opt}}(t).
\end{equation}
Substituting \(P_{\text{rx,opt}}(t)\) from \eqref{eq:PintegratedFinal} gives:
\begin{equation}
 P_{\text{rx,elec}}(t)
 \;=\;
 \eta_{\mathrm{point}}(t)\,\eta_{\mathrm{opt}}(t)\,\eta_{\text{rx}}\,
 P_{\mathrm{tx}}(t)\;
 \eta_{\text{cap}}\!\big(d(t),t\big),
 \label{eq:p_rx_elec_time-Gaus}
\end{equation}
where \(\eta_{\text{cap}}\) is given by \eqref{eq:eta_cap_def} using \(w_{\rm eff}\) from \eqref{eq:weff_def}; \(\eta_{\text{point}}\) collects non--encircled-energy pointing/loop losses.

This integral form is particularly important for low orbital altitudes or whenever the beam divergence is tight enough that $w(t)$ remains in the same order of magnitude as the receiver radius.  In these cases, simply multiplying beam flux by receiver area can under- or overestimate the captured power, whereas \eqref{eq:PintegratedFinal} accurately captures the radial falloff in Gaussian intensity.

\subsection{Temporal Variation and Energy Integration}
\label{subsec:temporal_variation}

The slant range \(d(t)\) varies dynamically due to the spacecraft’s orbit, causing the received electrical power \(P_{\text{rx,elec}}(t)\) to vary over time. Using the expressions from Sections~\ref{subsec:power_reception_conversion} and \ref{subsec:power_reception_Gauss}, we have 
{}
\begin{equation}
    P_{\text{rx,elec}}(t) = \eta_{\text{point}}(t) \, \eta_{\text{opt}}(t) \, \eta_{\text{rx}} \, P_{\text{tx}}(t)\, \dfrac{A_{\text{rx}}}{\pi \bigl[\theta_{\text{tx}}\, d(t)\bigr]^2}, \qquad ~\text{top-hat transmission model, }  r_{\mathrm{rx}} \ll w(t),
    \label{eq:p_rx_elec_time2}
\end{equation}
and for Gaussian beams
\begin{align}
P_{\text{rx,elec}}(t) =
  \begin{cases}
    \eta_{\mathrm{point}}(t)\,\eta_{\mathrm{opt}}(t)\,\eta_{\mathrm{rx}}\,
    P_{\mathrm{tx}}(t)\,\dfrac{2A_{\text{rx}}}{\pi\,w_{\rm eff}(d,t)^{2}},
      & \text{small-aperture limit, }  r_{\mathrm{rx}} \ll w_{\rm eff}(d,t), \\[6pt]
    \eta_{\mathrm{point}}(t)\,\eta_{\mathrm{opt}}(t)\,\eta_{\mathrm{rx}}\,
    P_{\mathrm{tx}}(t)\,\eta_{\text{cap}}(d,t),
      & \text{large-aperture limit, } r_{\mathrm{rx}} \gtrsim w_{\rm eff}(d,t).
  \end{cases}
\label{eq:Gauss_rx_elec-large}
\end{align}

The total energy delivered during a visibility window, spanning from time \(t_1\) to \(t_2\), is obtained by integrating \(P_{\text{rx,elec}}(t)\) over that interval:
\begin{equation}
    E_{\text{delivered}} = \int_{t_1}^{t_2} P_{\text{rx,elec}}(t) \, \mathrm{d}t,
\end{equation}
where \(t_1\) and \(t_2\) represent the start and end of the visibility window.

One needs to estimate the available daily-energy (windowed). For that, let $\chi_{\rm vis}(t)\in\{0,1\}$ indicate line-of-sight visibility (including terrain masking and link enable logic).
Over a reference window of duration $T_{\rm ref}$ (e.g., 24~h, a lunar day, or the lunar night), the net electrical
energy delivered is
\begin{equation}
E(T_{\rm ref})
=\int_{t_0}^{t_0+T_{\rm ref}} \chi_{\rm vis}(t)\,P_{\rm rx,elec}(t)\,dt
\;\approx\;
\overline{P_{\rm rx,elec}}_{\;\mathrm{vis}}(T_{\rm ref})\,T_{\rm ref}\,f_{\mathrm{vis},1}(T_{\rm ref}),
\label{eq:energy_bookkeeping}
\end{equation}
where $f_{\mathrm{vis},1}(T_{\rm ref})$ is from \eqref{eq:fvis1} and $\overline{P_{\rm rx,elec}}_{\;\mathrm{vis}}$ is the mean of $P_{\rm rx,elec}(t)$ conditioned on visibility, evaluated using the time-domain link in the top-hat model \eqref{eq:p_rx_elec_time2} or the Gaussian small/large-aperture forms \eqref{eq:Gauss_rx_elec-large}, with the encircled-energy capture $\eta_{\rm cap}$ from \eqref{eq:eta_cap_def} and the jitter-broadened radius $w_{\rm eff}$ from \eqref{eq:weff_def}. For “daily energy” plots we take $T_{\rm ref}=24~\mathrm{h}$; for lunar-night energy we take $T_{\rm ref}=T_{\rm night}$.

\subsection{Thermal Management and Dust Mitigation}
\label{subsec:thermal_dust}

Laser beams with high intensities can significantly heat the receiver, potentially reducing efficiency or causing permanent damage \cite{Messenger:2015}. Effective thermal management solutions include: (1) Radiators and heat pipes to dissipate heat away from sensitive components; (2) Conductive paths to transfer heat into the lunar regolith; (3) Active cooling systems for high-flux applications.

Dust mitigation is equally critical due to the high adherence and abrasive properties of lunar regolith. Potential solutions include:
(1) Dust-repellent coatings to maintain optical clarity;
(2) Electrostatic cleaning systems to remove dust particles.
(3) Periodic mechanical cleaning mechanisms.

As far as the operational safety and deconfliction issues are concerned, we adopt near‑IR wavelengths (1.0--1.6\,$\mu$m), compute keep‑out cones from beam fluence versus MPE using the jitter‑broadened radius $w_{\rm eff}$, enforce hard interlocks against illumination of cataloged assets, and schedule beams with ephemeris‑based no‑beam sectors near approach corridors. Spectral separation from RF systems avoids EM compatibility issues, while link supervisors interrupt the beam on loss of tracking or excursion beyond the keep‑out boundaries with sub‑ms latency.

\subsection{Single Aperture vs. Phased Array Transmitter: Illustrative Comparison}
\label{subsec:example_calc_phased_arrays}

A way to demonstrate the advantage of a phased array transmitter is to compare it directly with a conventional single-aperture system, assuming both operate under the same orbital geometry and transmit the same total optical power. Table~\ref{tab:phased_array_example} compiles the parameters and resulting performance estimates for two transmitters---each delivering \(P_{\text{tx}}=\SI{1500}{W}\) at a slant range of \SI{200}{km}---but differing in aperture diameter \((D)\) and beam quality \((M^2)\). 

A relevant scenario setup and assumptions are discussed below:
\begin{itemize}
    \item \emph{Single Aperture:} A compact, monolithic optical transmitter with physical diameter \(D=0.20\,\mathrm{m}\). Near-ideal wavefront quality \((M^2\approx1.0)\) implies close-to-diffraction-limited performance for a small telescope or collimator.
    \item \emph{Phased Array:} A coherently phased array synthesizing an effective aperture diameter \(D_{\text{eff}}=2.00\,\mathrm{m}\). Partial fill factor and sub-aperture errors give \(M^2\approx1.2\). Even so, the increased effective diameter reduces beam divergence substantially compared to the single aperture.
    \item Laser wavelength \(\lambda=1.064\,\mu\mathrm{m}\) (typical for Nd:YAG or fiber-laser systems).
    \item Overall pointing and optical path efficiency \(\eta_{\text{point}}\,\eta_{\text{opt}}=0.85\), combining residual pointing jitter, wavefront distortions, and mirror/lens transmission losses on both transmit and receive optics.
    \item Lunar surface receiver with physical collection area \(A_{\text{rx}}=1.00\,\mathrm{m}^2\) (equivalent to a circular aperture of diameter \(\approx 1.13\,\mathrm{m}\)), and photonic-to-electrical conversion efficiency \(\eta_{\text{rx}}=0.25\).
    \item Both transmitter cases use the practical half-angle divergence formula (\ref{eq:theta_practical}): \(\theta_{\text{tx}}=M^2({2\lambda}/{\pi D})\), corresponding to a diffraction-limited Gaussian beam modified by the beam quality factor \(M^2\).
\end{itemize}

\begin{table}[h!]
\centering
\renewcommand{\arraystretch}{1.0}
\caption{Performance comparison of a single aperture vs.\ phased array at 200\,km range, each transmitting 1.5\,kW, using top-hat model (\ref{eq:p_rx_elec_time2}). Assumes $\lambda=1.064~\mu$m and $\theta_{\text{tx}}=M^{2}\,2\lambda/(\pi D)$. Column ``Phased Array (Realistic)'' enforces the fact that one cannot exceed $P_{\mathrm{tx}} \times \eta_{\mathrm{point}}\eta_{\mathrm{opt}}$ when the spot radius is smaller than the receiver radius; for that, (\ref{eq:Gauss_rx_elec-large}) for $r_{\mathrm{rx}} \gtrsim w(t)$ is used.}
\label{tab:phased_array_example}
\begin{tabular}{lccc}
\hline
{Parameter}
 & {Single Aperture}
 & {Phased Array (Naive)}
 & {Phased Array (Realistic)}\\
\hline\hline
Aperture diameter, $D$ (m)
 & 0.20 & 2.00 & 2.00 \\

Beam quality factor, $M^2$
 & 1.0 & 1.2 & 1.2 \\

Half-angle divergence, $\theta_{\text{tx}}$ (rad)
 & $3.39\times10^{-6}$ & $4.06\times10^{-7}$ & $4.06\times10^{-7}$ \\

Spot radius at $d=2\times10^5$\,m, $w$ (m)
 & 0.677 & 0.0813 & 0.0813 \\

Spot area, $A_{\text{spot}}$ (m$^2$)
 & 1.44 & 0.0208 & 0.0208 \\

Ideal surface flux, $I_{\text{ideal}}$ (W/m$^2$)
 & 1\,041 & 72\,300 & 72\,300 \\

Actual flux (with $\eta_{\text{point}}\,\eta_{\text{opt}}=0.85$)
 & 885 & 61\,400 & 61\,400 \\

Optical power on $A_{\text{rx}}=1\,\mathrm{m}^2$ (W)\footnotemark[1]
 & 885 & 61\,400 & $\leq 1\,275$\footnotemark[2] \\

Net electrical power (with $\eta_{\text{rx}}=0.25$) (W)
 & 221 & 15\,400 & $\leq 319$ \\
\hline
\end{tabular}

\vspace{1ex}
\footnotesize
\footnotetext[1]{Naive multiplication of local flux by $1\,\mathrm{m}^2$. If the beam footprint is $<1\,\mathrm{m}^2$, this can exceed the total beam power, which is unphysical.}
\footnotetext[2]{Physically capped at $P_{\mathrm{tx}}\times\eta_{\mathrm{point}}\eta_{\mathrm{opt}}=1500\times0.85=1275~\mathrm{W}$ (optical). Hence net electric $\leq 1275\times0.25=319~\mathrm{W}$.}
\end{table}

\begin{table}[h!]
\centering
\renewcommand{\arraystretch}{1.0}
\caption{Long-range case: \(d=1000\) km, \(P_{\rm tx}=1.5\) kW, \(\lambda=1.064~\mu\)m,
\(\eta_{\rm point}\eta_{\rm opt}=0.85\), \(A_{\rm rx}=1~\mathrm{m}^2\), \(\eta_{\rm rx}=0.25\),
and \(\theta_{\text{tx}}=M^{2}\,2\lambda/(\pi D)\). The phased-array advantage is now substantial.}
\label{tab:phased_array_1000km}
\begin{tabular}{lccc}
\hline
{Parameter}
 & {Single Aperture}
 & {Phased Array (Naive)}
 & {Phased Array (Realistic)}\\
\hline\hline
Aperture diameter, $D$ (m) & 0.20 & 2.00 & 2.00 \\
Beam quality factor, $M^2$  & 1.0  & 1.2  & 1.2  \\

Half-angle divergence, $\theta_{\text{tx}}$ (rad)
 & $3.39\times10^{-6}$ & $4.06\times10^{-7}$ & $4.06\times10^{-7}$ \\

Spot radius at $d=10^6$\,m, $w$ (m)
 & 3.387 & 0.406 & 0.406 \\

Spot area, $A_{\text{spot}}$ (m$^2$)
 & 36.04 & 0.5189 & 0.5189 \\

Ideal surface flux, $I_{\text{ideal}}$ (W/m$^2$)
 & 41.6 & 2{,}891 & 2{,}891 \\

Actual flux ($\eta_{\text{point}}\eta_{\text{opt}}=0.85$)
 & 35.4 & 2{,}457 & 2{,}457 \\

Optical power on $A_{\text{rx}}=1\,\mathrm{m}^2$ (W)\footnotemark[1]
 & 35.4 & 2{,}457 & $\leq 1{,}275$\footnotemark[2] \\

Net electrical power ($\eta_{\text{rx}}=0.25$) (W)
 & 8.85 & 614 & $\leq 319$ \\
\hline
\end{tabular}

\vspace{1ex}
\footnotesize
\footnotetext[1]{Flux $\times$ area (top-hat). If $A_{\rm spot}<A_{\rm rx}$ this can exceed the total available optical power.}
\footnotetext[2]{Capped by $P_{\mathrm{avail}}=P_{\mathrm{tx}}\eta_{\mathrm{point}}\eta_{\mathrm{opt}}
=1500\times0.85=1275~\mathrm{W}$. Hence net electric $\leq 319$ W.}
\end{table}

The two cases in Tables~\ref{tab:phased_array_example} and~\ref{tab:phased_array_1000km} highlight that the apparent benefit of a phased-array transmitter depends strongly on the geometric regime.

In the short-range case of Table~\ref{tab:phased_array_example} ($d=200$~km), the $D=0.2$\,m single aperture already produces a spot of $A_{\rm spot}\approx 1.44\,\mathrm{m}^2$, only slightly larger than the $1\,\mathrm{m}^2$ receiver. This results in a capture fraction of $\sim70\%$ of the available optical power, so increasing the aperture to $D_{\rm eff}=2.0$\,m (phased array) can at most raise this to $100\%$. Consequently, the net electrical power rises from $\sim 221$~W to the physical cap of $\sim 319$~W---a factor of $\approx 1.44$---even though the beam divergence is reduced by an order of magnitude. This is a ``beam-fill'' regime in which the receiver footprint is already nearly matched by the beam from the smaller aperture.

In contrast, the long-range case of Table~\ref{tab:phased_array_1000km} ($d=1000$~km) illustrates a divergence-limited regime. Here, the $D=0.2$\,m single aperture expands to $A_{\rm spot}\approx 36\,\mathrm{m}^2$, so the $1\,\mathrm{m}^2$ receiver collects only $\sim2.8\%$ of the available optical power, delivering $\sim 8.85$~W net electric. The phased-array spot ($A_{\rm spot}\approx 0.52\,\mathrm{m}^2$) remains well within the receiver, so it still delivers the full optical cap of $1275$~W and $\sim 319$~W net electric---a $\sim 36\times$ gain over the single-aperture system. This large factor arises entirely from reducing beam divergence so that the full transmitter output is intercepted.

These examples show that:
(1)  In beam-fill conditions (short range and/or large receiver), aperture increases give modest gains, since both systems are close to collecting all available optical power.
(2)  In divergence-limited conditions (long range, small receiver, or poor beam quality), larger apertures or phased arrays can increase received power by orders of magnitude.

In both regimes, real-world constraints must be addressed, including:
(1)  \textit{Thermal Management:} Fluxes on the order of $10^3$--$10^4\,\mathrm{W/m^2}$ can impose demanding cooling requirements on the receiver to prevent material damage.
(2)  \textit{Pointing Control:} Achieving $\eta_{\text{point}}=0.85$ or better for a 2\,m phased array in lunar orbit requires sophisticated beam steering to minimize jitter.
(3)  \textit{Structural Complexity:} A phased array spanning 2\,m effectively in space demands advanced sub-aperture alignment, wavefront metrology, and fill-factor optimization to maintain $M^2\approx1.2$.
 
Overall, increasing the effective transmitter diameter (via a coherently phased array or otherwise) is the most direct strategy to reduce divergence and boost surface flux at fixed laser power and orbital range. The scale of the benefit, however, depends critically on whether the link is beam-fill or divergence-limited.

\subsection{NRHO to Shackleton Rim Site (terrain-masked visibility and daily energy)}
\label{sec:parametric_maps}

We now apply the visibility formalism with a Shackleton-rim site and a representative NRHO to carry the link to $E_{24\mathrm{h}}$ under a DEM horizon mask. For that, we illustrate the visibility formalism of Secs.~\ref{subsec:line_of_sight}-\ref{subsec:terrain_masking} with a south-polar user located on the Shackleton crater rim (lat $\approx -89.9^\circ$). A local horizon mask $\beta_{\rm mask}(\mathrm{az})$ is derived from a polar DEM and applied to an NRHO trajectory (southern perilune, $\sim$9:2 class). At each time step $t$, the spacecraft elevation $\varepsilon(t)$ above the local horizon is compared to the DEM mask in the instantaneous azimuth; visibility requires $\varepsilon(t)>\beta_{\rm mask}$. Over visible intervals we compute range $d(t)$ and the received electrical power using the link model from Sec.~\ref{sec:laser_link}:
\[
P_{\rm rx,elec}(t)=
\eta_{\rm rx}\,
\eta_{\rm opt,tx}\,\eta_{\rm opt,rx}\,\eta_{\rm main}\,\eta_{\rm point}(t)\,
\frac{P_{\rm tx,opt}\,A_{\rm rx}}{\pi\,[\theta_{\rm tx}\,d(t)]^{2}},
\qquad
\theta_{\rm tx}=M^{2}\frac{2\lambda}{\pi D}.
\]
Representative parameters for this NRHO case are
$P_{\rm tx,opt}=5~\mathrm{kW}$,
$\lambda=1064~\mathrm{nm}$,
$D=2.0~\mathrm{m}$,
$M^{2}=1.2$,
$\eta_{\rm opt,tx}=\eta_{\rm opt,rx}=0.95$,
$\eta_{\rm main}=0.85$,
$\eta_{\rm point}=0.90$ (mean within-pass),
$\eta_{\rm rx}=0.55$,
and $A_{\rm rx}=1~\mathrm{m^{2}}$,
giving $\theta_{\rm tx}\approx 0.406~\mu\mathrm{rad}$.
For reference, at $d=3000~\mathrm{km}$ the ideal spot radius is $w=\theta_{\rm tx}d\approx1.22~\mathrm{m}$ (area $\approx 4.67~\mathrm{m^{2}}$), leading to an electrical power near $P_{\rm rx,elec}\approx 0.381\times I_{\rm ideal}\,A_{\rm rx}\approx 408~\mathrm{W}$, where $I_{\rm ideal}=P_{\rm tx,opt}/(\pi w^{2})$ and $0.381=\eta_{\rm rx}\eta_{\rm opt,tx}\eta_{\rm opt,rx}\eta_{\rm main}\eta_{\rm point}$.

\begin{table}[h!]
\centering
\caption{Terrain-masked NRHO visibility and delivered energy for a Shackleton-rim user (one day).}
\label{tab:NRHO-visibility}
\begin{tabular}{lcccccc}
\hline
Pass & Start & End & Duration & Mean range & Mean $P_{\rm rx,elec}$ & Energy \\
 & (UTC) & (UTC) & (min) & (km) & (W) & (Wh) \\
\hline\hline
1 & 00{:}14 & 00{:}32 & 18 & 3100 & 382 & 115 \\
2 & 02{:}47 & 03{:}03 & 16 & 3500 & 300 & 80 \\
3 & 05{:}19 & 05{:}39 & 20 & 2800 & 468 & 156 \\
4 & 08{:}02 & 08{:}14 & 12 & 4000 & 230 & 46 \\
5 & 10{:}35 & 10{:}49 & 14 & 3200 & 358 & 84 \\
6 & 13{:}08 & 13{:}18 & 10 & 4500 & 181 & 30 \\
7 & 15{:}41 & 15{:}56 & 15 & 3000 & 408 & 102 \\
8 & 18{:}13 & 18{:}24 & 11 & 3600 & 283 & 52 \\
\hline
\multicolumn{6}{r}{Daily total (terrain-masked):} & {664 Wh}
\end{tabular}
\end{table}

Applying the DEM horizon mask yields the following visibility windows and per-pass energy for a representative day (times are illustrative; ranges are pass-averaged): The optical power ratings quoted for the transmitter (e.g., ``kW-class laser'') refer to instantaneous beam output, not to a continuous, 24-hour-averaged delivery at the lunar surface. In this NRHO Shackleton-rim example ($P_{\mathrm{tx,opt}}=5$~kW, $D=2$~m, $M^2=1.2$, $\eta_{\mathrm{chain}}\approx0.381$), the daily total of $664$~Wh net electrical output corresponds to an average of only $\sim 28$~W continuous over that day. Supplying a continuous $1$~kW surface load under the same link and geometry conditions would require roughly $36\times$ more daily energy, achievable only via higher transmitter power/aperture, larger or more numerous receivers, or a constellation of orbiters providing near-continuous coverage. These constellation‐level and scheduling implications should be considered early in mission design to avoid overestimating available continuous power.

\begin{figure}[t]
  \centering
    \includegraphics[width=0.50\linewidth]{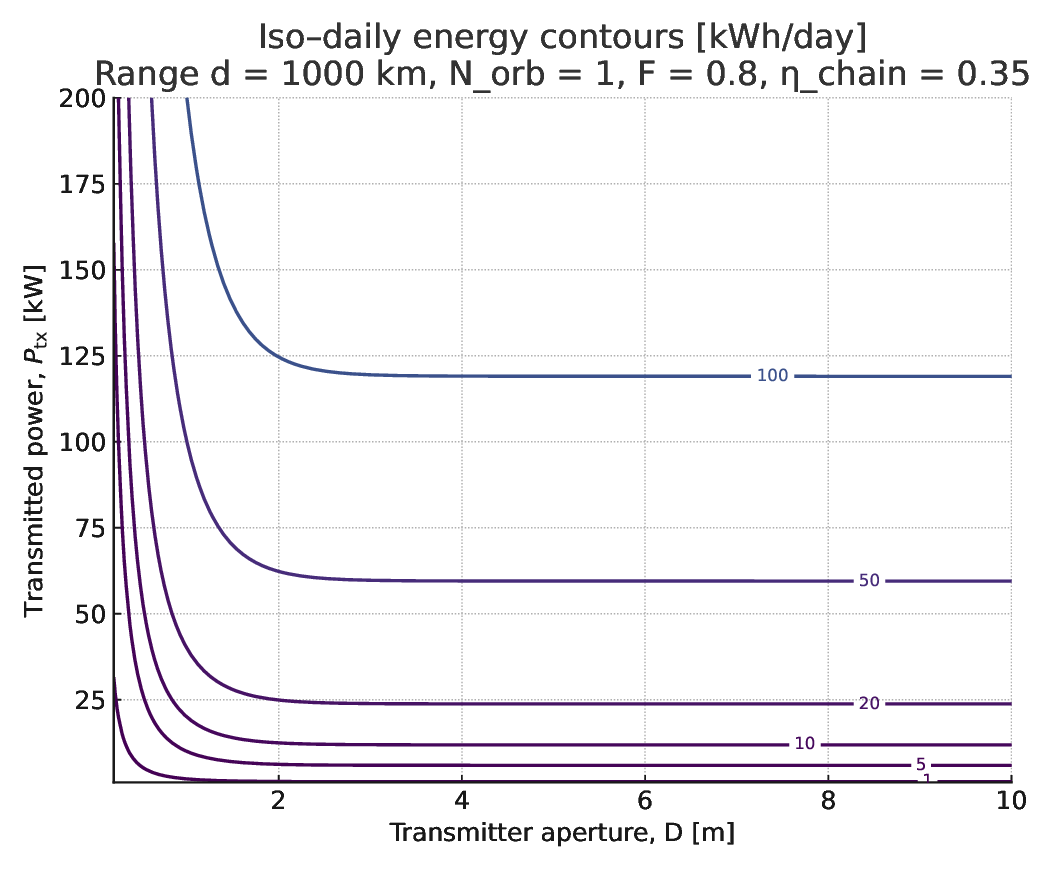}
  \caption{Iso--daily energy contours (kWh/day) as a function of transmitter
  aperture $D$ and transmitted power $P_{\rm tx}$ for range $d=1000$\,km and
  $N_{\rm orb}=1$. The model uses $\theta = M_{\rm eff}^2(F)\,2\lambda/(\pi D)$,
  $w_{\rm eff}=\sqrt{(\theta d)^2 + (d\,\sigma_{\rm point})^2}$, Gaussian
  encircled-energy capture $\eta_{\rm cap}=1-\exp[-2 r_{\rm rx}^2/w_{\rm eff}^2]$, 
  for a $1~\mathrm{m}^2$ receiver ($r_{\rm rx}=\sqrt{1/\pi}$). Daily energy is
$E_{\rm day} = P_{\rm tx}\,\eta_{\rm chain}\,\eta_{\rm cap}\, f_{\rm vis,1} \times 24$\,h.
Parameters here: $\lambda=1064$~nm, $\eta_{\rm chain}=0.35$,
$M_{\rm eff}^2(F)=1+k(1-F)$ with $F=0.8, k=1, \sigma_{\rm point}=0.2~\mu\mathrm{rad}$,
$f_{\rm vis,1}=0.10$.
}
  \label{fig:param-map}
\end{figure}

For this day, the DEM mask removes several low-elevation intervals and clips others, reducing daily energy by $\sim$12\% compared to a smooth-sphere horizon assumption (which would yield $\sim$750~Wh with the same link parameters). The result demonstrates how terrain masking propagates through visibility, range, and link budget to the operational metric of daily delivered energy. Mean daily Wh delivery scales approximately as:
\[
E_{\rm day} \propto N_{\rm orb} \Big(\frac{P_{\rm tx,opt} D^2}{d_{\rm eff}^2}\Big) A_{\rm rx} \eta_{\rm chain}(\phi_{\rm site}),
\]
where $N_{\rm orb}$ is the number of orbiters and $\phi_{\rm site}$ the target latitude. For polar sites in NRHO, doubling $N_{\rm orb}$ nearly doubles $E_{\rm day}$. A parametric map of iso‐Wh\,day$^{-1}$ in $(D, P_{\rm tx,opt})$ space for fixed $A_{\rm rx}$ would serve as a mission‐design chart (see Fig.~\ref{fig:param-map}).

Note that at multi-kW optical fluxes, coherent speckle patterns from phased-array sidelobes or atmospheric‐like modal structure can produce local hot‐spots on the receiver. Mitigation options include:  
(i) temporal phase dithering between sub-apertures,  
(ii) deliberate beamlet steering within the receiver aperture at $\ll$Hz rates, and  
(iii) insertion of mild diffusers or beam homogenizers ahead of the PV surface.  
These reduce spatial irradiance contrast while keeping the main-lobe on target.

On the transmitter side, high average powers per fiber/amplifier channel must remain below thermal-lensing and stimulated Brillouin scattering (SBS) thresholds. For silica fiber at $\lambda\!\approx\!1.06~\mu$m, SBS onset can occur at $\sim\!10$--20\,W CW in narrow-linewidth operation unless linewidth is broadened to $>$100 MHz; thermal lensing in bulk optics typically emerges at absorbed powers of a few watts/cm$^2$ without active cooling. These constraints inform the maximum per-channel power before coherent combination.

The NRHO case should be interpreted as a method anchor rather than a capability limit: for fixed geometry the delivered energy scales nearly linearly with $P_{\rm tx}$ and as $D_{\rm eff}^{2}$ through encircled-energy capture, and increases with the visibility fraction that can be raised by small constellations. In practice, the same scalings place the 10\,m/$100$\,kW case in the $\sim$30–50\,kWh\,day$^{-1}$ band under typical polar visibility, consistent with the design maps. These levers move the link from the demonstrative Wh\,day$^{-1}$ regime of this subsection into the practical kWh\,day$^{-1}$ band without altering the underlying bookkeeping. The same daily-energy values can then be compared to Eqs.~(\ref{eq:mpv})–(\ref{eq:mfis}) to determine which provisioning is mass-optimal for the stated site and load.

\subsection{Implementation insights from the parametric sizing maps}
\label{sec:parametric_discussion}

\begin{figure*}[h!]
\centering
\subfloat[Daily electrical energy delivered ($d=3000$\,km, $T_{\rm vis}=2$\,h/day, $A_{\rm rx}=1$\,m$^2$).]{%
  \includegraphics[width=0.45\textwidth]{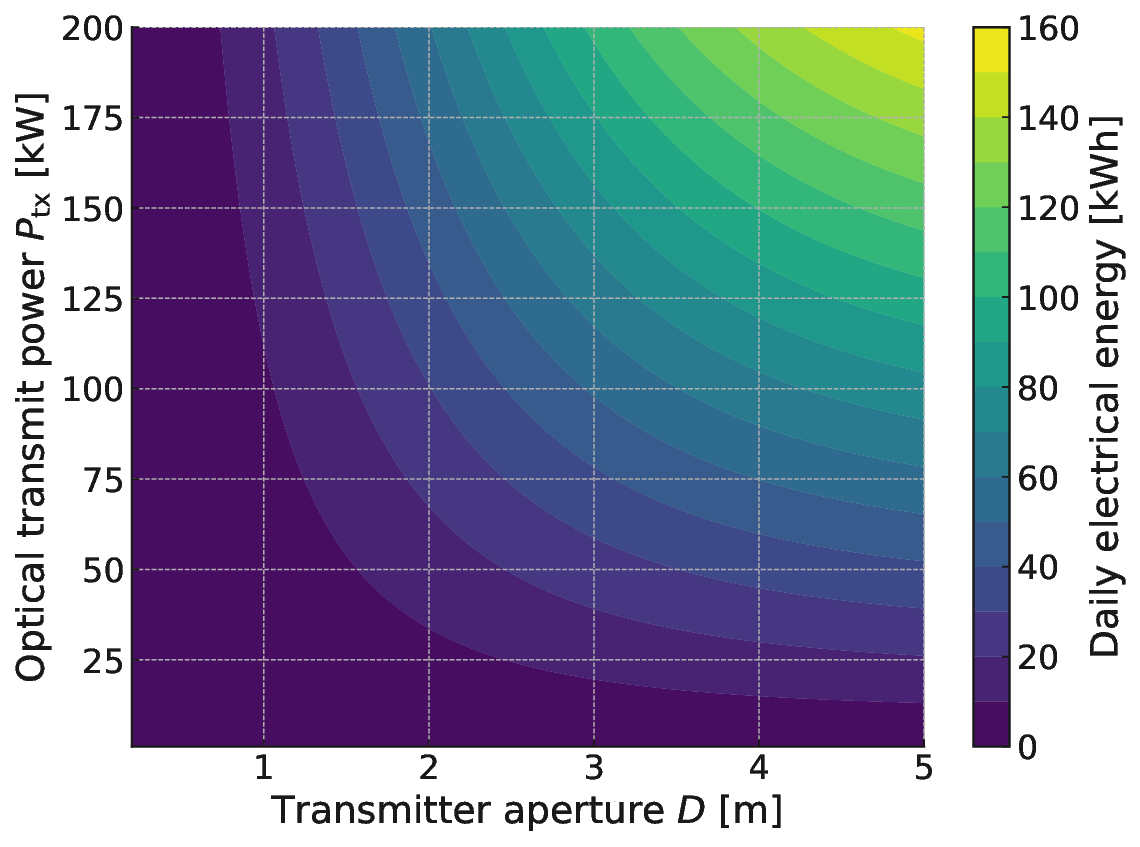}\label{fig:panel-a}}
\hfill
\subfloat[Transmitter PV area required vs.\ $D$ and $P_{\rm tx}$.]{%
  \includegraphics[width=0.45\textwidth]{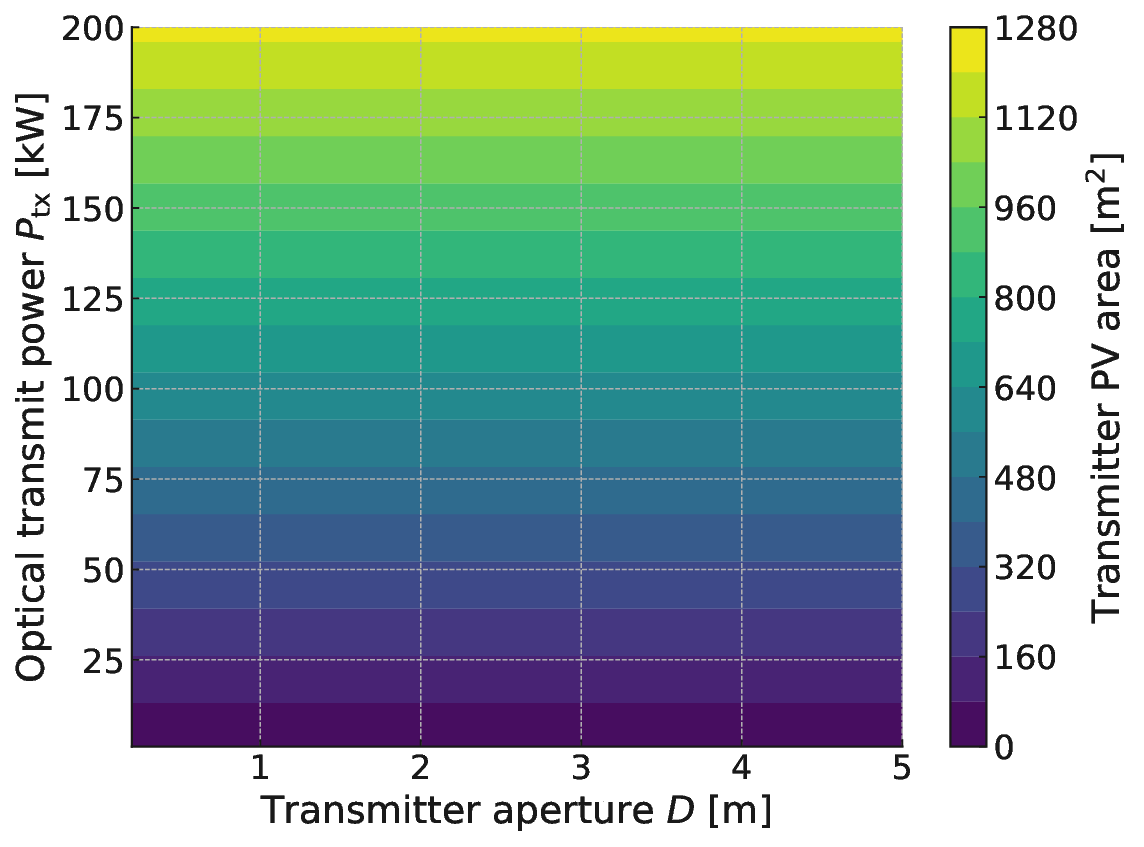}\label{fig:panel-b}}\\[0.8em]
\subfloat[Receiver radiator area sized for instantaneous waste heat.]{%
  \includegraphics[width=0.45\textwidth]{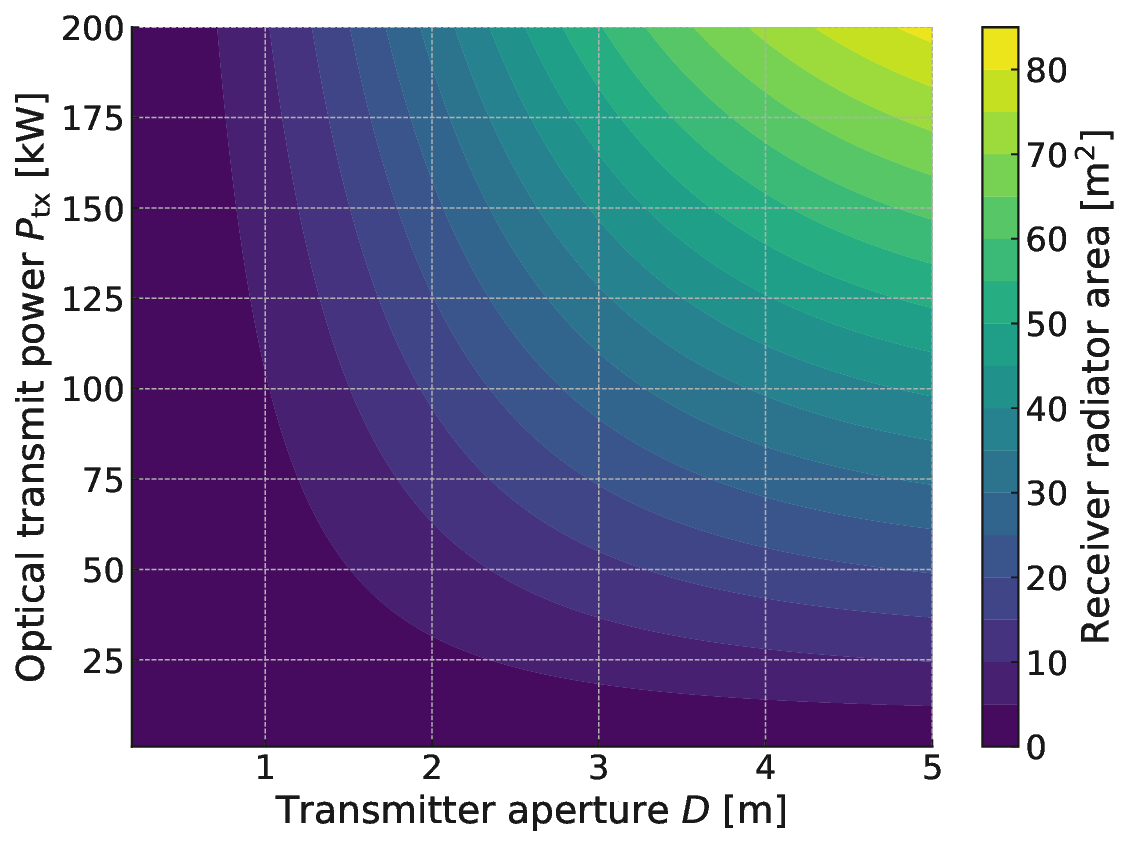}\label{fig:panel-c}}
\hfill
\subfloat[Transmitter radiator area for laser waste heat.]{%
  \includegraphics[width=0.45\textwidth]{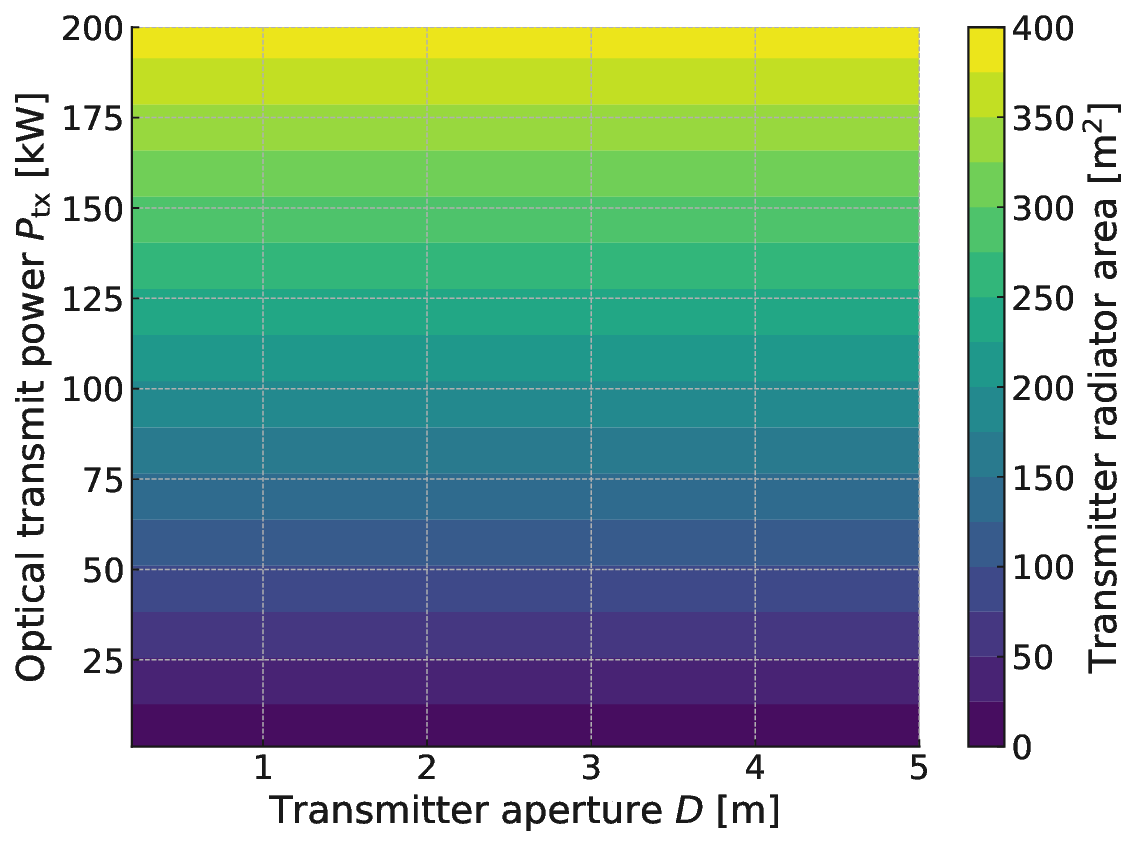}\label{fig:panel-d}}

\caption{Parametric sizing maps: (a) daily kWh delivered, (b) transmitter PV area, (c) receiver radiator area, (d) transmitter radiator area. Defaults: $\lambda=1064$\,nm, $M^2=1.2$, $\sigma_\theta=0.10\,\mu$rad, $d=3000$ km, $T_{\mathrm{vis},24\mathrm{h}}=2$ h in $24$ h (i.e., $f_{\mathrm{vis},1}^{24\mathrm{h}}=2/24\approx 0.083$), $A_{\mathrm{rx}}=1~\mathrm{m}^2$. The contours are intended to be read directly against Eqs.~(\ref{eq:mpv})–(\ref{eq:mfis}): select $(D_{\rm eff},P_{\rm tx},A_{\rm rx},N)$ to meet a site’s $E_{\rm day}$ target, then compare $M_{\mathrm{LPB}}$ to $M_{\mathrm{PV+bat}}$ and $M_{\mathrm{fis}}$ on the same $P_{\mathrm{eq}}$ basis.}
\label{fig:parametric-2x2}
\end{figure*}

Figure~\ref{fig:parametric-2x2} synthesizes the optical, electrical, and thermal constraints of the link by sweeping transmitter aperture $D$ and optical transmit power $P_{\rm tx}$ at the NRHO--to--polar reference point stated in Sec.~\ref{sec:parametric_maps}. The calculations use the divergence model (\ref{eq:theta_practical}), far-field beam radius (\ref{eq:slant-range}), jitter-broadened spot (\ref{eq:weff_def}), Gaussian encircled--energy capture fraction (\ref{eq:eta_cap_def}), and received power relations (\ref{eq:Prx_opt_main}), (\ref{eq:p_rx_elec_time}), (\ref{eq:p_rx_elec_time-Gaus}). Geometry and visibility are from Sec.~\ref{sec:orbit_geometry}, (\ref{eq:slant-dist})--(\ref{eq:slant-angle}), with temporal integration following Sec.~\ref{subsec:temporal_variation}. Upstream electrical power is set by (\ref{eq:p_tx}).

Panel~\ref{fig:panel-a} is set by the interplay of diffraction and jitter. Increasing $D$ reduces $\theta$ (\ref{eq:theta_practical}), shrinking the far--field radius $w$ from (\ref{eq:slant-range}); adding $\sigma_\theta$ yields $w_{\rm eff}$ (\ref{eq:weff_def}). The capture fraction $\eta_{\rm cap}$ follows (\ref{eq:eta_cap_def}), with the knee occurring at $w_{\rm eff} \approx r_{\rm rx}$. Left of the knee, $\eta_{\rm cap} < 1$ keeps $P_{\rm rx,opt}$ (\ref{eq:Prx_opt_main}) small, so increasing $P_{\rm tx}$ is inefficient. Right of the knee, $\eta_{\rm cap} \to 1$ and $P_{\rm rx,elec}$ scales linearly with $P_{\rm tx}$ through (\ref{eq:p_rx_elec_time})/(\ref{eq:p_rx_elec_time-Gaus}); further $D$ growth is limited by the jitter term in (\ref{eq:weff_def}). The knee location shifts with $d$ and $T_{\rm vis}$ (\ref{eq:slant-dist})--(\ref{eq:slant-angle}) and with pointing performance from (\ref{eq:psd_budget_master}), Sec.~\ref{subsec:pointing_budget_psd}.

Panel~\ref{fig:panel-b} derives from (\ref{eq:p_tx}), linking $P_{\rm tx}$ to $P_{\rm array}$. For fixed $\eta_\ell$, $I_\odot$, and $\eta_{\rm pv}$, $A_{\rm PV,tx} \propto P_{\rm tx}$, independent of $D$. High--energy points right of the knee in Panel~\ref{fig:panel-a} must be cross--checked against available $P_{\rm array}$ and duty cycle from Sec.~\ref{sec:orbit_geometry} and Sec.~\ref{subsec:temporal_variation}.

Panel~\ref{fig:panel-c} follows from $P_{\rm rx,opt}$ (\ref{eq:Prx_opt_main}) and $(1-\eta_{\rm rx})P_{\rm rx,opt}$ driving radiator sizing via (\ref{eq:thermal_balance}) in Sec.~\ref{sec:receiver_model}. As $\eta_{\rm cap}$ (\ref{eq:eta_cap_def}) approaches unity, $P_{\rm rx,opt}$ rises, and required area increases from modest (capture--limited) to tens of m$^2$/m$^2$ aperture (flux--limited), co--located with the optical knee.

Panel~\ref{fig:panel-d} is set by $P_{\rm tx}(\eta_\ell^{-1}-1)$ in Sec.~\ref{sec:spacecraft_power} and radiator sizing (\ref{eq:thermal_balance}). This term is independent of $D$ and scales linearly with $P_{\rm tx}$. Main--lobe efficiency $\eta_{\rm main}$ (\ref{eq:eta_main_bounded}) multiplies $P_{\rm rx,opt}$ (\ref{eq:Prx_opt_main}); degraded $\eta_{\rm main}$ or increased $M_{\rm eff}^2$ (\ref{eq:m2_bounded}) shifts operation leftward in Panel~\ref{fig:panel-a} and increases both transmitter and receiver thermal loads.

A viable point in Fig.~\ref{fig:parametric-2x2} must satisfy: (i) capture and pointing constraints (\ref{eq:theta_practical}), (\ref{eq:slant-range}), (\ref{eq:weff_def}), (\ref{eq:eta_cap_def}), (\ref{eq:psd_budget_master}); (ii) delivered energy (\ref{eq:Prx_opt_main}), (\ref{eq:p_rx_elec_time})/(\ref{eq:p_rx_elec_time-Gaus}) integrated per Sec.~\ref{subsec:temporal_variation}; (iii) PV supply (\ref{eq:p_tx}); and (iv) thermal limits (\ref{eq:thermal_balance}). Left of the knee, reduce $\sigma_\theta$ and increase $D$; right of the knee, $E_{\rm day} \propto P_{\rm tx}$ but PV and radiator areas (Panels~\ref{fig:panel-b}, \ref{fig:panel-d}) and receiver thermal capacity (Panel~\ref{fig:panel-c}) dominate.
 
\subsection{Constellation-Level Modeling}
\label{subsec:constellation}

The single--orbiter analyses in the preceding subsections quantify the daily net electrical energy $E_{\mathrm{day},1}$ as the time integral of the received power $P_{\mathrm{rx,elec}}(t)$ from \eqref{eq:p_rx_elec_time}, \eqref{eq:p_rx_elec_time-Gaus}, using the line-of-sight access geometry from Sec.~\ref{sec:orbit_geometry} and the time--dependent link model of Sec.~\ref{subsec:temporal_variation}. While these results establish realistic performance for a single transmitter, practical lunar power--beaming systems---particularly those intended to serve polar PSRs---will often require constellations of orbiters to reduce coverage gaps and thereby decrease the size and mass of surface energy storage systems.

Let $f_{\mathrm{vis},1}$ be the single--orbiter visibility fraction defined in \eqref{eq:fvis1} (with the appropriate $T_{\mathrm{ref}}$). For $N_{\mathrm{orb}}$ identical spacecraft in similar orbits, phased to maximize temporal coverage of a target site, the probability that none is in view at a given instant is $(1 - f_{\mathrm{vis},1})^{N_{\mathrm{orb}}}$. The aggregate coverage fraction is then
\begin{equation}
f_{\mathrm{vis},N} \;=\; 1 - \left( 1 - f_{\mathrm{vis},1} \right)^{N_{\mathrm{orb}}},
\label{eq:fvisN}
\end{equation}
which rises steeply for the first few spacecraft but asymptotically approaches unity. For small $f_{\mathrm{vis},1}$, the scaling is nearly linear: $f_{\mathrm{vis},N} \approx N_{\mathrm{orb}} f_{\mathrm{vis},1}$. When orbital geometry or terrain masking causes correlated visibility, \eqref{eq:fvisN} can be corrected with an overlap factor $0 < \kappa \leq 1$:
\begin{equation}
f_{\mathrm{vis},N} \;\approx\; 1 - \left( 1 - \kappa f_{\mathrm{vis},1} \right)^{N_{\mathrm{orb}}},
\label{eq:fvisN_corr}
\end{equation}
where $\kappa<1$ reduces the effective gain per added orbiter.

\begin{figure}[t]
  \centering
  \includegraphics[width=0.55\linewidth]{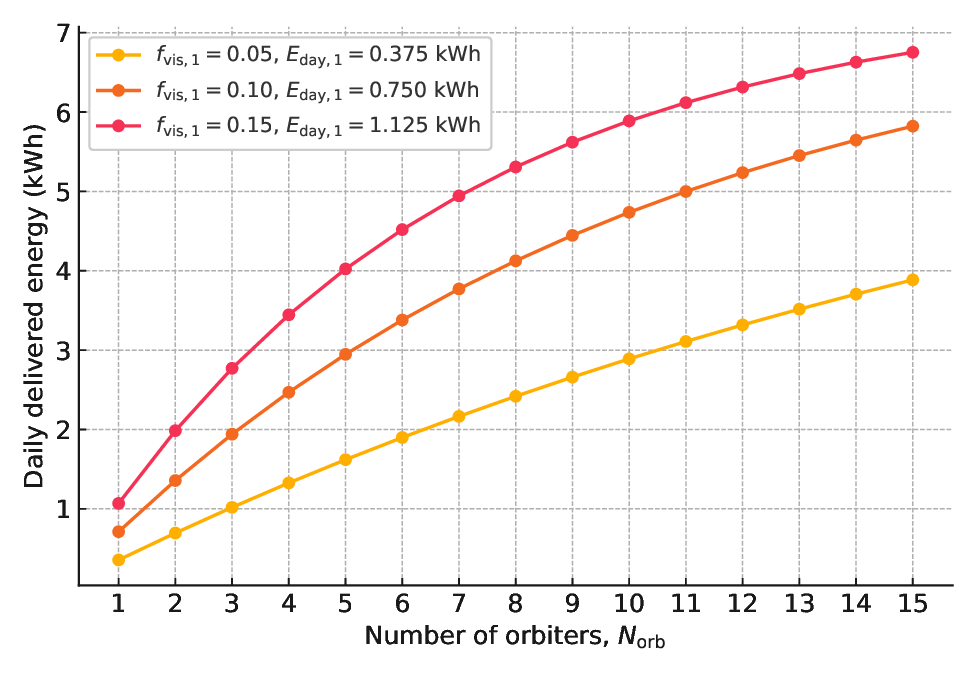}
  \caption{Daily energy delivery $E_{\mathrm{day}}(N_{\mathrm{orb}})$ for integer $N_{\mathrm{orb}}$ and representative single--orbiter visibility fractions $f_{\mathrm{vis},1} = 0.05,\ 0.10,\ 0.15$, each scaled to its corresponding $E_{\mathrm{day},1}$ assuming identical link performance. Curves use \eqref{eq:fvisN} with $\kappa=0.95$; correlation ($\kappa<1$) shifts them downward. Vertical separation between curves reflects the proportional increase in total energy for higher single--orbiter visibility.}
  \label{fig:eday_vs_norb}
\end{figure}

Since the instantaneous link budget does not change with constellation size, the daily energy delivered scales directly with the ratio of multi--orbiter to single--orbiter coverage:
\begin{equation}
E_{\mathrm{day}}(N_{\mathrm{orb}}) \;=\; \frac{f_{\mathrm{vis},N}}{f_{\mathrm{vis},1}} \, E_{\mathrm{day},1},
\label{eq:EdayN}
\end{equation}
where $E_{\mathrm{day},1}$ is the single--orbiter daily energy from Sec.~\ref{subsec:temporal_variation}. This relationship makes the saturation behaviour of \eqref{eq:fvisN} directly visible in energy terms: each additional spacecraft yields progressively smaller fractional gains as coverage approaches 100\%. For instance, for an NRHO case with $f_{\mathrm{vis},1} = 0.10$ and $E_{\mathrm{day},1} = 0.75$~kWh, \eqref{eq:EdayN} gives:
$\{N_{\rm orb}=1: E_{\rm day} = 0.75~{\rm kWh}\}; 
\{N_{\rm orb}=3: E_{\rm day} \approx 2.03~{\rm kWh}\}; 
\{N_{\rm orb}=6: E_{\rm day} \approx 3.52~{\rm kWh}\}; 
\{N_{\rm orb}=12: E_{\rm day} \approx 5.38~{\rm kWh}\}.$
If $f_{\mathrm{vis},1}$ is smaller (e.g., 0.05 in a low-inclination orbit), all energy values scale down proportionally; if it is larger (e.g., 0.15 with favourable geometry), they scale up. The vertical separation between curves in Fig.~\ref{fig:eday_vs_norb} directly reflects these differences in single--orbiter coverage.

These trends have direct architectural implications, namely:
(1) \emph{Constellation size vs.\ storage mass:} Higher $f_{\mathrm{vis},1}$ and/or $N_{\mathrm{orb}}$ increase average available power and reduce battery depth-of-discharge requirements. For sites with very low $f_{\mathrm{vis},1}$, even modest constellations can cut storage mass substantially.
(2) \emph{Orbit optimization:} Using mixed altitudes, inclinations, or orbital phasing reduces correlation ($\kappa \to 1$), improving the marginal return of each additional orbiter.
(3) \emph{Economic balance:} Because scaling is sub-linear, the cost-optimal $N_{\mathrm{orb}}$ often lies well short of continuous coverage, especially when spacecraft and launch costs scale nearly linearly.

Figure~\ref{fig:eday_vs_norb} shows $E_{\mathrm{day}}(N_{\mathrm{orb}})$ for three representative single--orbiter visibility fractions, each scaled to the corresponding $E_{\mathrm{day},1}$ assuming the same instantaneous link budget. Unlike normalized plots, this presentation directly shows that sites with larger $f_{\mathrm{vis},1}$ deliver proportionally more absolute energy at all $N_{\mathrm{orb}}$. The steep initial rise illustrates why the first few orbiters have the highest impact, while the curve flattening highlights the diminishing returns that set the practical limit for constellation size.

\subsection{Operational safety, regulatory framing, and cislunar interference}

Operations adopt hardwired “keep-out cones” about the bore-sight and a no-illumination altitude band near crewed assets. Operation near $1.06\,\mu$m maximizes PV efficiency but imposes tighter radiological/ocular exclusion than eye-safer $\sim 1.5\,\mu$m; the latter relaxes hazard distances by a few$\times$ at the cost of $\sim$10--20\% lower conversion efficiency at the receiver, and reduces stray-light risk to star trackers and optical sensors near crewed hardware.

The transmitter includes independent inhibit paths that drop optical power on (i) ATP loss, (ii) boresight crossing of a protect list (crewed vehicles, optics), and (iii) exceedance of a pointing-jitter threshold derived from the closed-loop budget. Protective interlocks are  at sub-Hz rates to bound time-integrated irradiance on any off-nominal intercept.

Beams are scheduled to avoid line-of-sight overlap with optical comm terminals at the surface or on orbit; spectrum and pointing coordination is brokered through a “flight plan” that specifies (start time, duration, azimuth/elevation raster, wavelength). The main-lobe/sidelobe allocation of phased arrays is controlled (via phase dithers and mild diffusers at the receiver plane) to suppress persistent speckle hot-spots and to keep sidelobe levels below saturating thresholds for nearby optical sensors. Keep-out sectors and time windows are published for GNSS-like PNT beacons to minimize interference to time-transfer and navigation observations.

Elevated scattering from lofted dust or glazed regolith is mitigated by keeping the steady-state flux within the receiver’s thermal envelope and by using electrostatic or mechanical cleaning for the receiver entrance pupil. The dust loss term in the link is parameterized as a secular fouling plus a flux-triggered instantaneous penalty and is exercised explicitly in the delivered-energy audits.

Operations conform to existing space-laser safety frameworks by (i) preventing illumination of crewed vehicles and specified infrastructure; (ii) managing hazard distances with wavelength choice and beam truncation; and (iii) validating keep-out geometry pre-mission via end-to-end demonstrations with representative ATP and interlocks.

\section{Summary of Key Equations}
\label{sec:summary_equations}

This section consolidates the derivations from previous sections into a concise set of equations that trace the solar power incident on a spacecraft in lunar orbit to the electrical power ultimately available at a receiver on the Moon's surface. Table~\ref{tab:notation} summarizes the principal notation used throughout the analysis. 

\begin{table}[ht]
\centering
\caption{Summary of parameters used in modeling power flow from spacecraft solar arrays to lunar surface receivers.}
\label{tab:notation}
\begin{tabular}{lp{0.69\textwidth}}
\hline
 {Parameter} &  {Definition/Description} \\
\hline\hline
\(I_{\text{sun}}\) & Solar irradiance in near-lunar space (nominally \SI{1361}{W/m^2}) \\
\(A_{\text{SA}}\) & Total area of the spacecraft's solar arrays (\si{m^2}) \\
\(\theta_{\text{inc}}(t)\) & Incidence angle between solar rays and the array normal (radians or degrees) \\
\(\eta_{\text{pv}}\) & Photovoltaic efficiency of the spacecraft's solar cells \\
\(P_{\text{SC,aux}}\) & Ancillary spacecraft power consumption (\si{W}) \\
\(\eta_{\ell}\) & Electrical-to-laser conversion efficiency of the onboard laser system \\
\(P_{\text{tx}}(t)\) & Optical power emitted by the laser transmitter (\si{W}); see \eqref{eq:p_tx} \\
\(\lambda\) & Laser wavelength used for power beaming (\si{\micro m} or \si{nm}) \\
\(D\), \(D_{\mathrm{eff}}\) & Physical (monolithic) or effective (phased-array) transmit aperture diameter (\si{m}) \\
\(M^2\) & Beam-quality factor of the transmitter optics (dimensionless) \\
\(\theta_{\text{tx}}\) & Practical half-angle beam divergence of the transmitted laser (radians); see \eqref{eq:theta_practical} \\
\(\sigma_{\theta}(t)\) & RMS line-of-sight pointing jitter (radians); enters \(w_{\mathrm{eff}}\) via \eqref{eq:weff_def} \\
\(\theta_{\mathrm{eff}}(t)\) & Effective half-angle including jitter, \(\theta_{\mathrm{eff}}(t) \equiv w_{\mathrm{eff}}(d,t)/d(t)\) \\
\(d(t)\) & Slant range between the spacecraft and the lunar surface receiver at time \(t\) (\si{m}) \\
\(w(t)\), \(w_{\mathrm{eff}}(d,t)\) & \(1/e^2\) beam radius at range, \(w(t)=\theta_{\text{tx}} d(t)\); jitter-broadened \(w_{\mathrm{eff}}^2=w^2+[2\sigma_\theta d]^2\) \eqref{eq:weff_def} \\
\(\eta_{\text{point}}(t)\) & Time-dependent pointing efficiency (accounts for beam steering and residual loop losses) \\
\(\eta_{\text{opt}}(t)\) & Time-dependent optical path efficiency (accounts for transmission losses and dust) \\
\(\eta_{\mathrm{main}}\) & Main-lobe efficiency (array factor for phased/sparse arrays); see \eqref{eq:eta_main_bounded} \\
\(A_{\text{rx}}\), \(r_{\text{rx}}\) & Receiver aperture (effective collection area) and radius \(r_{\text{rx}}=\sqrt{A_{\text{rx}}/\pi}\) \\
\(\eta_{\text{cap}}(d,t)\) & Encircled-energy capture fraction for a Gaussian beam; see \eqref{eq:eta_cap_def} \\
\(\eta_{\text{rx}}\) & Photonic-to-electrical conversion efficiency of the receiver \\
\(I_{\text{eff}}(t)\) & Effective irradiance at the receiver plane (W/m\(^2\)); see \eqref{eq:Ieff_def} or \eqref{eq:I_eff} \\
\(P_{\text{rx,elec}}(t)\) & Electrical power received at the surface receiver (\si{W}) \\
\hline
\end{tabular}
\end{table}

\subsection{Power Generation on the Spacecraft}
\label{subsec:spacecraft_generation_summary}

The instantaneous power generated by the spacecraft's solar arrays depends on solar irradiance \(I_{\text{sun}}\), array area \(A_{\text{SA}}\), incidence angle \(\theta_{\text{inc}}(t)\), and photovoltaic efficiency \(\eta_{\text{pv}}\):
\begin{equation}
    P_{\text{array}}(t) = I_{\text{sun}} A_{\text{SA}}\eta_{\text{pv}} \cos\bigl[\theta_{\text{inc}}(t)\bigr].
    \label{eq:P_array}
\end{equation}
Over an orbit, \(\theta_{\text{inc}}(t)\) can vary, and eclipses cause \(I_{\text{sun}}\) to drop to zero, so onboard energy storage is often required to smooth out power generation.

Subtracting the spacecraft’s ancillary power load \(P_{\text{SC,aux}}\) yields the net electrical power available for laser generation:
\begin{equation}
    P_{\text{net}}(t) \;=\; P_{\text{array}}(t) \;-\; P_{\text{SC,aux}}.
    \label{eq:P_net}
\end{equation}
Applying the laser’s electrical-to-optical efficiency \(\eta_{\ell}\) converts this net power into optical power (from (\ref{eq:p_tx})):
\begin{equation}
    P_{\text{tx}}(t) \;=\; P_{\text{net}}(t)\,\eta_{\ell}
    =\Big(P_{\text{array}}(t) \;-\; P_{\text{SC,aux}}\Big)\eta_{\ell}.
    \label{eq:p_tx2}
\end{equation}
If \(P_{\text{array}}(t) < P_{\text{SC,aux}}\) (e.g., during partial eclipse), the system may draw power from on-board energy storage (batteries, supercapacitors) to maintain a given laser output or temporarily reduce the laser transmission.

\subsection{Transmission and Reception}
\label{subsec:transmission_reception_summary}

As the laser beam propagates from orbit to the surface, it diverges according to the half-angle \(\theta_{\mathrm{tx}}\), and additional losses arise from pointing (jitter/misalignment) and optical path attenuation (e.g., imperfections in transmitter optics or scattering by lunar dust). The practical, diffraction-limited-plus-quality divergence used throughout the paper is given by (\ref{eq:theta_practical}) as below
\begin{equation}
\theta_{\text{tx}} = M^2 \,\frac{2\lambda}{\pi D_{\mathrm{eff}}},
\label{eq:theta_practical2}
\end{equation}
where \(D_{\mathrm{eff}}\) is the physical or synthesized aperture and \(M^2\) is the beam-quality factor. For phased arrays, the main-lobe efficiency \(\eta_{\mathrm{main}}(F)\) follows (see (\ref{eq:eta_main_bounded}))
\begin{equation}
\eta_{\mathrm{main}}(F) = \eta_{\infty} - \big(\eta_{\infty} - \eta_{0}\big)\exp\!\Big[-\Big(\frac{F}{F_{c}}\Big)^{p}\Big],
\label{eq:eta_main_bounded2}
\end{equation}
where \(F\) is the fill factor and \(\eta_{\infty},\eta_{0},F_{c},p\) are fit parameters. Jitter broadens the spot in quadrature (from (\ref{eq:weff_def})):
\begin{equation}
w_{\mathrm{eff}}^2(d,t) = w^2(d,t)+\left[2\sigma_\theta(t)\,d(t)\right]^2, \qquad w(t) = \theta_{\text{tx}}\,d(t),
\label{eq:weff_def2}
\end{equation}
with capture fractions for Gaussian beams given by
\begin{equation}
\eta_{\mathrm{cap}}(d,t) = 1 - \exp\Big[-\,2\,\frac{r_{\mathrm{rx}}^{2}}{w_{\mathrm{eff}}^{2}(d,t)}\Big].
\label{eq:eta_cap_def2}
\end{equation}

The effective power flux at the surface in the scalar top-hat model is
\begin{equation}
    I_{\text{eff}}(t) =
    \eta_{\text{point}}(t)\,\eta_{\text{opt}}(t)
    \frac{P_{\text{tx}}(t)}{\pi \left[\theta_{\text{tx}}d(t)\right]^{2}},
    \label{eq:I_eff}
\end{equation}
while in the Gaussian capture model it is
\begin{equation}
I_{\mathrm{eff}}(t) =
\eta_{\mathrm{main}}\,\eta_{\text{point}}(t)\,\eta_{\text{opt}}(t)\,
\frac{P_{\text{tx}}(t)\,\eta_{\mathrm{cap}}(d,t)}{A_{\rm rx}}.
\label{eq:Ieff_def2}
\end{equation}

Let \(A_{\text{rx}}\) be the receiver’s projected area in the beam (e.g., the physical aperture of a PV panel or an optical collector). A baseline, ``top-hat'' approximation then gives the intercepted optical power as
\begin{equation}
    P_{\text{rx,opt}}(t)
    \;=\;
    I_{\text{eff}}(t)\;\times\;A_{\text{rx}}.
    \label{eq:P_rx_opt}
\end{equation}
If \(A_{\text{rx}}\) is smaller than the nominal beam footprint, the receiver collects only a fraction of the total transmitted power. Conversely, if the beam footprint is fully contained within \(A_{\text{rx}}\), the captured optical power cannot exceed \(P_{\text{tx}}(t)\,\eta_{\text{point}}(t)\,\eta_{\text{opt}}(t)\) (i.e., all of the beam minus pointing/optical losses).

Multiplying by the receiver's photonic-to-electrical conversion efficiency \(\eta_{\text{rx}}\) gives the net electrical output:
\begin{equation}
    P_{\text{rx,elec}}(t)
    =
    \eta_{\text{rx}}P_{\text{rx,opt}}(t)
    =
    \eta_{\text{point}}(t)\,\eta_{\text{opt}}(t)\eta_{\text{rx}}P_{\text{tx}}(t)
    \frac{A_{\text{rx}}}{\pi\bigl[\theta_{\text{tx}}d(t)\bigr]^{2}}
    \quad
    \bigl(\text{top-hat approximation}\bigr).
    \label{eq:P_rx_elec_time}
\end{equation}
Substituting \(P_{\text{tx}}(t)\) from Eq.~(\ref{eq:p_tx}) (or the relevant power budget expression) then gives a complete link from spacecraft solar array output to surface power.

In practice, the beam intensity profile may be closer to a 2D Gaussian. (i)~If the receiver aperture is much smaller than the beam waist, only a small on-axis fraction is collected; (ii)~If the receiver is much larger than the beam footprint, it captures nearly all of \(P_{\text{tx}}(t)\,\eta_{\text{point}}(t)\,\eta_{\text{opt}}(t)\). A more general expression, valid for any ratio of receiver radius \(r_{\mathrm{rx}}\) to beam radius \(w(t)=\theta_{\text{tx}}\,d(t)\), is:
\begin{equation}
    P_{\mathrm{rx,opt}}(t)
    =
    \eta_{\mathrm{point}}(t)\eta_{\mathrm{opt}}(t)P_{\mathrm{tx}}(t)
    \Bigl[
      1-\exp\Big(\!-2\frac{r_{\mathrm{rx}}^{2}}{w(t)^{2}}\Big)
    \Bigr],
    \label{eq:GaussianCaptureGeneral}
\end{equation}
where \(r_{\mathrm{rx}}=\sqrt{A_{\mathrm{rx}}/\pi}\) and \(w(t)=\theta_{\text{tx}}d(t)\) is the \(1/e^2\) radius. Equivalently, using the encircled-energy definition \(\eta_{\text{cap}}(d,t)=1-\exp[-2r_{\rm rx}^2/w_{\rm eff}^2(d,t)]\) from \eqref{eq:eta_cap_def} with \(w_{\rm eff}\) given by \eqref{eq:weff_def}, one may write the link in the compact form \eqref{eq:Ieff_def}. After including \(\eta_{\mathrm{rx}}\) for photonic-to-electrical conversion, one obtains
\( P_{\mathrm{rx,elec}}(t) = \eta_{\mathrm{rx}}\times P_{\mathrm{rx,opt}}(t)\), yielding (\ref{eq:p_rx_elec_time-Gaus})
\begin{equation}
 P_{\text{rx,elec}}(t)=
\eta_{\mathrm{point}}(t)\eta_{\mathrm{opt}}(t)\eta_{\text{rx}}
P_{\mathrm{tx}}(t)
\Bigl[
1 - \exp\Bigl(-\,2\,\dfrac{r_{\mathrm{rx}}^{2}}{w(t)^{2}}\Bigr)
\Bigr].
\label{eq:p_rx_elec_time-Gaus-fin}
\end{equation}
Thus, whether a simple top-hat approximation (\ref{eq:P_rx_elec_time}) or the integrated Gaussian capture (\ref{eq:p_rx_elec_time-Gaus-fin}) is used, the final result depends on the system geometry, beam divergence, optical and pointing efficiencies, and the effective aperture of the receiver. For phased arrays, \(\eta_{\mathrm{main}}\) from \eqref{eq:eta_main_bounded} is applied consistently via \eqref{eq:Ieff_def} so that main-lobe power allocation, geometric capture, and residual pointing are not double-counted. The 1064\,nm receiver efficiencies used here are in family with demonstrated laser power converters and legacy PV receiver design analyses~\cite{Gou2022OE1064,Algora2022Joule,Landis1993JPP}.

\subsection{Chaining from Solar Array to Receiver}
\label{subsec:chaining_equations}

Combining (\ref{eq:P_array})--(\ref{eq:P_rx_elec_time}) gives the full time-dependent chain from solar collection to net power on the lunar surface:
\begin{equation}
    P_{\text{rx,elec}}(t)
   =
    \underbrace{\eta_{\text{point}}(t) \eta_{\text{opt}}(t)\eta_{\text{rx}}}_{
      \text{Link efficiencies}}
    \Bigl\{\Big(P_{\text{array}}(t) - P_{\text{SC,aux}}\Big)\eta_{\ell}\Big\}
   \frac{A_{\text{rx}}}{\pi \bigl[\theta_{\text{tx}} d(t)\bigr]^2},
    \label{eq:P_rx_elec_chained}
\end{equation}
with
\begin{equation}
    P_{\text{array}}(t)=
    I_{\text{sun}} A_{\text{SA}} \eta_{\text{pv}}
    \cos\bigl[\theta_{\text{inc}}(t)\bigr].
\end{equation}
Eq.~\eqref{eq:P_rx_elec_chained} explicitly shows how each subsystem's efficiency and the system's geometry influence the final power received on the lunar surface, succinctly illustrating how the solar array, onboard consumption, laser efficiency, beam divergence, pointing accuracy, and receiver performance all contribute to final delivered power.

In (\ref{eq:P_rx_elec_chained}), multiplying flux by \(A_{\text{rx}}\) assumes a top-hat beam spot. If the laser beam follows a 2D Gaussian profile and its radius \(w(t) = \theta_{\text{tx}}\;d(t)\) is \emph{fully} contained by the receiver, then the maximum intercepted optical power is limited to (with \(P_{\mathrm{tx}}(t)\) from (\ref{eq:p_tx}))
\[
  P_{\mathrm{tx}}(t)\,\eta_{\text{point}}(t)\,\eta_{\text{opt}}(t)\equiv \eta_{\text{point}}(t)\,\eta_{\text{opt}}(t)\Big\{\Big(P_{\text{array}}(t) - P_{\text{SC,aux}}\Big)\eta_{\ell}\Big\}.
\]
For partial interception, a more accurate approach integrates the radial Gaussian intensity out to \(r_{\mathrm{rx}}=\sqrt{A_{\mathrm{rx}}/\pi}\). A typical form is \( P_{\mathrm{rx,opt}}(t)
  \;=\;
  \bigl[\eta_{\text{point}}(t)\,\eta_{\text{opt}}(t)\bigr]\,
  P_{\mathrm{tx}}(t)
  \;\bigl[
    1 - \exp\bigl(-\,2\,{r_{\mathrm{rx}}^2}/{w(t)^2}\bigr)
  \bigr]\), yielding received electric power
\begin{align}
  P_{\mathrm{rx,elec}}(t)
  &\;=\;    \underbrace{\eta_{\text{point}}(t)\,\eta_{\text{opt}}(t)\,\eta_{\text{rx}}}_{
      \text{Link efficiencies}}\; \Big\{\Big(P_{\text{array}}(t) - P_{\text{SC,aux}}\Big)\eta_{\ell}\Big\} \Bigl[
    1 - \exp\Bigl(-\,2\,\dfrac{r_{\mathrm{rx}}^2}{w(t)^2}\Bigr)
  \Bigr],
\end{align}
which smoothly covers the regime \(r_{\mathrm{rx}}\ll w(t)\) (small receiver) up to \(r_{\mathrm{rx}}\gg w(t)\) (receiver fully captures the beam). Multiplying by \(\eta_{\text{rx}}\) then yields the net electrical power for the Gaussian-capture scenario.

{Key observations are as follows:}
\begin{itemize}
    \item \textit{Time Dependence:} Both \(d(t)\) and \(\theta_{\text{inc}}(t)\) may change over an orbital period or multiple orbits. Additionally, \(I_{\text{sun}}\) can drop to zero if the spacecraft enters eclipse behind the Moon or the Earth.
    \item \textit{Instantaneous vs. Integrated Power:} To determine the energy delivered over a pass of duration \(\Delta t\), one needs to integrate \(P_{\text{rx,elec}}(t)\) over that interval:
    \[
        E_{\text{delivered}}(t_0,\Delta t) = \int_{t_0}^{t_0+\Delta t} P_{\text{rx,elec}}(t)\,\mathrm{d}t.
    \]
    \item \textit{Storage Smoothing:} In practice, \(P_{\text{tx}}(t)\) may be held nearly constant by drawing from on-board energy storage, even if \(P_{\text{array}}(t)\) is intermittent. This allows a more stable power delivery to the receiver.
    \item \textit{Phased-Array Levers:} Increasing \(D_{\mathrm{eff}}\) reduces \(\theta_{\text{tx}}\) via \eqref{eq:theta_practical}, directly shrinking \(w(t)\) and boosting surface flux. The main-lobe fraction \(\eta_{\mathrm{main}}(F)\) in \eqref{eq:eta_main_bounded} controls how much of \(P_{\text{tx}}\) is available in the synthesized beam core, while \(\eta_{\text{cap}}(d,t)\) from \eqref{eq:eta_cap_def} captures geometric interception (with jitter through \eqref{eq:weff_def}). Residual loop losses remain in \(\eta_{\text{point}}(t)\) as in \eqref{eq:Ieff_def}.
 \item \textit{Daily--energy shorthand:} Over any reference window $T_{\rm ref}$,
\[
E(T_{\rm ref}) \;\approx\; \overline{P_{\rm rx,elec}}_{\;\mathrm{vis}}(T_{\rm ref})\,T_{\rm ref}\,f_{\mathrm{vis},1}(T_{\rm ref}),
\]
with $f_{\mathrm{vis},1}$ from \eqref{eq:fvis1} and $\overline{P_{\rm rx,elec}}_{\;\mathrm{vis}}$ evaluated via the time-domain link \eqref{eq:p_rx_elec_time2} or \eqref{eq:Gauss_rx_elec-large} (using $\eta_{\rm cap}$ from \eqref{eq:eta_cap_def} and $w_{\rm eff}$ from \eqref{eq:weff_def}).

\end{itemize}

\subsection{Parameter Ranges and Operational Insights}
\label{subsec:parameter_ranges}

Typical performance ranges for the main efficiencies and geometric parameters in (\ref{eq:P_rx_elec_chained}) appear in Table~\ref{tab:typical_parameters}. In particular, note that larger transmitter apertures (whether a large monolithic design or a \emph{phased array} synthesizing an equivalent diameter) can sharply reduce the half-angle divergence \(\theta_{\text{tx}}\) and thereby boost power density on the surface. The Gaussian capture and the scalar irradiance bookkeeping should not be mixed in a way that double-counts geometric interception: use either the \(\eta_{\text{cap}}\) form with \eqref{eq:Ieff_def} and \(w_{\rm eff}\) from \eqref{eq:weff_def}, or the top-hat form \eqref{eq:I_eff}--\eqref{eq:P_rx_elec_time}.

\begin{table}[ht]
\vskip -10pt 
\centering
\caption{Typical ranges for key parameters in laser power beaming systems.}
\label{tab:typical_parameters}
\begin{tabular}{ll}
\hline
{Parameter} & {Typical Range or Value} \\
\hline\hline
Solar array efficiency, \(\eta_{\text{pv}}\) & 0.20--0.35 (up to 0.40 for advanced cells) \\
Laser efficiency, \(\eta_{\ell}\) & 0.30--0.50 \\
Pointing efficiency, \(\eta_{\text{point}}\) & 0.70--0.95 \\
Optical path efficiency, \(\eta_{\text{opt}}\) & 0.80--0.99 \\
Receiver efficiency, \(\eta_{\text{rx}}\) & 0.20--0.60 \\
Practical half-angle divergence, \(\theta_{\text{tx}}\) & 1--10 \(\mu\)rad (can be lower with large aperture)\\
\hline
\end{tabular}
\end{table}
The following range parameter behavior is expected: 
\begin{itemize}
    \item \textit{Solar Array Efficiency, \(\eta_{\text{pv}}\):}  
    Ranges from 0.20 to 0.35 for common space‐qualified multi‐junction cells. Advanced high‐concentration cells can exceed 0.40 under optimal conditions.
    \item \textit{Laser Conversion Efficiency, \(\eta_{\ell}\):}  
    Typically 0.30 to 0.50 at the system level for diode or fiber lasers. Laboratory values may be higher, but one must account for driver electronics and thermal management.
    \item \textit{Pointing Efficiency, \(\eta_{\text{point}}\):}  
    Varies from $\approx$ 0.70 (if pointing/tracking is poor) to above 0.95 (with precise beam steering).
    \item \textit{Optical Path Efficiency, \(\eta_{\text{opt}}\):}  
    Typically ranges from 0.80 to 0.99, accounting for lens/mirror reflectivity or transmissivity, dust effects, and minor scattering.
    \item \textit{Receiver Conversion Efficiency, \(\eta_{\text{rx}}\):}  
    Ranges from 0.20 to 0.60 for laser-optimized photovoltaic receivers, depending on spectral matching, operating temperature, and irradiance levels.
    \item \textit{Practical Half-Angle Beam Divergence, \(\theta_{\text{tx}}\):}  
    On the order of 1--10 \(\mu\)rad for medium-aperture space-based transmitters; larger apertures reduce \(\theta_{\text{tx}}\) but add mass/complexity.
\end{itemize}

Overall, \eqref{eq:p_tx}--\eqref{eq:P_rx_elec_chained} provide the analytic backbone for modeling an orbital laser power beaming system. By specifying the orbital profile, subsystem efficiencies, beam divergence, and receiver characteristics, one can compute both instantaneous and integrated power available for lunar operations. Increasing the effective aperture---via a \emph{phased array} or otherwise---is among the most powerful levers to reduce divergence and achieve significantly higher flux at the surface, especially when paired with high \(\eta_{\mathrm{main}}(F)\) from \eqref{eq:eta_main_bounded}, careful jitter control through \eqref{eq:weff_def}, and consistent bookkeeping using \eqref{eq:Ieff_def}.

Alternative architectures---surface PV with storage or compact fission---admit a simple comparison at the “daily-energy to continuous-power” level. For a target $P_{\rm cont}=E_{24\mathrm{h}}/24\,$h at a polar site with low sunlight fraction during night, PV+battery mass scales as $\sigma_{\rm PV}A_{\rm PV}+E_{\rm store}/e_{\rm bat}$ with $E_{\rm store}\!\approx\!(1-f^{24\mathrm{h}}_{\odot})\,24\,{\rm h}\,P_{\rm cont}$. For compact fission, $M_{\rm fis}\!\approx\!\alpha_{\rm fis} P_{\rm cont}$. The beamed‑power system mass aggregates transmitter structure, sources, and radiators sized by the same chain used here. Break‑even contours occur when $M_{\rm beam}\!\lesssim\!M_{\rm PV+bat}$ or $M_{\rm beam}\!\lesssim\!M_{\rm fis}$ at the site’s visibility fraction and range.

\section{Conclusion}
\label{sec:conclusion}

The analyses and equations presented in this work provide a rigorous framework for conceptualizing and evaluating a lunar-orbit laser power beaming system. By integrating time-dependent models of power generation, transmission, and reception, the framework enables accurate predictions of system performance and supports informed design decisions for lunar power delivery missions.

Evaluating a laser power beaming system from lunar orbit to the lunar surface entails five interdependent steps:

\begin{enumerate}
    \item \textit{Spacecraft Solar Power Generation:}  
    The instantaneous solar array output, \(P_{\text{array}}(t)\), stems from the incident solar flux \(I_{\text{sun}}\), array area \(A_{\text{SA}}\), photovoltaic efficiency \(\eta_{\text{pv}}\), and the incidence angle \(\theta_{\text{inc}}(t)\). Variations in spacecraft attitude, eclipses, and distance-related effects introduce time-dependent power fluctuations that must be managed---often using onboard energy storage---to ensure nearly continuous power availability.

    \item \textit{Spacecraft Power Budget and Laser Output:}  
    Subtracting the spacecraft’s ancillary load \(P_{\text{SC,aux}}\) from \(P_{\text{array}}(t)\) yields the net electrical power available for laser generation. Multiplying by the electrical-to-laser conversion efficiency \(\eta_{\ell}\) provides the transmitted optical power \(P_{\text{tx}}(t)\). During periods of diminished solar input, onboard batteries or supercapacitors can buffer and smooth out \(P_{\text{tx}}(t)\).

    \item \textit{Laser Beam Propagation and Link Budget:}  
    The laser beam’s half-angle divergence \(\theta_{\text{tx}}\) depends on the transmitter aperture diameter, beam quality factor \((M^2)\), and pointing stability. In particular, \emph{phased-array transmitters} can combine multiple sub-apertures coherently, effectively enlarging the aperture diameter and reducing beam divergence. Over the slant range \(d(t)\), the expanding beam footprint dictates the local power flux, while additional losses arise from pointing errors \(\eta_{\text{point}}(t)\) and optical path inefficiencies \(\eta_{\text{opt}}(t)\). Consequently, the delivered flux scales as 
 \(P_{\text{tx}}(t) / \bigl(\pi [\theta_{\text{tx}}\, d(t)]^2\bigr)\).

    \item \textit{Surface Reception and Conversion:}  
    The receiver intercepts a portion of the beam according to its effective area \(A_{\text{rx}}\). After interception, photonic-to-electrical conversion at efficiency \(\eta_{\text{rx}}\) yields the net electrical output \(P_{\text{rx,elec}}(t)\). Advanced receiver designs---such as using optical concentrators or wavelength-tailored PV cells---can enhance \(\eta_{\text{rx}}\). Concurrently, robust thermal management and dust mitigation are essential to ensuring stable, long-term performance in the lunar environment.

    \item \textit{Time-Dependent Integration:}  
    As parameters like \(d(t)\), \(\theta_{\text{inc}}(t)\), and subsystem efficiencies vary over an orbital cycle, so does \(P_{\text{rx,elec}}(t)\). Integrating the instantaneous power over each visibility window quantifies the total energy delivered:
    \[ E_{\text{delivered}} = \int_{t_1}^{t_2} P_{\text{rx,elec}}(t) \mathrm{d}t.   \]
    This step captures not only the hardware capabilities but also the real-time orbital and operational constraints, forming the basis for mission-level energy planning.
\end{enumerate}

System performance is highly sensitive to several key parameters:
\begin{itemize}
\item Increasing the transmitter aperture \(D_{\rm eff}\) reduces the practical beam divergence \(\theta_{\text{tx}}\), thereby enhancing the power flux density at the receiver (roughly scaling as \((D_2 / D_1)^2\)). Larger apertures may also be implemented as phased arrays, providing a more flexible path to large effective diameters but at the expense of an increased mass and greater system structural complexity.
    \item Enhancing the laser system efficiency \((\eta_{\ell})\) and the receiver conversion efficiency \((\eta_{\text{rx}})\) directly improves overall power throughput, allowing higher energy delivery for the same solar input.
    \item Reducing the slant range \(d(t)\) increases the flux density, but this may impose more restrictive orbital constraints.
    \item Advanced beam steering and adaptive optics can mitigate losses due to pointing jitter and misalignment, particularly during rapid orbital motion.

\end{itemize}

Balancing these factors ensures a robust system design capable of meeting power requirements for  lunar applications.

Several avenues remain for further refining and deploying laser power beaming systems in lunar orbit:

\begin{itemize}
    \item \textit{Orbital Mechanics and Coverage Optimization:} 
    High-fidelity orbital modeling of near-rectilinear halo orbits (NRHOs) and high-eccentricity trajectories can identify prolonged intervals of uninterrupted sunlight for polar regions. Such modeling should account for multi-body gravitational effects, lunar libration, and eclipse durations to maximize coverage of permanently shadowed areas.

    \item \textit{Real-Time Beam Steering and Adaptive Optics:}
    Fast-feedback beam steering (e.g., 1--10\,kHz update rates) combined with adaptive optics can mitigate jitter and wavefront distortions caused by thermal expansion, structural vibrations, and spacecraft attitude changes. By maintaining near-diffraction-limited performance, these methods significantly reduce pointing losses and improve overall power throughput.

    \item \textit{Advanced Thermal Management:}
    Both on-board laser diodes and high-intensity photovoltaic receivers require detailed thermal control strategies to ensure stable operation under the Moon’s extreme temperature variations. Techniques could include conduction into the regolith, radiative cooling panels, or thermal storage systems capable of dissipating up to 1--2\,kW of waste heat per kW of optical output.

    \item \textit{Scalability and Deployment Strategies:}
    Comprehensive trade studies factoring in launch costs, array mass, and potential in-situ manufacturing (for receivers or large aperture structures) will clarify the cost--benefit ratio of different power levels and aperture sizes. Larger phased arrays deliver higher flux densities and enable greater stand-off distances, but at the expense of increased structural complexity and mass.

    \item \textit{Multi-Point Beaming and Constellations:}
    Providing power to multiple lunar sites (e.g., for rover recharging, habitat support, or ISRU operations) demands sophisticated scheduling algorithms and agile pointing controls, potentially requiring sub-arcminute targeting accuracy at distances of hundreds or thousands of kilometers. Either a single orbiter with multiple phased-array beams or a constellation of dedicated “power satellites” could deliver increased coverage and redundancy.

    \item \textit{Safety and Interference Mitigation:}
    Phased-array transmitters operating at high power present additional safety considerations---such as laser eye-safety zones, radiation exposure limits for crewed missions, and electromagnetic compatibility with other spacecraft systems. Ensuring reliable fail-safes, advanced beam-interruption protocols, and robust spacecraft-to-spacecraft coordination will be essential to prevent inadvertent illumination or interference, particularly near critical infrastructure or human-tended habitats.
\end{itemize}

To translate our framework into mission-ready hardware, three focused developments are recommended: 
\begin{enumerate}
\item A flight-like end-to-end demonstration using an effective transmitter aperture \(D_{\mathrm{eff}}\!\sim\!1\)--\(2~\mathrm{m}\) and \(P_{\mathrm{tx}}\!\sim\!0.5\)--\(1~\mathrm{kW}\) optical from LLO or NRHO to a \(\sim\!1~\mathrm{m}^2\) surface receiver would directly validate geometric capture efficiency \(\eta_{\mathrm{cap}}\), quantify the dependence on pointing jitter and residual misalignment, assess dust-induced throughput degradation, and verify radiator performance under relevant thermal boundary conditions.
\item  The parametric sizing approach developed here should be extended into an integrated mass--energy trade, combining photovoltaic generation, laser modules and drivers, radiator area, structural penalties, energy storage, and GNC into a single relation of total system mass versus continuous-equivalent surface power \(P_{\mathrm{cont}}\equiv E_{\mathrm{delivered}}/24~\mathrm{h}\) at a specified site and visibility geometry.
\item  A \(\lambda=1.55~\mu\mathrm{m}\) performance should be produced by recomputing the key performance surfaces with receiver conversion efficiencies \(\eta_{\text{rx}}\) appropriate to extended-bandgap devices; this will quantify the delivered-power penalty relative to the \(1064~\mathrm{nm}\) baseline while capturing eye-safety and stray-light advantages that are critical near crewed assets.
\end{enumerate}

Laser power beaming stands poised to transform lunar surface exploration and settlement by alleviating the longstanding challenges of extended lunar nights and permanently shadowed regions. In particular, phased-array transmitter architectures enable larger effective apertures that can deliver high-density, precisely targeted energy across varied orbital distances, supporting continuous operations in areas that would otherwise be power-starved. This scalable technology---from single satellites to multi-element constellations---provides a direct pathway to sustaining missions involving robotic prospecting, resource extraction, and human habitation. 

Placed against the alternatives, the present link provides quantitative decision rules rather than a universal preference. PV dominates at well-insolated sites where storage depth is shallow; compact fission dominates when continuous multi-kW baseload is required at fixed locations; laser beaming is preferred where darkness or shadow makes $\Delta E$ large, where loads are distributed or duty-cycled, or where a shared transmitter can be amortized across multiple consumers. On the same daily-energy scale the maps reported here identify $(D_{\rm eff},P_{\rm tx},A_{\rm rx},N)$ that achieve a target $E_{\rm day}$ for a given terrain mask and pointing budget; Eqs.~(\ref{eq:mpv})–(\ref{eq:mfis}) then provide a transparent check of whether the resulting $M_{\mathrm{LPB}}$ is competitive with $M_{\mathrm{PV+bat}}$ or with a reactor characterized by $\alpha_{\mathrm{fis}}$. This framing turns the abstract scaling ($\propto D_{\rm eff}^{2}$ and $\propto P_{\rm tx}$, with a multiplicative visibility factor) into site-specific design choices and clarifies when laser beaming is the mass-efficient option for near-term lunar operations.

As both government agencies and commercial stakeholders intensify their focus on the Moon, the technical and economic insights presented here will be pivotal in guiding future mission architectures, technology roadmaps, and operational planning. By demonstrating robust end-to-end feasibility and spotlighting phased arrays' ability to mitigate beam divergence, this research underscores laser power beaming’s potential to become a cornerstone of cislunar development and long-duration human presence on the lunar surface.

\section*{Acknowledgments}

The work described here was carried out the Jet Propulsion Laboratory, California Institute of Technology, Pasadena, California, under a contract with the National Aeronautics and Space Administration. Pre-decisional information for planning and discussion purposes only.
 

%

\appendix

\section{Illustrative Example: Low Lunar Orbit (LLO)}
\label{sec:illustrative_example}

Our LLO geometry and duty-cycle assumptions are comparable to prior lunar PSR beaming scenarios~\cite{Landis2020LunarLaser}.
To illustrate the interplay of our equations and parameters in determining the final electrical power delivered to the lunar surface, we present a numerical example. Table~\ref{tab:example_params} lists the key parameters used.

\begin{table}[ht]
\centering
\caption{Key parameters for the illustrative laser power beaming scenario in Low Lunar Orbit (LLO).}
\label{tab:example_params}
\begin{tabular}{ll}
\hline
{Parameter} & {Value / Description} \\
\hline\hline
Spacecraft solar array area, \(A_{\text{SA}}\) & \SI{10}{m^2} \\
Solar array efficiency, \(\eta_{\text{pv}}\) & 0.30 \\
Spacecraft ancillary power, \(P_{\text{SC,aux}}\) & \SI{200}{W} \\
Laser conversion efficiency, \(\eta_{\ell}\) & 0.40 \\
Laser wavelength, \(\lambda\) & \(1\,\mu\text{m}\) \\
Aperture diameter, \(D\) & \SI{0.1}{m} \\
Beam quality factor, \(M^2\) & 1.2 \\
Practical half-angle beam divergence, \(\theta_{\text{tx}}\) & \(7.64\times10^{-6}\,\text{rad}\) (\ref{eq:theta_practical}) \\
Pointing efficiency, \(\eta_{\text{point}}\) & 0.90 \\
Optical path efficiency, \(\eta_{\text{opt}}\) & 0.95 \\
Receiver area, \(A_{\text{rx}}\) & \SI{0.2}{m^2} \\
Receiver photonic conversion efficiency, \(\eta_{\text{rx}}\) & 0.50 \\
Lunar orbit altitude, \(h\) & \SI{100}{km} \\
Approximate slant range, \(d(t)\) & \(\sim\SI{200}{km}\) (near overhead) \\
\hline
\end{tabular}
\end{table}

Assume a nominal solar irradiance of 
\(I_{\text{sun}} = \SI{1361}{W/m^2}\) 
and normal incidence (\(\theta_{\text{inc}}(t)=0^\circ\)) on the spacecraft’s solar array.  
From Eq.~\eqref{eq:P_array}, the instantaneous power produced by the array is
\[
P_{\text{array}}(t) = I_{\text{sun}} \, A_{\text{SA}} \, \eta_{\text{pv}} 
= 1361 \times 10 \times 0.30 \approx \SI{4083}{W}.
\]
This value represents the ideal full-sun output, without accounting for eclipses or partial Sun angles.

The net power available for laser transmission is found by subtracting the spacecraft’s ancillary load \(P_{\text{SC,aux}}\) and applying the laser conversion efficiency \(\eta_\ell\).  
Using Eq.~\eqref{eq:p_tx}, the transmitted optical power is
\[
P_{\text{tx}}(t) = \big(P_{\text{array}}(t) - P_{\text{SC,aux}}\big)\,\eta_{\ell} 
= (4083 - 200)\times 0.40 \approx \SI{1553}{W}.
\]
Thus, the spacecraft can emit approximately \SI{1.55}{kW} of optical power under these conditions.

Assume the spacecraft is at a slant range of 
$
d(t) = \SI{200}{km} = 2\times10^5\,\text{m}.
$
With a practical beam divergence of 
$
\theta_{\text{tx}} = 7.64\times10^{-6}\,\text{rad}
$ from Eq.~\eqref{eq:theta_practical},
the beam radius at the receiver (Eq.~(\ref{eq:weff_def}) without jitter) is
$
w(t) = \theta_{\text{tx}}\, d(t) \approx 7.64\times10^{-6}\,\text{rad} \times 2\times10^5\,\text{m} \approx 1.53\,\text{m}.
$
The corresponding far-field spot area is
$
A_{\text{spot}}(t) = \pi \,[w(t)]^2 \approx \pi \times (1.53\,\text{m})^2 \approx \SI{7.34}{m^2}.
$
If we initially ignore losses, the raw flux density would be
\[
\frac{P_{\text{tx}}}{A_{\text{spot}}(t)} = \frac{1553}{7.34} \approx \SI{211.6}{W/m^2}.
\]
Accounting for the pointing efficiency \(\eta_{\text{point}}(t)\) and optical path efficiency \(\eta_{\text{opt}}(t)\), the effective power flux becomes
\[
I_{\text{eff}}(t) = \eta_{\text{point}}(t) \,\eta_{\text{opt}}(t) \,\frac{P_{\text{tx}}(t)}{A_{\text{spot}}(t)} \approx 0.9 \times 0.95 \times 211.6 \approx \SI{181}{W/m^2}.
\]
Thus, roughly \(\SI{181}{W/m^2}\) reaches the surface within the \(\SI{7.34}{m^2}\) beam footprint.

The receiver intercepts only a fraction of the beam proportional to its area, \(A_{\text{rx}} = \SI{0.2}{m^2}\). From Eq.~\eqref{eq:P_rx_elec_time}, the optical power collected is
$
 I_{\text{eff}}(t) \times A_{\text{rx}}.
$
After conversion by the receiver (with efficiency \(\eta_{\text{rx}} = 0.50\)), the net received electrical power is
\[
P_{\text{rx,elec}}(t) = \eta_{\text{rx}} \, P_{\text{rx,opt}}(t) \approx 0.50 \times 36.2 \approx \SI{18.1}{W}.
\]
Under these nominal conditions, the user on the lunar surface would receive approximately \SI{18}{W} of electrical power.

Assuming the \SI{18.1}{W} of electrical power is available continuously during a visibility window of \(\SI{15}{minutes} = 900\,\text{s}\), the total energy accumulated in the batteries is
\[
E_{\text{accumulated}} = P_{\text{rx,elec}} \times t_{\text{visibility}} \approx 18.1\,\text{W} \times 900\,\text{s} \approx \SI{16290}{J} \approx \SI{4.5}{Wh}.
\]
This stored energy can supply power during intervals when the laser is not directly targeting the receiver.

Although \(\SI{18.1}{W}\) of electrical power may seem modest, several approaches can scale the delivered power upward, namely
(1) Increasing the area  \(A_{\text{SA}}\) or efficiency \(\eta_{\text{pv}}\) of the spacecraft’s solar array.
(2) Improving  the laser’s conversion efficiency \(\eta_{\ell}\)  or using a larger transmit aperture \(D_{\mathrm{eff}}\) (which reduces \(\theta_{\text{tx}}\)).
(3) Improving pointing accuracy (\(\eta_{\text{point}}(t)\)) or optical transmission (\(\eta_{\text{opt}}(t)\)).
(4) Expanding the surface receiver area \(A_{\text{rx}}\)  or incorporating optical concentration methods.
(5) Decreasing the orbital altitude or slant range \(d(t)\), thereby reducing the spot size  \(w(t)\) and increasing flux density.
Furthermore, the power received over an entire visibility window (e.g., 10--15 minutes per orbital pass) can be accumulated in on-site batteries or capacitors for later use. By dynamically adjusting system parameters and leveraging advancements in efficiency, the delivered power can be scaled to meet mission-specific demands.

\begin{table}[ht]
\centering
\caption{Illustrative comparison of a single 0.10\,m aperture vs.\ a 1.0\,m phased array, 
         both emitting \(\sim1.55\,\mathrm{kW}\) at a 200\,km slant range (top-hat model). 
         A ``Phased Array (Realistic)'' column ensures the captured optical power 
         cannot exceed 
         $P_{\mathrm{tx}}\times\eta_{\mathrm{point}}\times\eta_{\mathrm{opt}}$ 
         when the beam spot is fully contained by the 0.20\,m\(^2\) receiver. 
         For that column, Eq.~(\ref{eq:Gauss_rx_elec-large}) is used to compute 
         the relevant values for $r_{\mathrm{rx}}\gtrsim w(t)$.}
\label{tab:single_vs_array_updated}
\renewcommand{\arraystretch}{1.1}
\begin{tabular}{lccc}
\hline
Parameter 
 & Single Aperture
 & Phased Array (Naive)
 & Phased Array (Realistic) \\
\hline\hline
Laser power, \(P_{\mathrm{tx}}\) (W) 
 & 1550 
 & 1550 
 & 1550 \\

Aperture diameter, \(D_{\mathrm{eff}}\) (m) 
 & 0.10 
 & 1.00 
 & 1.00 \\

Beam quality factor, \(M^2_{\mathrm{eff}}\) 
 & 1.2 
 & 1.2 
 & 1.2 \\

Half-angle divergence, \(\theta_{\mathrm{tx}}\) (rad) 
 & \(7.64\times10^{-6}\) 
 & \(7.64\times10^{-7}\) 
 & \(7.64\times10^{-7}\) \\

Slant range, \(d\) (m) 
 & \(2.0\times10^5\) 
 & \(2.0\times10^5\) 
 & \(2.0\times10^5\) \\

Spot radius, \(w = \theta_{\mathrm{tx}}\,d\) (m) 
 & 1.53 
 & 0.153 
 & 0.153 \\

Spot area, \(A_{\mathrm{spot}} = \pi\,w^2\) (m\(^2\)) 
 & 7.34 
 & 0.0734 
 & 0.0734 \\

Pointing/optical efficiency, \(\eta_{\mathrm{point}}\times\eta_{\mathrm{opt}}\) 
 & 0.855 
 & 0.855 
 & 0.855 \\

Effective flux, \(I_{\mathrm{eff}}\) (W/m\(^2\))\(^\dagger\) 
 & 180 
 & \(1.8\times10^4\) 
 & \(1.8\times10^4\) \\

Receiver area, \(A_{\mathrm{rx}}\) (m\(^2\)) 
 & 0.20 
 & 0.20 
 & 0.20 \\

Optical power on \(A_{\mathrm{rx}}\) (W)\footnotemark[1]
 & 36 
 & 3600 
 & \(\leq 1324\)\footnotemark[2] \\

Receiver efficiency, \(\eta_{\mathrm{rx}}\) 
 & 0.50 
 & 0.50 
 & 0.50 \\

Net electrical power, \(P_{\mathrm{rx,elec}}\) (W) 
 & 18 
 & 1800 
 & \(\leq 662\) \\
\hline
\end{tabular}
\vspace{1ex}
\begin{tabular}{p{0.96\textwidth}}
\footnotesize
\(^\dagger\)\,Effective flux is 
\(
I_{\mathrm{eff}} 
= (\eta_{\text{point}}\times\eta_{\text{opt}})
\,P_{\text{tx}}\,/\,A_{\mathrm{spot}}.
\)
We take \(\eta_{\mathrm{point}}=0.90\) and \(\eta_{\mathrm{opt}}=0.95\), giving 0.855 overall.\\[-11ex]
\footnotetext[1]{Naive multiplication of local flux by 0.20\,m\(^2\). If the beam footprint is $<0.20$\,m\(^2\), this would exceed the total beam power, which is physically impossible.}\\
\footnotetext[2]{Capped at $(1550\,\mathrm{W})\times0.855=1324\,\mathrm{W}$ optical. Multiplying by \(\eta_{\mathrm{rx}}=0.50\) yields net \(\leq662\,\mathrm{W}\).}
\end{tabular}
\end{table}

To highlight the impact of a larger effective aperture on the received power, 
Table~\ref{tab:single_vs_array_updated} compares two transmitters that each deliver about 1.55\,kW of optical power at a slant range of 200\,km. One system uses a small single 0.10\,m aperture, whereas the other is a phased array with an effective diameter of 1.0\,m. Both systems share the same beam quality factor ($M^2_{\mathrm{eff}}=1.2$) and identical  pointing/optical path efficiencies. In the “Phased Array (Naive)” column, we simply multiply the local flux by the receiver area -- even if that artificially exceeds the total beam power when the beam spot is smaller than the receiver. By contrast, the “Phased Array (Realistic)” column caps the optical power at $P_{\mathrm{tx}}\,(\eta_{\mathrm{point}}\eta_{\mathrm{opt}})$  if the beam is entirely contained within the 0.20\,m\(^2\) receiver.

{Key observations include:}
(1) A tenfold increase in transmitter diameter shrinks the beam spot area by a factor of 100, boosting the surface flux more than a hundredfold at the same orbital distance.
(2) Even after pointing/optical losses, the phased array can deliver up to hundreds of watts or even kilowatts to a small receiver, versus only tens of watts for the 0.10\,m aperture.
(3) This demonstrates how increasing \(D_{\mathrm{eff}}\) is among the most powerful ways to reduce beam divergence and fulfill higher power demands on the lunar surface.

Thus, simply moving from a small single-aperture telescope to a larger phased array can yield a substantial net electrical power gain at the same laser output, primarily by cutting down the beam footprint.

\section{Illustrative Example: Spacecraft at L1}
\label{sec:illustrative_example_L1}

For distant orbits, delivered power is diffraction-limited unless aperture and transmitter power scale substantially, consistent with~\cite{Brandhorst2009Acta}.
To illustrate how the model equations apply, we consider a spacecraft operating in a quasi-static halo/Lissajous orbit about the Earth--Moon L1 point, beaming power to a receiver on the lunar equator. While such an orbit provides continuous solar illumination and a stable line-of-sight geometry, the large slant range ($\sim 5.8\times 10^{7}$\,m) makes the link strongly diffraction-limited, so significant delivered flux requires very large apertures. Table~\ref{tab:example_params_L1} lists the key parameters for this scenario.

\begin{table}[ht]
\centering
\caption{Key parameters for the laser power beaming scenario at L1.}
\label{tab:example_params_L1}
\begin{tabular}{ll}
\hline
Parameter & Value / Description \\
\hline\hline
Spacecraft solar array area, \(A_{\text{SA}}\) 
& \SI{50}{m^2} \\
Solar array efficiency, \(\eta_{\text{pv}}\) 
& 0.35 \\
Spacecraft ancillary power, \(P_{\text{SC,aux}}\) 
& \SI{300}{W} \\
Laser conversion efficiency, \(\eta_{\ell}\) 
& 0.50 \\
Laser wavelength, \(\lambda\) 
& \(1\,\mu\text{m}\) \\
Aperture diameter, \(D\) 
& \SI{0.2}{m} \\
Beam quality factor, \(M^2\) 
& 1.2 \\
Practical half-angle divergence, \(\theta_{\text{tx}}\) 
& \(3.82\times10^{-6}\,\text{rad}\) \\
Pointing efficiency, \(\eta_{\text{point}}\) 
& 0.95 \\
Optical path efficiency, \(\eta_{\text{opt}}\) 
& 0.97 \\
Receiver area, \(A_{\text{rx}}\) 
& \SI{1.0}{m^2} \\
Receiver photonic conversion efficiency, \(\eta_{\text{rx}}\) 
& 0.60 \\
Slant range, \(d(t)\) 
& \(5.8\times10^7\,\text{m}\) \\
\hline
\end{tabular}
\end{table}

Assume a nominal solar irradiance of 
$
I_{\text{sun}} = \SI{1361}{W/m^2},
$
with normal incidence \(\theta_{\text{inc}}(t)=0^\circ\) on the spacecraft’s solar array. From Eq.~\eqref{eq:P_array}, the instantaneous power produced is
\[
P_{\text{array}}(t) = I_{\text{sun}}\, A_{\text{SA}}\, \eta_{\text{pv}}
= 1361 \times 50 \times 0.35
\;\approx\; \SI{23817.5}{W}.
\]
This value represents the ideal full-sun output, assuming optimal orientation and no eclipses.

The net power available for laser transmission is found by subtracting the spacecraft’s ancillary load \(P_{\text{SC,aux}}\) and applying the laser conversion efficiency \(\eta_{\ell}\). From Eq.~\eqref{eq:p_tx}:
\begin{equation}
P_{\text{tx}}(t)
= \Big(P_{\text{array}}(t) - P_{\text{SC,aux}}\Big)\eta_{\ell}
= (23817.5 - 300)\times 0.5
\;\approx\; \SI{11758.75}{W}.
\end{equation}
Hence, the spacecraft can emit \(\sim\SI{11.76}{kW}\) of optical power under these conditions.

At L1, the spacecraft is at a slant range of
$
d(t) = 58\times10^3 \,\text{km} 
= 5.8\times10^7\,\text{m}.
$
With a practical beam divergence of
$
\theta_{\text{tx}} \approx 3.82\times10^{-6}\,\text{rad},
$
the beam radius at the receiver is
$
w(t) = \theta_{\text{tx}}\, d(t)
\approx 3.82\times10^{-6}\,\text{rad}\,\times 5.8\times10^7\,\text{m} 
\;\approx\; 221.56\,\text{m}.
$
The corresponding far-field spot area is
$
A_{\text{spot}}(t) 
= \pi\,\bigl[w(t)\bigr]^2 
\;\approx\;\pi \times \bigl(221.56\,\text{m}\bigr)^2
\;\approx\;\SI{1.55e5}{m^2}.
$
Ignoring further losses, the raw flux density is
\[
\frac{P_{\text{tx}}}{A_{\text{spot}}(t)}
\;\approx\; \frac{11758.75}{1.55\times10^5}
\;\approx\; \SI{0.076}{W/m^2}.
\]
Accounting for the pointing and optical path efficiencies,
\[
I_{\text{eff}}(t) 
= \eta_{\text{point}}(t)\,\eta_{\text{opt}}(t)\,\frac{P_{\text{tx}}(t)}{A_{\text{spot}}(t)}
\;\approx\; 0.95 \times 0.97 \times 0.076
\;\approx\; \SI{0.070}{W/m^2}.
\]

With a receiver area \(A_{\text{rx}}= \SI{1.0}{m^2}\),
the optical power arriving on the receiver is
$
P_{\text{rx,opt}}(t) 
= I_{\text{eff}}(t)\,\times\,A_{\text{rx}}
\;\approx\; 0.070\,\text{W/m}^2 \times 1.0\,\text{m}^2
\;\approx\;\SI{0.070}{W}.
$
After conversion by the receiver (\(\eta_{\text{rx}}=0.60\)), the net electrical power is
\[
P_{\text{rx,elec}}(t) 
= \eta_{\text{rx}}\times P_{\text{rx,opt}}(t)
\;\approx\; 0.60 \times 0.070
\;\approx\;\SI{0.042}{W}.
\]

Assume the \SI{0.042}{W} of electrical power is delivered continuously during a visibility window of
$
t_{\text{visibility}}
= 12\,\text{hours}
= 43200\,\text{s}.
$
The total energy accumulated is then
\[
E_{\text{accumulated}}
= P_{\text{rx,elec}}(t)\,\times\,t_{\text{visibility}}
\;\approx\; 0.042\,\text{W}\times43200\,\text{s}
\;\approx\;\SI{1814}{J}\;\approx\;\SI{0.50}{Wh}.
\]
This stored energy can supply power during periods when direct beaming is unavailable. 

Although \(\SI{0.042}{W}\) (42\,mW) of electrical power is low, several strategies can enhance performance: 
(1)  Increasing the transmitter aperture \(D\) (or employing a phased array) to reduce beam divergence and thus raise flux density.
(2)  Improving the laser conversion efficiency \(\eta_{\ell}\) or optical path efficiency \(\eta_{\text{opt}}\) to minimize losses.
(3) Expanding the receiver area \(A_{\text{rx}}\) or using optical concentrators to intercept more of the beam’s energy.
(4)  Reducing the slant range \(d(t)\) by choosing an orbit closer to the lunar surface (e.g., NRHO) to mitigate beam spreading.
Even at L1’s large distance, a sufficiently large aperture or phased array design could substantially increase the power delivered to the lunar surface, albeit at higher engineering complexity.

Table~\ref{tab:L1_single_vs_array} compares a small single-aperture transmitter (0.20\,m diameter) with a larger phased-array transmitter (2.00\,m effective diameter), each emitting approximately 11.76\,kW under the same beam-quality and efficiency assumptions at L1. While the underlying calculations are straightforward, they highlight how increasing the transmitter aperture by an order of magnitude reduces the far-field spot area by two orders of magnitude, thereby boosting the flux and resulting net power at the receiver.

\begin{table}[ht]
  \centering
  \renewcommand{\arraystretch}{1.0}
  \caption{Illustrative comparison of single-aperture vs.\ phased-array transmitters at L1. 
           Both deliver \(\sim 11.76\)\,kW but differ in effective diameter. 
           The beam quality factor is \(M^2 = 1.2\). 
           Pointing and optical efficiencies multiply to \(\eta_{\rm point} \times \eta_{\rm opt} \approx 0.92\).}
  \label{tab:L1_single_vs_array}
  \begin{tabular}{lcc}
    \hline
    Parameter & Single Aperture & Phased Array \\
    \hline\hline
    Transmitted power, \(P_{\mathrm{tx}}\) (W) &
    11760 &
    11760 \\
    Aperture diameter, \(D_{\mathrm{eff}}\) (m) &
    0.20 &
    2.00 \\
    Beam quality factor, \(M^2_{\rm eff}\) &
    1.2 &
    1.2 \\
    Half-angle divergence, \(\theta_{\mathrm{tx}}\)\(^\dagger\) (rad) &
    \(3.82\times10^{-6}\) &
    \(3.82\times10^{-7}\) \\
    Slant range, \(d\) (m) &
    \(5.8\times10^{7}\) &
    \(5.8\times10^{7}\) \\
    Spot radius, \(w = \theta_{\mathrm{tx}} \, d\) (m) &
    \(222\) &
    \(22.2\) \\
    Spot area, \(A_{\mathrm{spot}} = \pi\,w^2\) (m\(^2\)) &
    \(1.55\times10^5\) &
    \(1.55\times10^3\) \\
    Effective flux, \(I_{\mathrm{eff}}\)\(^\ddagger\) (W/m\(^2\)) &
    \(0.07\) &
    \(7.0\) \\
    Receiver area, \(A_{\mathrm{rx}}\) (m\(^2\)) &
    1.0 &
    1.0 \\
    Optical power on \(A_{\mathrm{rx}}\) (W) &
    \(0.07\) &
    \(7.0\) \\
    Receiver efficiency, \(\eta_{\mathrm{rx}}\) &
    0.60 &
    0.60 \\
    Net electrical power, \(P_{\mathrm{rx,elec}}\) (W) &
    \(0.042\) &
    \(4.2\) \\
    \hline
  \end{tabular}
  \vspace{1ex}
  \begin{tabular}{p{0.96\textwidth}}
    \footnotesize
    \(^\dagger\)\,Using \(\theta_{\mathrm{tx}} = M^2 \,\bigl(2\,\lambda\bigr)\big/\bigl(\pi\,D_{\mathrm{eff}}\bigr)\) with \(\lambda=1.0\times 10^{-6}\,\mathrm{m}\). \\
    \(^\ddagger\)\,Calculated as \(\bigl(\eta_{\rm point} \times \eta_{\rm opt}\bigr)\times \bigl(P_{\mathrm{tx}} / A_{\mathrm{spot}}\bigr)\).
  \end{tabular}
\end{table}

Key observations for the E-M L1 case are as follows:
(1) A 0.20\,m aperture delivers a flux of only \(\sim0.07\,\mathrm{W/m^2}\) at L1, whereas scaling \(D_{\mathrm{eff}}\) to 2.00\,m raises that flux to about \(7.0\,\mathrm{W/m^2}\) (a factor of 100).
(2) Although L1’s distance is very large, increasing the aperture from 0.20\,m to 2.00\,m elevates the net electrical power on a 1\,m\(^2\) receiver from milliwatts to a few watts.
(3) While maintaining a coherent phased array at 2\,m diameter in L1 orbit is a significant engineering task, the fundamental advantage of reduced beam divergence---and thus higher flux on the surface---remains clear.

This example underscores the key role of aperture size for long-range power beaming. Even when absolute power levels are relatively small at such extreme distances, a larger effective diameter dramatically improves the final power delivered to a lunar receiver.

\section{Illustrative Example: Near-Rectilinear Halo Orbit (NRHO)}
\label{sec:illustrative_example_nrho}

Orbit-to-surface feasibility and networked small-lander concepts have been examined in~\cite{OlesonLandis2024OrbitBeaming}; the present model extends these with time-resolved delivered-energy maps. To demonstrate how the various equations and parameters interact to determine the final electrical power delivered to the lunar surface, we examine a concrete numerical scenario for a spacecraft in a NRHO transmitting power to a receiver located near the lunar South Pole. Table~\ref{tab:example_params_nrho} summarizes the key parameters assumed in this example.

\begin{table}[ht]
\centering
\caption{Key parameters for the illustrative laser power beaming scenario in an NRHO.}
\label{tab:example_params_nrho}
\begin{tabular}{ll}
\hline
{Parameter} & {Value / Description} \\
\hline\hline
Spacecraft solar array area, \(A_{\text{SA}}\) & \SI{20}{m^2} \\
Solar array efficiency, \(\eta_{\text{pv}}\) & 0.35 \\
Spacecraft ancillary power, \(P_{\text{SC,aux}}\) & \SI{300}{W} \\
Laser conversion efficiency, \(\eta_{\ell}\) & 0.45 \\
Laser wavelength, \(\lambda\) & \(1\,\mu\text{m}\) \\
Aperture diameter, \(D\) & \SI{0.2}{m} \\
Beam quality factor, \(M^2\) & 1.2 \\
Practical half-angle divergence, \(\theta_{\text{tx}}\) & \(3.82\times10^{-6}\,\text{rad}\) \\
Pointing efficiency, \(\eta_{\text{point}}\) & 0.92 \\
Optical path efficiency, \(\eta_{\text{opt}}\) & 0.97 \\
Receiver area, \(A_{\text{rx}}\) & \SI{0.5}{m^2} \\
Receiver photonic conversion efficiency, \(\eta_{\text{rx}}\) & 0.55 \\
Lunar orbit altitude, \(h\) & \SI{1000}{km} \\
Approximate slant range, \(d(t)\) & \SI{1200}{km} (near the South Pole) \\
\hline
\end{tabular}
\end{table}

Assume a nominal solar irradiance of 
$
I_{\text{sun}} = \SI{1361}{W/m^2},
$
with normal incidence (\(\theta_{\text{inc}}(t)=0^\circ\)) on the spacecraft’s solar array. From Eq.~\eqref{eq:P_array}, the instantaneous power produced by the array is
\[
P_{\text{array}}(t) = I_{\text{sun}} \times A_{\text{SA}} \times \eta_{\text{pv}} = 1361 \times 20 \times 0.35 \approx \SI{9527}{W}.
\]
This value represents the ideal full-Sun output.

Subtracting the spacecraft’s ancillary load,
$
P_{\text{net}}(t) = P_{\text{array}}(t) - P_{\text{SC,aux}} = 9527 - 300 \approx \SI{9227}{W},
$
and applying the laser conversion efficiency \(\eta_{\ell}\) (from Eq.~\eqref{eq:p_tx}), the transmitted optical power is
\[
P_{\text{tx}}(t) = \Big(P_{\text{array}}(t) - P_{\text{SC,aux}}\Big)\eta_{\ell} \approx 9227 \times 0.45 \approx \SI{4152}{W}.
\]
Thus, the spacecraft emits approximately \SI{4.15}{kW} of optical power.

Assume the spacecraft is at a slant range of 
$
d(t) = 1200~{\rm km} = 1.2\times10^6\,\text{m}.
$
With a practical divergence of 
$
\theta_{\text{tx}} = 3.82\times10^{-6}\,\text{rad},
$
the beam radius at the receiver is
$
w(t) = \theta_{\text{tx}}\, d(t)  \approx 4.58\,\text{m}.
$
The far-field spot area is then
$
A_{\text{spot}}(t) = \pi\, [w(t)]^2 \approx \pi \times (4.58\,\text{m})^2 \approx \SI{66.0}{m^2}.
$
Ignoring losses, the raw flux density is
\[
\frac{P_{\text{tx}}}{A_{\text{spot}}(t)} \approx \frac{4152}{66.0} \approx \SI{62.9}{W/m^2}.
\]
After accounting for the pointing and optical path efficiencies,
\[
I_{\text{eff}}(t) = \eta_{\text{point}}(t) \,\eta_{\text{opt}}(t) \, \frac{P_{\text{tx}}(t)}{A_{\text{spot}}(t)} \approx 0.92 \times 0.97 \times 62.9 \approx \SI{56.1}{W/m^2}.
\]
Thus, roughly \SI{56.1}{W/m^2} reaches the surface within the \SI{66.0}{m^2} beam footprint.

The receiver, with \(A_{\text{rx}} = \SI{0.5}{m^2}\), intercepts only a fraction of the beam. The optical power collected is
$
P_{\text{rx,opt}}(t) = I_{\text{eff}}(t) \times A_{\text{rx}} \approx \SI{28.1}{W}.
$
After conversion by the receiver (\(\eta_{\text{rx}} = 0.55\)), the net electrical power is
\[
P_{\text{rx,elec}}(t) = \eta_{\text{rx}} \times P_{\text{rx,opt}}(t) \approx 0.55 \times 28.1 \approx \SI{15.4}{W}.
\]
Under these nominal conditions, the user on the lunar surface would receive approximately \SI{15.4}{W} of electrical power.

Assuming the \(\SI{15.4}{W}\) of electrical power is delivered continuously during a visibility window of \(t_{\text{visibility}} = 12\,\text{hours} = 43200\,\text{s}\), the total energy accumulated is
\[
E_{\text{accumulated}} = P_{\text{rx,elec}}(t) \times t_{\text{visibility}} \approx 15.4\,\text{W} \times 43200\,\text{s} \approx \SI{665280}{J} \approx \SI{184.8}{Wh}.
\]
This stored energy can power systems during spacecraft outages or eclipses.

As a result, although \SI{15.4}{W} of electrical power may seem modest, several strategies can enhance performance:
(1) Increasing the transmitter aperture \(D\) (or using multiple transmitters) to reduce beam divergence and increase the flux density at the receiver.
(2)  Improving the laser conversion efficiency \(\eta_{\ell}\) and optical path efficiency \(\eta_{\text{opt}}(t)\) to minimize losses.
(3)  Expanding the receiver area \(A_{\text{rx}}\) or using optical concentrators to capture more energy.
(4) Reducing the slant range \(d(t)\) by adjusting the orbital configuration or targeting surface locations with a more favorable geometry.

Furthermore, power received over an entire visibility window (e.g., 12 hours for polar regions in an NRHO) can be integrated and stored in on-site batteries or capacitors, providing a continuous power supply during periods of non-beaming. By dynamically optimizing system parameters and leveraging advancements in efficiency, the delivered power can be scaled to meet mission-specific demands.

Table~\ref{tab:NRHO_single_vs_array} compares a small single-aperture transmitter (\(D_{\text{eff}} = 0.20\,\text{m}\)) with a larger phased-array transmitter (\(D_{\text{eff}} = 2.00\,\text{m}\)), each emitting  \(\approx4.15\,\text{kW}\) of optical power from a NRHO at a slant range \(d = 1.2\times10^{6}\,\text{m}\). Both transmitters share identical beam quality (\(M^{2}_{\rm eff} = 1.2\)) and combined pointing and optical efficiencies (\(\eta_{\text{point}} \times \eta_{\text{opt}} \approx 0.89\)). Increasing the effective aperture diameter by a factor of ten reduces the beam's far-field spot area by two orders of magnitude, significantly enhancing the flux density and net electrical power at the lunar surface receiver.

\begin{table}[ht]
  \centering
  \caption{Comparison of single-aperture vs phased-array transmitters in NRHO. Both systems transmit  \(\simeq4.15\,\text{kW}\), but differ in aperture size. Beam quality factor is \(M^{2}_{\rm eff} = 1.2\), with combined pointing and optical efficiencies \(\eta_{\text{point}}\times\eta_{\text{opt}}\approx0.89\).}
  \label{tab:NRHO_single_vs_array}
  \begin{tabular}{lcc}
    \hline
    Parameter & Single Aperture & Phased Array \\
    \hline\hline
    Transmitted power, \(P_{\text{tx}}\) [W] & \(4150\) & \(4150\) \\
    Effective aperture diameter, \(D_{\text{eff}}\) [m] & \(0.20\) & \(2.00\) \\
    Beam quality factor, \(M^{2}_{\rm eff}\) & \(1.2\) & \(1.2\) \\
    Half-angle divergence, \(\theta_{\text{tx}}\) [rad]\(^\dagger\) & \(3.82\times10^{-6}\) & \(3.82\times10^{-7}\) \\
    Slant range, \(d\) [m] & \(1.2\times10^{6}\) & \(1.2\times10^{6}\) \\
    Spot radius, \(w = \theta_{\text{tx}}\,d\) [m] & \(4.58\) & \(0.458\) \\
    Spot area, \(A_{\text{spot}} = \pi w^{2}\) [m\(^2\)] & \(65.9\) & \(0.659\) \\
    Effective flux, \(I_{\text{eff}}\) [W/m\(^2\)]\(^\ddagger\) & \(56.2\) & \(5619\) \\
    Receiver area, \(A_{\text{rx}}\) [m\(^2\)] & \(0.5\) & \(0.5\) \\
    Optical power on receiver, \(P_{\text{rx,opt}}=I_{\text{eff}}A_{\text{rx}}\) [W] & \(28.1\) & \(2810\) \\
    Receiver efficiency, \(\eta_{\text{rx}}\) & \(0.55\) & \(0.55\) \\
    Net electrical power, \(P_{\text{rx,elec}}=\eta_{\text{rx}}P_{\text{rx,opt}}\) [W] & \(15.5\) & \(1546\) \\
    \hline
  \end{tabular}

  \vspace{1ex}
  \footnotesize
  \begin{tabular}{p{0.95\textwidth}}
    \(^\dagger\)\,Half-angle divergence calculated as \(\theta_{\text{tx}} = M^{2}\bigl({2\lambda}/{\pi D_{\text{eff}}}\bigr)\), with laser wavelength \(\lambda=1.0\times10^{-6}\,\text{m}\).\\[0.5ex]
    \(^\ddagger\)\,Effective flux calculated by \(I_{\text{eff}}=\left(\eta_{\text{point}}\times \eta_{\text{opt}}\right)\times \bigl({P_{\text{tx}}}/{A_{\text{spot}}}\bigr)\), using \(\eta_{\text{point}}=0.92\), \(\eta_{\text{opt}}=0.97\).
  \end{tabular}
\end{table}

Key observations for the NRHO case include:
(1)  A \(0.20\,\text{m}\) aperture produces an effective flux of approximately \(56\,\text{W/m}^{2}\), delivering roughly \(15\,\text{W}\) of net electrical power to a lunar receiver of area \(0.5\,\text{m}^{2}\).
(2)  Increasing the effective aperture to \(2.00\,\text{m}\) reduces the far-field spot area by  a factor of \(100\). Thus, the lunar surface flux increases dramatically to  \(\sim5600\,\text{W/m}^{2}\), enhancing the net electrical power received to  \(\sim1540\,\text{W}\).
(3)  While implementing a coherent phased array with a \(2.00\,\text{m}\) effective diameter at NRHO involves substantial engineering complexity (e.g., maintaining beam coherence and alignment across multiple sub-apertures), the fundamental advantage of reduced divergence---and thus increased received flux---remains significant.

This illustrative example highlights the considerable advantage of employing phased-array transmitters with large effective apertures for lunar laser power beaming missions, greatly surpassing the performance achievable by conventional single-aperture designs.

\end{document}